\documentclass[prc,preprint,superscriptaddress,showpacs,nofootinbib,tightenlines,amsmath]{revtex4}
\usepackage{graphics}
\usepackage{epsfig}
\usepackage{slashed}

\begin{document}
\title{Nucleon-to-Delta transition form factors in chiral effective field theory using the complex-mass scheme}
\author{M.~Hilt}
\author{T.~Bauer}
\author{S.~Scherer}
\author{L.~Tiator}
\affiliation{Institut f\"ur Kernphysik, Johannes Gutenberg-Universit\"at, D-55099 Mainz, Germany}

\date{December 24, 2017}
\preprint{MITP/17-105}
\begin{abstract}
   We calculate the form factors of the electromagnetic nucleon-to-$\Delta$-resonance
transition to third chiral order in manifestly Lorentz-invariant chiral effective field theory.
   For the purpose of generating a systematic power counting, the complex-mass scheme is applied
in combination with the small-scale expansion.
   We fit the results to available empirical data.
\end{abstract}

\maketitle
\section{Introduction}
   The $\Delta$(1232) resonance was discovered in $\pi^+p$ scattering in the early 1950s \cite{Anderson:1952nw},
and is the most prominent nucleon excitation.
   It is the lowest-lying resonance with spin and isospin quantum numbers 3/2.
   The $\Delta$(1232) is only about 300 MeV heavier than the nucleon and, due to the strong
coupling to the $\pi N$ channel, has a broad Breit-Wigner width of around 117 MeV \cite{Agashe:2014kda},
resulting in a lifetime of the order of $10^{-23}$ s.
   This makes direct measurements of its properties more complex than in the case of the nucleon.
   Above the pion production threshold, the $\Delta$(1232) dominates many processes involving the strong
interactions.
   Besides pion-nucleon scattering, important examples are pion photo- and electroproduction,
where the $\Delta$(1232) can be created as an intermediate state through the excitation of
the nucleon by a real or virtual photon.

   The form factors of the electromagnetic $\gamma N \rightarrow\Delta$ transition have been
the subject of numerous investigations from both the experimental side
\cite{Bartel:1968tw,Baetzner:1972bg,Stein:1975yy,Beck:1999ge,Pospischil:2000ad,Mertz:1999hp,Joo:2001tw,Sparveris:2004jn,
Elsner:2005cz,Kelly:2005jy,Stave:2008aa,Aznauryan:2009mx,Blomberg:2015zma}
and the theoretical side \cite{Dufner:1967yj,Jones:1972ky,Davidson:1985wb,Wirzba:1986sc,Bermuth:1988ms,Leinweber:1992pv,Butler:1993ht,Cardarelli:1995ug,
Buchmann:1996bd,Lu:1996rj,Gellas:1998wx,Tiator:2003xr,Tiator:2003uu,Alexandrou:2004xn,Pascalutsa:2005ts,Gail:2005gz,Braun:2005be,
Pascalutsa:2006up,Ramalho:2008dp,Alexandrou:2010uk,Tiator:2016btt} (see also the reviews on resonance excitations by
Tiator et al.~\cite{Tiator:2011pw} and by Aznauryan and Burkert \cite{Aznauryan:2011qj}).
   In this work, we will present a calculation of the $\gamma N\rightarrow\Delta$ transition
form factors in the framework of chiral effective field theory (ChEFT).
   In contrast to previous work in ChEFT, we put particular emphasis on the determination of
the form factors at the complex pole, i.e., we treat the $\Delta(1232)$ as an unstable particle \cite{Gegelia:2009py}.
   For that purpose, we combine a covariant description of the $\Delta(1232)$ resonance \cite{Hacker:2005fh,Wies:2006rv}
with the complex-mass scheme (CMS) (see Refs.~\cite{Stuart:1990,Denner:1999gp,Denner:2006ic,Actis:2006rc,Actis:2008uh}).
   The CMS was originally developed for deriving properties of $W$, $Z$, and Higgs bosons obtained from resonant
processes.
   More recently, it has also been applied in the context of effective field theories of the strong interactions
\cite{Djukanovic:2009zn,Djukanovic:2009gt,Bauer:2012at,Djukanovic:2013mka,Bauer:2014cqa,Djukanovic:2014rua,Djukanovic:2015gna,
Epelbaum:2015vea,Yao:2016vbz,Gegelia:2016pjm,Bauer:2016czc}.
   The CMS has the important property of perturbative unitarity, which was first demonstrated in Ref.~\cite{Bauer:2012gn}
in terms of a simple model and was later proven on general grounds in Ref.~\cite{Denner:2014zga}.
   In our calculation, a consistent power counting is implemented in terms of the small-scale expansion \cite{Hemmert:1997ye} in combination
with the CMS, treating the width (in the chiral limit) as a small quantity.
   For the interaction of pions with nucleons and the Delta resonance, we make use of the manifestly
consistent interaction Lagrangian of Ref.~\cite{Wies:2006rv}, which was obtained from a Dirac constraint analysis
\cite{Dirac,Gitman:1990qh,teitelboim}.

   This article is organized as follows.
   In Sec.~II, we introduce the $\gamma N \rightarrow\Delta$
transition process and discuss how it is related to pion electroproduction.
   In Sec.~III, we present the effective Lagrangians we used.
   Section~IV contains a discussion of the complex-mass scheme and the power counting.
   In Sec.~V, we calculate the transition form factors and show our
results.
   Section V contains a short summary.

\section{From pion electroproduction to the $\gamma N\rightarrow\Delta$ transition form factors}
\begin{figure}[htbp]
    \centering
        \includegraphics[width=0.8\textwidth]{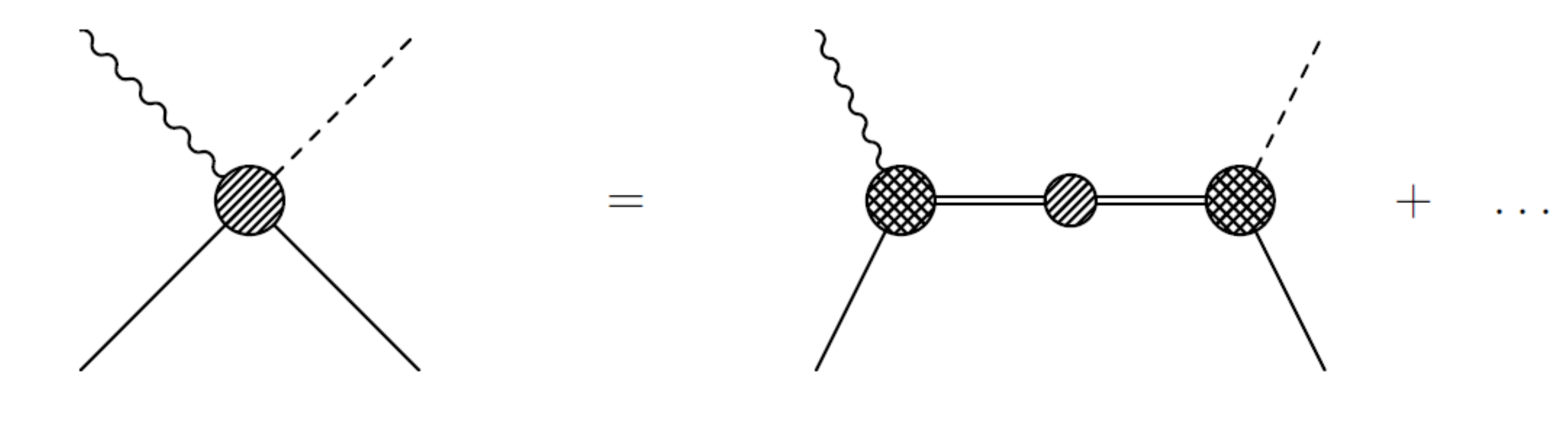}
    \caption{At $s\approx m_\Delta^2$,
the process is dominated by the $s$-channel pole diagram due to the propagator of the $\Delta$(1232).
The nucleon is represented by a single line, the $\Delta$(1232) by a double line, the photon by a wiggly line, and the pion by a dashed line.
The circles represent dressed vertices.}
    \label{fig:pionelectro}
\end{figure}

   The $\Delta$(1232) is an unstable particle with a very short lifetime of the order of $10^{-23}$ s.
   Therefore, a process like
\begin{equation}
\gamma+N\rightarrow\Delta
\end{equation}
cannot be described by an ``ordinary'' matrix element, because the $\Delta$(1232) is not
an asymptotic state of the strong interactions.
   This means that stable one-particle states $|\Delta(p)\rangle$
with $p^2=m_\Delta^2$ do not exist  \cite{Bjorken}.
(Of course, it is possible to study a
{\it theoretical} situation, where the sum of the nucleon and pion masses is larger than
the $\Delta$ mass, resulting in a stable $\Delta$ state.)
   However, one can investigate a complete scattering amplitude, where the $\Delta$(1232) contributes
as an intermediate ``state.''
   To be specific, we consider pion electroproduction on the nucleon below the two-pion production
threshold.
   There, the only possible process triggered by the nucleon
absorbing a (virtual) photon and involving the strong interactions is
\begin{equation}
\gamma^\ast+N\rightarrow N+\pi,
\end{equation}
where $\pi$ represents the corresponding pion.
   For kinematical conditions such that the square root of the Mandelstam variable $s$ is in the vicinity of
the complex pole position,
$$z_\Delta=m_\Delta-i\,\frac{\Gamma_\Delta}{2},$$
the process is dominated by the propagation of a $\Delta$ resonance in the $s$ channel
(see Fig.~\ref{fig:pionelectro}).

   One now parametrizes the contribution of the unstable
$\Delta$(1232) and defines the form factors in analogy to a stable
particle.
   For an unstable particle such as the $\Delta$(1232), ``on-shell kinematics'' are given by the complex pole position.

   Before addressing the Lorentz structure of the vertex, let us discuss the
isopin structure.
   Neglecting the contributions due to heavier quarks, the
electromagnetic current operator is given by
\begin{equation}
J^\mu(x)=\frac{2}{3}\bar{u}(x)\gamma^\mu
u(x)-\frac{1}{3}\bar{d}(x)\gamma^\mu
d(x)=\bar{q}(x)\left(\frac{1}{6}+\frac{\tau_3}{2}\right)\gamma^\mu
q(x)=J^{\mu(0)}_0+J^{\mu(1)}_0,
\end{equation}
where $J^{\mu(0)}_0$ and $J^{\mu(1)}_0$ denote the isoscalar and
isovector components, respectively.
   The interaction with an external electromagnetic four-vector potential
$\mathcal{A}_\mu$ reads
\begin{equation}
\mathcal{L}_{\textrm{e.m.}}=-e J^\mu \mathcal{A}_\mu,
\end{equation}
where $e>0$ denotes the proton charge.
   The isoscalar current cannot induce a transition from isospin $1/2$ to isospin $3/2$.
   Thus, the isospin structure of the transition matrix element is given by \cite{Nozawa:1989gy}
\begin{displaymath}
\langle 3/2, \tau_\Delta|J^\mu|1/2,\tau\rangle=(1/2,\tau;1,0|3/2,\tau_\Delta)
\langle 3/2||J^{\mu(1)}||1/2\rangle.
\end{displaymath}
   Here, $\langle 3/2||J^{\mu(1)}||1/2\rangle$ denotes the reduced matrix element, and
the value of the Clebsch-Gordan coefficient $(1/2,\tau;1,0|3/2,\tau_\Delta)$ is $\sqrt{2/3}$ for both $\gamma p\to \Delta^+$
and $\gamma n\to \Delta^0$ transitions.

\begin{figure}[htbp]
    \centering
        \includegraphics[width=0.2\textwidth]{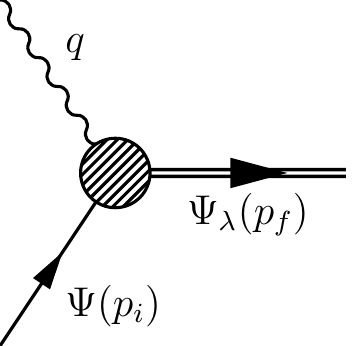}
    \caption{$\gamma N\rightarrow\Delta$ vertex:
    The momentum of the incoming nucleon is called $p_i$, $q$ and $p_f$ denote the momenta of the photon and of the outgoing $\Delta$(1232),
respectively.}
    \label{fig:diagrammklasse}
\end{figure}

  Let us now turn to the Lorentz structure.
  According to Ref.~\cite{Gegelia:2009py}, which describes a method for extracting from the general
vertex only those pieces surviving at the pole, the matrix element of the $\gamma N\rightarrow\Delta$
transition (see Fig.\ \ref{fig:diagrammklasse}) can be written as
\begin{equation}
{\cal M}=-ie\langle \Delta(p_f,s_f)|J^\mu(0)|N(p_i,s_i)\rangle\epsilon_\mu
=-ie\sqrt{\frac{2}{3}}\bar{w}_\lambda(p_f,s_f)\Gamma^{\lambda\mu}u(p_i,s_i)\epsilon_\mu.
\label{eqn:ffmatrix}
\end{equation}
   Here, the initial nucleon is described by the Dirac spinor $u(p_i,s_i)$ with real
mass $m_N$ and $p_i^2=m_N^2$, the final $\Delta$(1232) is described via the Rarita-Schwinger
vector-spinor $\bar{w}_\lambda(p_f,s_f)$  \cite{Rarita:1941mf,Kusaka} with a complex mass $z_\Delta$ and $p_f^2=z_\Delta^2$,
and the photon via the polarization vector $\epsilon_\mu$.
   In the following, it is always understood that the ``tensor'' $\Gamma^{\lambda\mu}$ is evaluated
between on-shell spinors $u$ and $\bar{w}_\lambda$, satisfying
\begin{align}
\label{dirac}
\slashed{p}_i u(p_i,s_i)&=m_N u(p_i,s_i),\\
\label{rarita_schwinger}
\bar{w}_\lambda(p_f,s_f)\slashed{p}_f&=z_\Delta \bar{w}_\lambda(p_f,s_f),\quad
\bar{w}_\lambda(p_f,s_f)\gamma^\lambda=0,\quad
\bar{w}_\lambda(p_f,s_f)p_f^\lambda=0.
\end{align}
   The expressions for a stable $\Delta$ resonance are obtained via the replacement
$z_\Delta\to m_\Delta$.
   The ``tensor'' $\Gamma^{\lambda\mu}$ contains a superposition of four Lorentz
tensors, which we choose as \cite{Mathews:1965zz}
\begin{equation}
\Gamma^{\lambda\mu}=i\left[D_1(q^2)g^{\lambda\mu}+D_2(q^2)q^\lambda p_i^\mu
+D_3(q^2) q^\lambda q^\mu+D_4(q^2)q^\lambda\gamma^\mu\right]\gamma_5,
\label{eqn:ff4tensors}
\end{equation}
where $q=p_f-p_i$.\footnote{For the $\gamma$ matrices we make use of the convention
of Ref.~\cite{Itzykson:1980rh}, in particular, $\gamma_5=\gamma^5=i\gamma^0\gamma^1\gamma^2\gamma^3$.
The overall factor of $i$ in Eq.~(\ref{eqn:ff4tensors}) is introduced for convenience to compensate for the different convention
for $\gamma_5$ used in Ref.~\cite{Jones:1972ky}.}
   At the pole, current conservation leads to
\begin{equation}
q_\mu\Gamma^{\lambda\mu}=0,
\label{eqn:cc}
\end{equation}
providing the additional constraint
\begin{equation}
\label{eqn:constrain}
D_1(q^2)+p_i\cdot q D_2(q^2)+q^2D_3(q^2)+(z_\Delta+m_N)D_4(q^2)=0.
\end{equation}
   This equation has been used both numerically and analytically as an important check of the explicit calculation.
   Using current conservation, the tensor
$\Gamma^{\lambda\mu}$ can be expressed in terms of three invariant functions,
\begin{equation}
\Gamma^{\lambda\mu}=G_1(q^2)L_1^{\lambda\mu}+G_2(q^2)L_2^{\lambda\mu}+G_3(q^2)L_3^{\lambda\mu}.
\label{eqn:mathews}
\end{equation}
   A possible set of independent structures is given by \cite{Jones:1972ky}
\begin{align}
L_1^{\lambda\mu}&=i\left(q^\lambda\gamma^\mu-\slashed{q}g^{\lambda\mu}\right)\gamma_5,\nonumber\\
L_2^{\lambda\mu}&=i\left(q^\lambda P^\mu-q\cdot P g^{\lambda\mu}\right)\gamma_5,\nonumber\\
L_3^{\lambda\mu}&=i\left(q^\lambda q^\mu-q^2 g^{\lambda\mu}\right)\gamma_5,
\label{eqn:lorentzstructure}
\end{align}
where $P=(p_i+p_f)/2$ and $q_\mu L_i^{\lambda\mu}=0$.\footnote{
The structures of Ref.~\cite{Jones:1972ky} differ from those of Ref.~\cite{Mathews:1965zz}
by the use of $P$ instead of $p_f$.}
   The invariant functions $G_i$ ($i=1,2,3$) are related to the functions $D_j$ ($j=1,2,3,4$)
by
\begin{align*}
\label{GDrel}
G_1&=D_4,\\
G_2&=D_2,\\
G_3&=D_3-\frac{1}{2}D_2.
\end{align*}
   Recall that the four $D_j$ are constrained by Eq.~(\ref{eqn:constrain}).
   The parametrization most widely used for $\Gamma^{\lambda\mu}$ is the one by Jones and Scadron \cite{Jones:1972ky}.
   Introducing
\begin{displaymath}
\epsilon_{\mu\nu}(a,b)=\epsilon_{\mu\nu\rho\sigma}a^\rho b^\sigma,\quad\epsilon_{0123}=1,
\end{displaymath}
the tensor $\Gamma^{\lambda\mu}$ is written as
\begin{equation}
\Gamma^{\lambda\mu}=G_M(Q^2)K_M^{\lambda\mu}+G_E(Q^2)K_E^{\lambda\mu}+G_C(Q^2)K_C^{\lambda\mu},
\label{eqn:ffjs}
\end{equation}
with\footnote{When replacing $z_\Delta$ by $m_\Delta$ for a stable particle, our convention agrees with Eqs.~(3.3a)--(3.3c)
of Ref.~\cite{Leinweber:1992pv}. Note, in particular, the imaginary unit in the second and third equation \cite{Leinweber:1992pv}.}
\begin{align}
K_M^{\lambda\mu}&=-3\left[(z_\Delta+m_N)^2-q^2\right]^{-1}\epsilon^{\lambda\mu}(P,q)\frac{z_\Delta+m_N}{2m_N},\\
K_E^{\lambda\mu}&=-K_M^{\lambda\mu}-12i\Delta^{-1}(q^2)\epsilon^\lambda{}_\sigma(P,q)\epsilon^{\mu\sigma}(p_f,q)\gamma_5\frac{z_\Delta+m_N}{2m_N},\\
K_C^{\lambda\mu}&=-6i\Delta^{-1}(q^2)q^\lambda\left[q^2P^\mu-q\cdot Pq^\mu\right]\gamma_5\frac{z_\Delta+m_N}{2m_N},
\end{align}
where
\begin{equation}
\label{Deltaq2}
\Delta(q^2)=[(z_\Delta+m_N)^2-q^2][(z_\Delta-m_N)^2-q^2].
\end{equation}
   The form factors $G_M$, $G_E$, and $G_C$ are referred to as magnetic dipole, electric quadrupole, and
Coulomb quadrupole form factors, respectively.\footnote{Following
common practice, we choose $Q^2\equiv -q^2$ as the argument of the form factors.}
   These form factors are related to the $G_i$ of Eq.~(\ref{eqn:mathews}) by \cite{Jones:1972ky} (see App.~\ref{relation_form_factors})
\begin{align}
\label{eqn:formfactorsGM}
G_M(Q^2)&=\frac{m_N}{3(z_\Delta+m_N)}\Big\{\left[(3z_\Delta+m_N)(z_\Delta+m_N)-q^2\right]\frac{G_1(q^2)}{z_\Delta}\nonumber\\
&\quad+\left(z_\Delta^2-m_N^2\right)G_2(q^2)+2q^2G_3(q^2)\Big\},\\
\label{eqn:formfactorsGE}
G_E(Q^2)&=\frac{m_N}{3(z_\Delta+m_N)}\Big[\left(z_\Delta^2-m_N^2+q^2\right)\frac{G_1(q^2)}{z_\Delta}+\left(z_\Delta^2-m_N^2\right)G_2(q^2)+2q^2G_3(q^2)\Big],\\
\label{eqn:formfactorsGC}
G_C(Q^2)&=\frac{m_N}{3(z_\Delta+m_N)}\Big[4z_\Delta G_1(q^2)+\left(3z_\Delta^2+m_N^2-q^2\right)G_2(q^2)+2\left(z_\Delta^2-m_N^2+q^2\right)G_3(q^2)\Big].
\end{align}
    We emphasize that equations such as (\ref{eqn:formfactorsGM})--(\ref{eqn:formfactorsGC}) involve the complex
$\Delta(1232)$ pole position $z_\Delta$ rather than the real (Breit-Wigner) masses.

\section{Effective Lagrangian}
   The effective Lagrangian $\mathcal{L}_{\rm eff}$ consists of a purely mesonic, a pion-nucleon, a pion-$\Delta$, and
a $\pi N\Delta$ Lagrangian,
\begin{equation}
\mathcal{L}_{\rm eff}=\mathcal{L}_\pi+\mathcal{L}_{\pi N}+ \mathcal{L}_{\pi\Delta}+\mathcal{L}_{\pi N\Delta},
\end{equation}
each of which is organized in a combined derivative and quark-mass expansion
(see, e.g., Refs. \cite{Scherer:2002tk,Scherer:2012zzd} for an introduction).
   In fact, from the mesonic Lagrangian we only need the lowest-order term \cite{Gasser:1983yg}:
\begin{equation}
\mathcal{L}^{(2)}_{\pi}=\frac{F^2}{4}\textnormal{Tr}\left[D_\mu U(D^\mu U)^\dagger\right]+\frac{F^2}{4}\textnormal{Tr}\left(\chi U^\dagger+U\chi^\dagger\right).
\label{eqn:pionlagrangian}
\end{equation}
   Here and in the following equations, superscripts refer to the chiral order of the respective Lagrangians.
   The pion fields are contained in the unimodular, unitary, $(2\times 2)$ matrix $U$:
\begin{equation}
\begin{split}
U(x)&=u^2(x)=\textnormal{exp}\left(i\frac{\Phi(x)}{F}\right),\\
\Phi(x)&=\sum_{i=1}^3\tau_i \phi_i(x)=
\left(\begin{array}{cc}\pi^0(x) & \sqrt{2}\pi^+(x)\\ \sqrt{2}\pi^-(x)&-\pi^0(x)\end{array}\right),
\end{split}
\label{eqn:pionmatrix}
\end{equation}
   where $F$ denotes the pion-decay constant in the chiral limit: $F_\pi=F[1+\mathcal{O}(\hat{m})]=92.2$ MeV
with $\hat m=m_u=m_d$ being the isospin-symmetric limit of the light-quark masses.
   Furthermore, $\chi=2B(s+ip)$ includes the quark masses as
$\chi=2B\hat{m}=M^2$, where $M^2$ is the squared pion mass at leading order in the
quark-mass expansion, and $B$ is related to the scalar singlet
quark condensate $\left\langle \bar{q}q\right\rangle_0$
in the chiral limit \cite{Gasser:1983yg,Colangelo:2001sp}.
   Finally, the interaction with an external electromagnetic four-vector potential ${\mathcal A}_\mu$ is generated
through the covariant derivative
\begin{displaymath}
D_\mu U=\partial_\mu U +i\frac{e}{2}{\cal A}_\mu[\tau_3, U].
\end{displaymath}

   Defining the nucleon isospin doublet
\begin{displaymath}
\Psi=\begin{pmatrix} p\\n\end{pmatrix}
\end{displaymath}
in terms of the two four-component Dirac fields $p$ and $n$ of the proton and the neutron,
respectively, the lowest-order pion-nucleon Lagrangian is given by (see Ref.~\cite{Gasser:1987rb} for details)
\begin{equation}
{\cal L}_{\pi N}^{(1)}=\bar{\Psi}\left(i\slashed{D}-m+\frac{\texttt{g}_A}{2}\gamma^\mu
\gamma_5 u_\mu \right)\Psi,
\label{eq:LpiN}
\end{equation}
with
\begin{equation}
\begin{split}
D_\mu \Psi&=\left(\partial_\mu+\Gamma_\mu-iv_\mu^{(s)}\right)\Psi,\\
\Gamma_\mu&=\frac{1}{2}\left[u^\dagger(\partial_\mu-ir_\mu)u+u(\partial_\mu-il_\mu)u^\dagger\right],\\
u_\mu&=i\left[u^\dagger\left(\partial_\mu-ir_\mu\right)u-u\left(\partial_\mu-il_\mu\right)u^\dagger\right],\\
\end{split}
\label{eqn:nuktrafo}
\end{equation}
where $ v_{\mu}^{(s)}=-e \mathcal{A}_\mu/2$ and $r_\mu=l_\mu=-e\tau_3{\cal A}_\mu/2$.
   In Eq.~(\ref{eq:LpiN}), $m$ and $\texttt{g}_A$ denote the chiral limit
of the physical nucleon mass and the axial-vector coupling constant,
respectively.

   The technical details concerning how we include the $\Delta$(1232) in BChPT can be found in Refs.~\cite{Hacker:2005fh,Wies:2006rv}.
   Here, we give only a short summary (see section 4.7 of Ref.~\cite{Scherer:2012zzd} for more details).
   As the $\Delta$(1232) is a particle with both spin and isospin equal to 3/2, it can be described via
a vector-spinor isovector-isospinor field with 96 components $\Psi_{\lambda,\alpha,i,r}$, where $\lambda$ denotes the
Lorentz-vector index, $\alpha$ the Dirac-spinor index, $i$ the isovector index, and $r$ the isospinor index.
   The most general first-order interaction Lagrangian for the $\Delta$(1232) in the chiral expansion depends on three
coupling constants $\texttt{g}_i$ \cite{Hemmert:1997ye} and a so-called ``off-shell parameter'' $A$ \cite{Moldauer:1956zz}.
   As one deals with a higher-spin system, one automatically introduces unphysical degrees of freedom due to the
coupling of spins ($\frac{1}{2}\otimes1=\frac{1}{2}\oplus\frac{3}{2}$).
   When analyzing the constraints to obtain the correct number of degrees of freedom, one ends up with relations among
the coupling constants, involving the parameter $A$.
   The Lagrangian is invariant under so-called ``point transformations'' (see Refs.~\cite{Hacker:2005fh} and \cite{Wies:2006rv}
for further details).
   As a result of the invariance property, physical quantities do not depend on $A$.
   Choosing $A=-1$ makes, e.g., the propagator of the $\Delta$(1232) simpler to deal with.
   For this particular choice, the leading-order Lagrangian reads
\begin{equation}
\mathcal{L}^{(1)}_{\pi\Delta}=\bar{\Psi}_\lambda\xi^\frac{3}{2}\Lambda^{(1)\lambda\nu}\xi^\frac{3}{2}\Psi_\nu,
\label{LpiDelta}
\end{equation}
where $\xi^\frac{3}{2}$ is a matrix representation of the projection operator for the isospin-$\frac{3}{2}$ component of the fields
with $\xi^\frac{3}{2}_{ij}=\delta_{ij}-\frac{1}{3}\tau_i\tau_j$, and
\begin{align}
\Lambda^{(1)\lambda\nu}&=-\Big[(i\slashed{D}-m_{\Delta})g^{\lambda\nu}
-i(\gamma^\lambda D^\nu+\gamma^\nu D^\lambda)\nonumber\\
&\quad+i\gamma^\lambda \slashed{D}\gamma^\nu
+m_{\Delta}\gamma^\lambda\gamma^\nu
+\frac{\texttt{g}_1}{2}\left(\slashed{u}g^{\lambda\nu}-\gamma^\lambda u^\nu-u^\lambda\gamma^\nu
+\gamma^\lambda \slashed{u}\gamma^\nu\right)\gamma_5\Big].\label{eqn:ldelta}
\end{align}
   The covariant derivative of the $\Delta$(1232) field is given by
\begin{equation}
D_\mu\Psi_i=\partial_\mu\Psi_i-2i\epsilon_{ijk}\Gamma_{\mu,k}\Psi_j+\Gamma_\mu\Psi_i-iv_\mu^{(s)}\Psi_i,
\end{equation}
where we have suppressed the Lorentz-vector and Dirac spinor indices as well as the isospinor index.
   Here, again, the pion fields and external sources are hidden in the definition of $v_\mu^{(s)}$, $u_\mu$ and
$\Gamma_\mu=\sum_{k=1}^3\Gamma_{\mu,k}\tau_k$.
   For a detailed discussion of $\mathcal{L}^{(1)}_{\pi\Delta}$, see Refs.~\cite{Hacker:2005fh,Wies:2006rv,Scherer:2012zzd}.
   The $\pi\Delta\Delta$ interaction is generated by the last term of Eq.~(\ref{eqn:ldelta}).
   For the $\pi N\Delta$ interaction at leading order, the Lagrangian reads
\begin{equation}
\mathcal{L}_{\pi N\Delta}^{(1)}=\texttt{g}\bar{\Psi}_{\lambda,i}\xi^\frac{3}{2}_{ij}(g^{\lambda\nu}-\gamma^\lambda\gamma^\nu)u_{\nu,j}\Psi+\text{H.c.},
\label{LpiNDelta}
\end{equation}
where H.c.~refers to the Hermitian conjugate.
   The Lagrangians of Eqs.~(\ref{eqn:pionlagrangian}), (\ref{eq:LpiN}), (\ref{LpiDelta}), and
(\ref{LpiNDelta}) contain in total seven low-energy constants: $F$ and $B$ from the mesonic
sector, $\texttt{g}_A$ from ${\cal L}_{\pi N}^{(1)}$, $\texttt{g}_1$ and $m_\Delta$ from $\mathcal{L}^{(1)}_{\pi\Delta}$,
and $\texttt{g}$ from the $\pi N\Delta$ interaction Lagrangian $\mathcal{L}_{\pi N\Delta}^{(1)}$.
   Strictly speaking, before renormalization all the fields and parameters should be regarded as
bare quantities which should be denoted by a symbol $B$ for {\it bare}.
   However, to keep the notation simple, we have deliberately omitted such an index.

\section{The complex-mass scheme and power counting}

   To have a consistent power counting, we apply the CMS, which may be regarded as an extension of the extended
on-mass-shell renormalization scheme \cite{Gegelia:1999gf,Gegelia:1999qt,Fuchs:2003qc} to unstable particles.
   This renormalization scheme is achieved by splitting the bare parameters
(and fields) of the Lagrangian into complex renormalized parameters
and counter terms.
   We choose the renormalized masses as the poles of the dressed propagators in the chiral limit:
\begin{align}
m_B & =  m+\delta m,\nonumber\\
m_{\Delta B} & = z_\Delta+\delta z_\Delta=m_{\Delta\chi}-i\,\frac{\Gamma_{\Delta\chi}}{2}+\delta z_\Delta.
\label{barerensplit}
\end{align}
   Here, $m_B$ and $m_{\Delta B}$ refer to the bare masses of the nucleon and $\Delta$ fields,
whereas $m$ is the mass of the nucleon in the chiral limit, and $z_{\Delta}$ is the complex pole
of the $\Delta$(1232) propagator in the chiral limit.
   We define the pole mass $m_{\Delta\chi}$ and the width
$\Gamma_{\Delta\chi}$ of the $\Delta$(1232) as the real part and $(-2)$ times the
imaginary part of the pole and assume $\Gamma_{\Delta\chi}$ to be small
in comparison to both $m_{\Delta\chi}$ and the scale of spontaneous
chiral symmetry breaking, $\Lambda_\chi=4\pi F$.
   We include the renormalized parameters $z_\Delta$ and $m$ in the free propagators
and treat the counter terms perturbatively.
   The renormalized couplings are chosen such that the corresponding counter terms
exactly cancel the power-counting-violating parts of the loop
diagrams.

   While the starting point is a Hermitian Lagrangian in terms of bare
parameters and fields, the CMS involves complex parameters in the
basic Lagrangian and complex counter terms.
   Applying generalized cutting rules for loop integrals involving propagators with
complex masses, it can be shown that unitarity is satisfied order by order in
perturbation theory \cite{Bauer:2012gn,Denner:2014zga}.
   In agreement with Ref.~\cite{Veltman:1963th}, the unitarity conditions are valid for an
$S$-matrix connecting stable states only.

   We organize our perturbative calculation by adopting the standard
power counting of Refs.~\cite{Weinberg:1991um,Ecker:1995gg} in
combination with the small-scale expansion of Ref.~\cite{Hemmert:1997ye} to the
renormalized diagrams, i.e., an interaction vertex obtained from an
${\cal O}(q^n)$ Lagrangian counts as order $q^n$, a pion propagator
as order $q^{-2}$, nucleon and $\Delta$(1232) propagators as order
$q^{-1}$, and the integration of a loop as order $q^4$.
   In addition, we assign the order $q$ to the difference between the $\Delta$(1232) mass
and the nucleon mass.
  In practice, we implement this scheme by subtracting the loop diagrams at complex ``on-mass-shell''
points in the chiral limit.

\begin{figure}[htbp]
    \centering
        \includegraphics[width=0.75\textwidth]{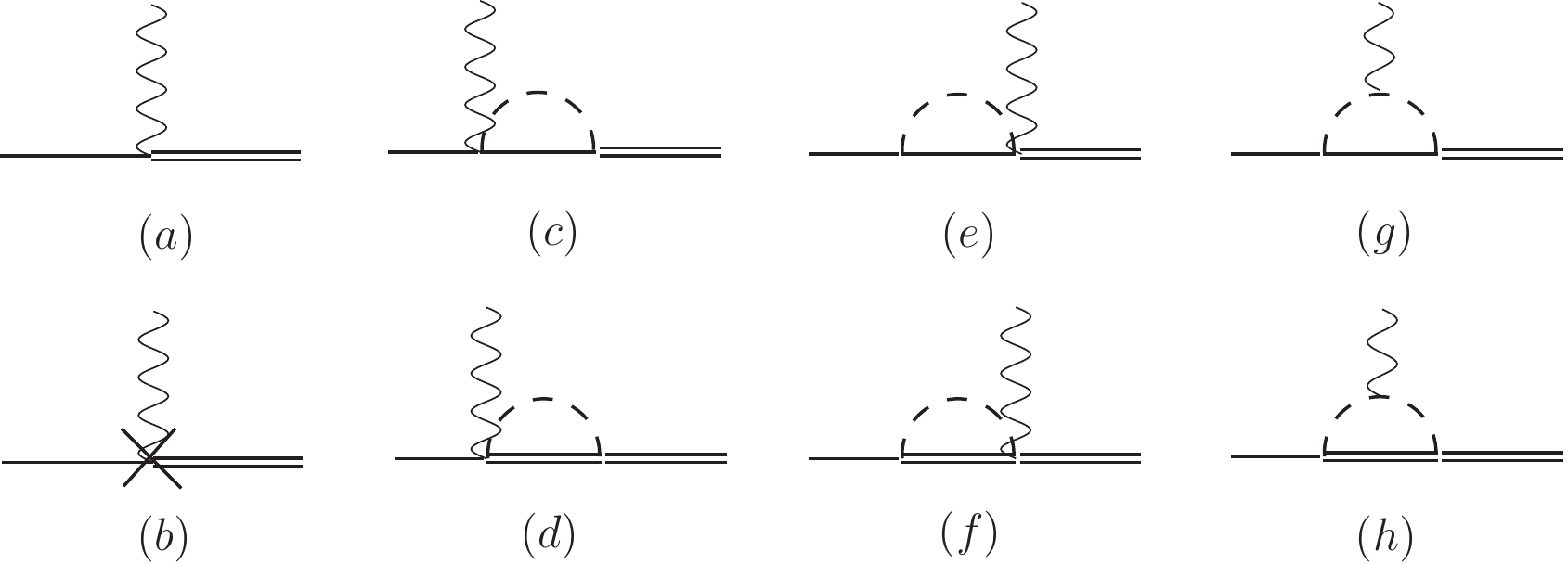}
    \caption{Contributions to the $\gamma N\rightarrow\Delta$ transition form factors up to and including
$\mathcal{O}(q^3)$.
   The vertices of the diagrams $(c)$--$(j)$ are always of the lowest possible order.
   Diagram $(a)$ represents the contact contributions and diagram $(b)$ the counter term.}
    \label{fig:diagrams}
\end{figure}

\begin{figure}[htbp]
    \centering
        \includegraphics[width=0.2\textwidth]{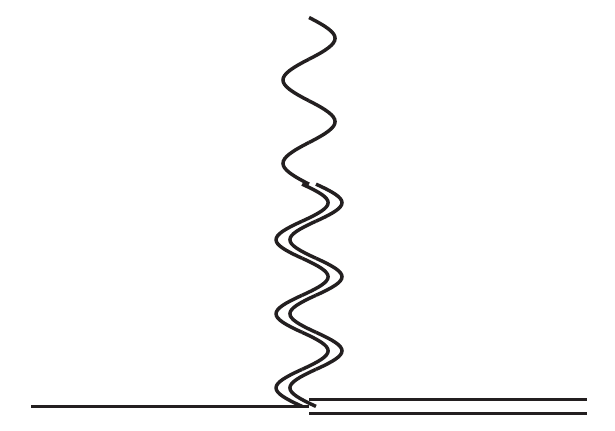}
    \caption{Contribution of the $\rho$ meson (wiggly double line) to the transition form factors in a semi-phenomenological approach.}
    \label{fig:rho}
\end{figure}

   Figure \ref{fig:diagrams} shows all diagrams contributing to the
$\gamma N\rightarrow\Delta$ transition form factors up to and
including chiral order three.
   At tree level, there is no diagram of ${\cal O}(q)$.
   Therefore, for a calculation at ${\cal O}(q^3)$, it is not necessary
to consider the wave-function-renormalization constants, because they are
of the form $\sqrt{Z}=1+{\cal O}(q^2)$ for both the nucleon and the $\Delta$.
   The product with the diagrams of Fig.~\ref{fig:diagrams} generates additional
terms at ${\cal O}(q^4)$, which are beyond the accuracy of our calculation.
   At tree level, only diagrams at chiral order two and
three contribute to the given process (see App.~\ref{tree1} for details).
   Our tree-level diagram contains three free parameters ($C_1^\gamma$, $C_2^\gamma$, and $C_3^\gamma$),
which we fit to experimental data.
   This procedure will be explained in the next section.
   After we calculated the diagrams of Fig.~\ref{fig:diagrams}, checked current conservation,
and fitted the free parameters to experimental data, it turned out that the results only
poorly described the data.
   To improve our results, we included a contribution of the $\rho$ meson at tree level
(see Fig.~\ref{fig:rho}) in a semi-phenomenological approach.
   For the details of this step, we refer to App.~\ref{tree2}.

\section{results\label{results}}

   Before addressing the numerical results of the one-loop calculation, let us discuss
some general features of the chiral expansion of the $\gamma N\Delta$ transition.
   At chiral order one, the Lagrangian does not contribute to the transition.
   Therefore, the tree-level contribution starts at ${\cal O}(q^2)$, and the renormalized
loop diagrams contribute at ${\cal O}(q^3)$.

      At ${\cal O}(q^2)$, the form factors are given in terms of a single coupling constant,
namely, $C_1^\gamma$:
\begin{align}
\label{eqn:formfactorsGM_treeLO}
G_M^{\text{tree}\,(2)}(Q^2)&=\frac{m_N}{3(z_\Delta+m_N)}\frac{(3z_\Delta+m_N)(z_\Delta+m_N)+Q^2}{z_\Delta}C_1^\gamma,\\
\label{eqn:formfactorsGE_treeLO}
G_E^{\text{tree}\,(2)}(Q^2)&=\frac{m_N}{3(z_\Delta+m_N)}\frac{z_\Delta^2-m_N^2-Q^2}{z_\Delta}C_1^\gamma,\\
\label{eqn:formfactorsGC_treeLO}
G_C^{\text{tree}\,(2)}(Q^2)&=\frac{m_N}{3(z_\Delta+m_N)}
4z_\Delta C_1^\gamma.
\end{align}
   The superscripts $(2)$ refer to chiral order 2.
   At the real-photon point, $Q^2=0$, Eqs.~(\ref{eqn:formfactorsGM_treeLO})--(\ref{eqn:formfactorsGC_treeLO})
entail model-independent predictions for the pole ratios $R_{EM}^{\text{pole}}$
and $R_{SM}^\text{pole}$ \cite{Tiator:2016btt}, namely,
\begin{align}
\label{REMO2}
R_{EM}^{\text{pole}\,(2)}(0)&=-\frac{G_E^{(2)}(0)}{G_M^{(2)}(0)}=-\frac{z_\Delta^2-m_N^2}{(3z_\Delta+m_N)(z_\Delta+m_N)},\\
\label{RSMO2}
R_{SM}^{\text{pole}\,(2)}(0)&=-\frac{z_\Delta^2-m_N^2}{4z_\Delta^2}\frac{G_C^{(2)}(0)}{G_M^{(2)}(0)}=R_{EM}^\text{pole\,(2)}(0).
\end{align}
   Note that these results remain intact even after the inclusion of the $\rho$ meson [see Eq.~(\ref{Cireplacement})].
   Using $z_\Delta=(1210-i\,50)$ MeV and $m_N=938$ MeV, one obtains from Eq.~(\ref{REMO2})
\begin{equation}
R_{EM}^{\text{pole}\,(2)}(0)=(-5.98+i\,0.90)\,\%.
\label{REMO2result}
\end{equation}

   The explicit expressions for the tree-level contributions to the form factors up to and including ${\cal O}(q^3)$
are given in Eqs.~(\ref{eqn:formfactorsGM_tree})--(\ref{eqn:formfactorsGC_tree}) and involve three parameters,
$C_1^\gamma$, $C_2^\gamma$, and $C_3^\gamma$.
   Given the fact that the loop contributions are fixed, once the coupling constants $\texttt{g}_A$,
$\texttt{g}_1$, and $\texttt{g}$ have been fixed (see discussion below and Table \ref{tab:constants}),
one might expect that the three $C_i^\gamma$ can be determined in terms of the empirical values
of the form factors at the real-photon point.
   However, this is not the case, as $C_1^\gamma$ and $C_3^\gamma$ always contribute in the
linear combination
\begin{displaymath}
C_1^\gamma+\frac{1}{2}\left(z_\Delta^2-m_N^2+Q^2\right)C_3^\gamma.
\end{displaymath}

   When calculating the loop contributions involving a Delta line in the
loop [see Figs.~3 $(d)$, $(f)$, and $(h)$], we neglect the width.
   This amounts to neglecting terms of ${\cal O}(\hbar^2)$, which are
beyond the accuracy of a one-loop calculation.

   In the following, we will distinguish between the transition form
factors at the Breit-Wigner position $W=m_\Delta=1232$~MeV on the
real (physical) energy axis and at the pole position
$W=(1210-i\,50)$~MeV in the lower half-plane of the second Riemann
sheet.
   The Breit-Wigner form factors are denoted by $G_M^*$,
$G_E^*$, and $G_C^*$, where the latter two are usually given as
ratios to the dominant magnetic form factor,
\begin{eqnarray}
\label{Eq:REM}
R_{EM}(Q^2) &=& -\frac{G_E^*(Q^2)}{G_M^*(Q^2)}\,,\\
\label{Eq:RSM}
R_{SM}(Q^2) &=&
-\frac{|\vec q_{\rm cm}|}{2m_\Delta}\frac{G_C^*(Q^2)}{G_M^*(Q^2)},
\end{eqnarray}
   where $\vec q_{\rm cm}$ denotes the three-momentum of the virtual photon
in the center-of-momentum frame.
   These form factors are real quantities and positive for $Q^2=0$, and
are related to the electromagnetic pion production multipoles
$M_{1+}^{(3/2)}$, $E_{1+}^{(3/2)}$, and $S_{1+}^{(3/2)}$ at the resonance
position.
   To determine the transition form factors at the Breit-Wigner position,
we make use of Eqs.~(\ref{eqn:formfactorsGM})--(\ref{eqn:formfactorsGC}) as
follows.
   We replace $z_\Delta$ by $m_\Delta$, make use of real coupling constants,
and consider only the real parts of the so-obtained expressions, i.e.,
we omit the imaginary parts of the loops.

   At the pole position in the complex plane, the form factors are
denoted by $G_M$, $G_E$, and $G_C$ and have complex values.
   Recently, data for such complex form factors have been determined from
the partial wave analyses of MAID and SAID~\cite{Tiator:2016btt}.
   In our calculation, these form factors are obtained by using the
complex Delta mass (pole position) and complex coupling constants.

   In Table \ref{tab:constants}, we collect the masses and coupling constants
which have been fixed from other sources and which are not considered as free
parameters in our calculation.
   The values for $M_\pi$, $m_N$, $m_\Delta$, $z_\Delta$,  $M_\rho$, $F_\pi$, and $\texttt{g}_A$ are taken
from the {\it Review of Particle Physics} \cite{Agashe:2014kda}.
   For $\texttt{g}$ we take $\texttt{g}=1.13$ as obtained from a fit to the $\Delta\to\pi N$ decay width \cite{Hacker:2005fh}.
   Furthermore, we make use of the quark-model estimate $\texttt{g}_1=\frac{9}{5} {\texttt g}_A=2.29$ \cite{Hemmert:1997ye}.
   Note that the quark-model estimate for $\texttt{g}$, namely, $\texttt{g}=\frac{3}{5}\sqrt{2} {\texttt g}_A=1.08$, is slightly smaller
than the empirical value.
\begin{table}[htbp]
    \centering
        \begin{tabular}{|c|c|c|c|c|c|c|c|c|}
        \hline
$M_\pi$ [GeV] & $m_N$ [GeV] &  $m_\Delta$[GeV] &  $z_\Delta$[GeV] & $M_\rho$[GeV] &$F_\pi$ [GeV] & $\texttt{g}_A$ & $\texttt{g}$ & $\texttt{g}_1$\\
\hline
0.140  & 0.938 & $1.232$ &$1.21-i\,0.05$ & 0.77 & 0.0922 & 1.27 & $1.13$ & 2.29\\
\hline

        \end{tabular}
        \caption{Masses and coupling constants which are not considered as free parameters in our calculation.}
        \label{tab:constants}

\end{table}

   To determine the unknown parameters of the tree-level diagrams, we
perform a simultaneous fit of all available experimental
data of $G_M^*$, $G_E^*$, and $G_C^*$, where the latter two were
taken from the ratios $R_{EM}$ and $R_{SM}$ (for values of
$Q^2=-q^2\leq 0.3\ \textnormal{GeV}^2$, i.e., the spacelike region).
   We refer to the results without the $\rho$ meson as Fits I and II, and to
the results including the $\rho$ as Fit III.
   In Fits I and III, we set $C_3^\gamma=0$.
   The results for the fitted constants ($C_i^\gamma$ and $C_i^\rho$)
are shown in Table \ref{tab:parameters}.
   Note that the coupling constants $C_i^\rho$ enter the calculation in the combination $C_i^\rho/g$,
with $g=5.91$ in terms of the Kawarabayashi-Suzuki-Riazuddin-Fayyazuddin relation \cite{Kawarabayashi:1966kd,Riazuddin:1966sw}
(see App.~\ref{tree2}).
   Equation (\ref{Cireplacement}) suggests that we should compare the values of $C_i^\gamma$ without the $\rho$ meson
with the combination $\tilde{C}_i=C_i^\gamma-C_i^\rho/g$ including the $\rho$ meson.
   In the present case we obtain $\tilde{C}_1=(1.91\pm 0.34)$~GeV$^{-1}$ and $\tilde{C}_2=(1.12\pm 0.36)$~GeV$^{-2}$.
   Taking the expansion scale to be of ${\cal O}(1\,\text{GeV})$, we find that the parameters
$C_i^\gamma$ and $\tilde{C}_i$ turn out to be of a natural size of order 1 GeV$^{-1}$ and 1 GeV$^{-2}$, respectively.

\begin{table}[htbp]
    \centering
        \begin{tabular}{|l|c|c|c|c|c|}
        \hline
 & $C_1^\gamma$ [GeV$^{-1}$] & $C_2^\gamma$ [GeV$^{-2}$]& $C_3^\gamma$ [GeV$^{-3}$]& $C_1^\rho$ [GeV$^{-1}$] & $C_2^\rho$ [GeV$^{-2}$]\\
\hline
Fit I & $1.01\pm0.07$ & $1.57\pm0.06$ & 0 & -- & -- \\
\hline
Fit II & $3.01\pm0.28$ & $1.59\pm0.04$ & $-4.73\pm0.65$ & -- & -- \\
\hline
Fit III &$-1.69\pm0.20$ & $2.91\pm0.21$ & 0 & $-21.3\pm1.6$ & $10.6\pm1.7$  \\
\hline

        \end{tabular}
        \caption{The results of the fitting procedure for the parameters of the tree-level contributions of diagram $(a)$ of Fig.~\ref{fig:diagrams}
are labeled $C_i^\gamma$. The results including the $\rho$ meson
contain, in addition, the parameters $C_i^\rho$ (see App.~\ref{tree} for
definitions). The errors are obtained from the fit to the form
factor data with $Q^2\le 0.3$~GeV$^2$.}
        \label{tab:parameters}

\end{table}

\begin{figure}[htbp]
    \centering
\includegraphics[height=4.5cm]{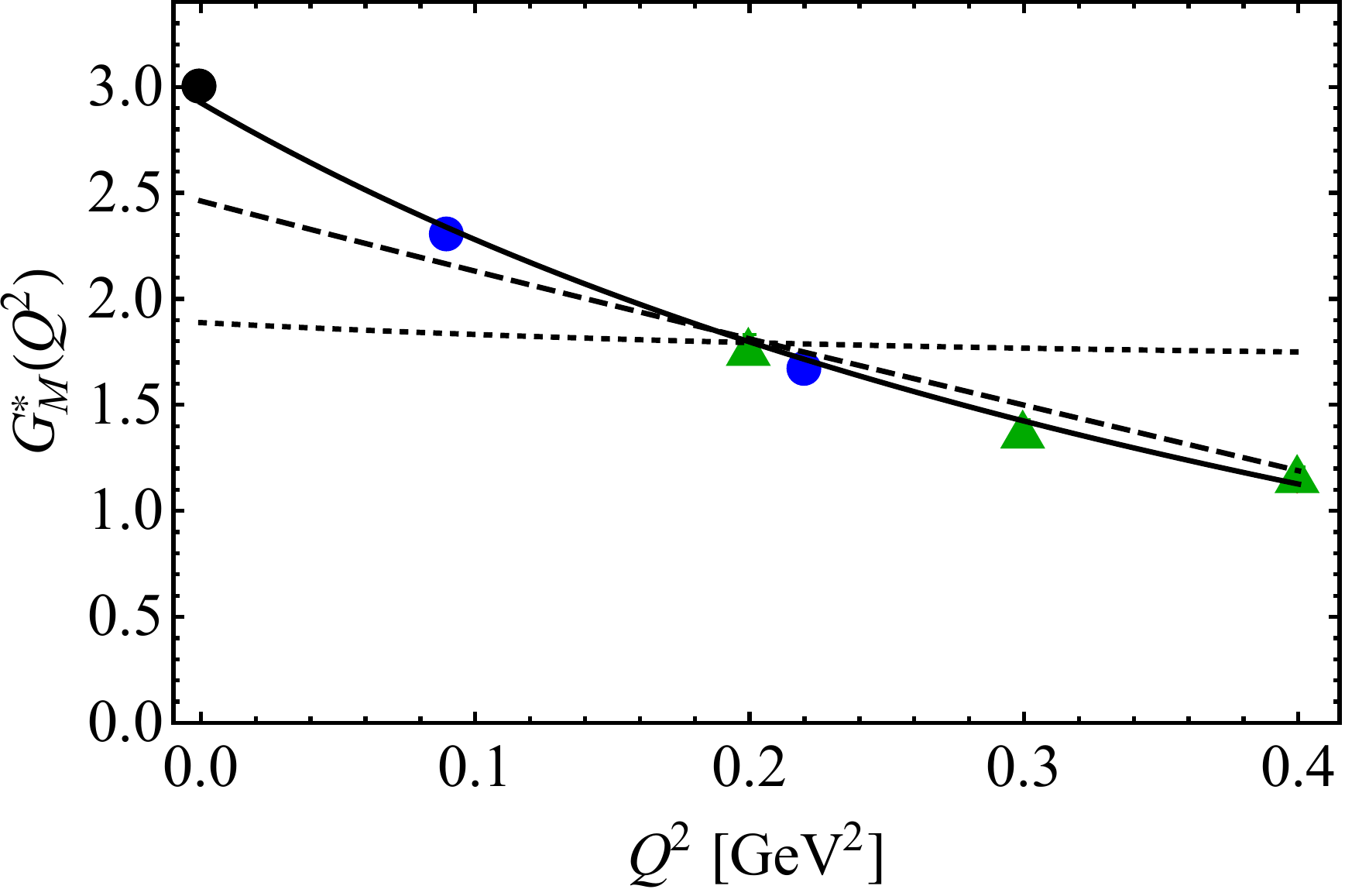}
\hspace*{0.5cm}
\includegraphics[height=4.5cm]{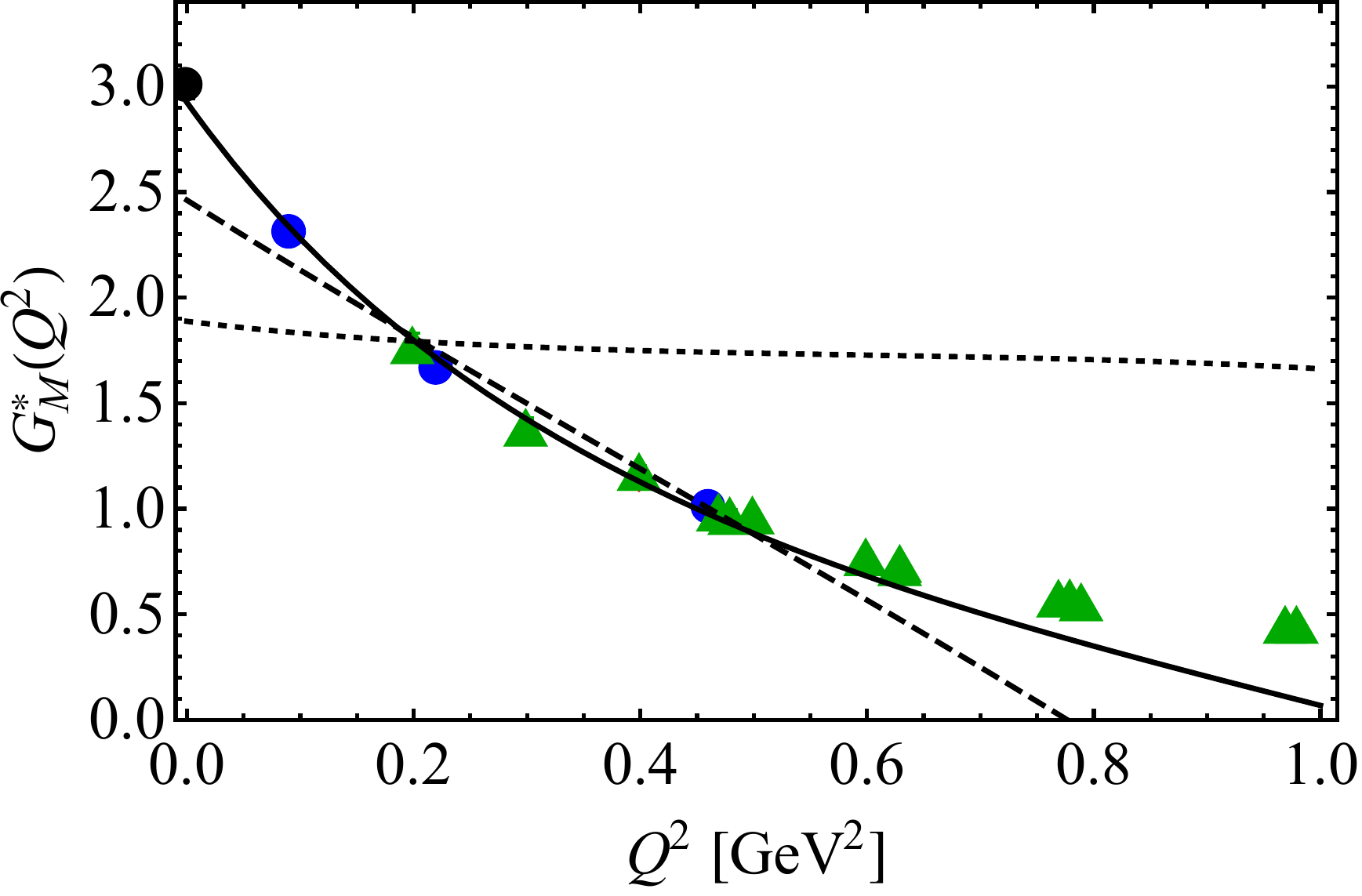}

\vspace*{0.2cm}
\includegraphics[height=4.5cm]{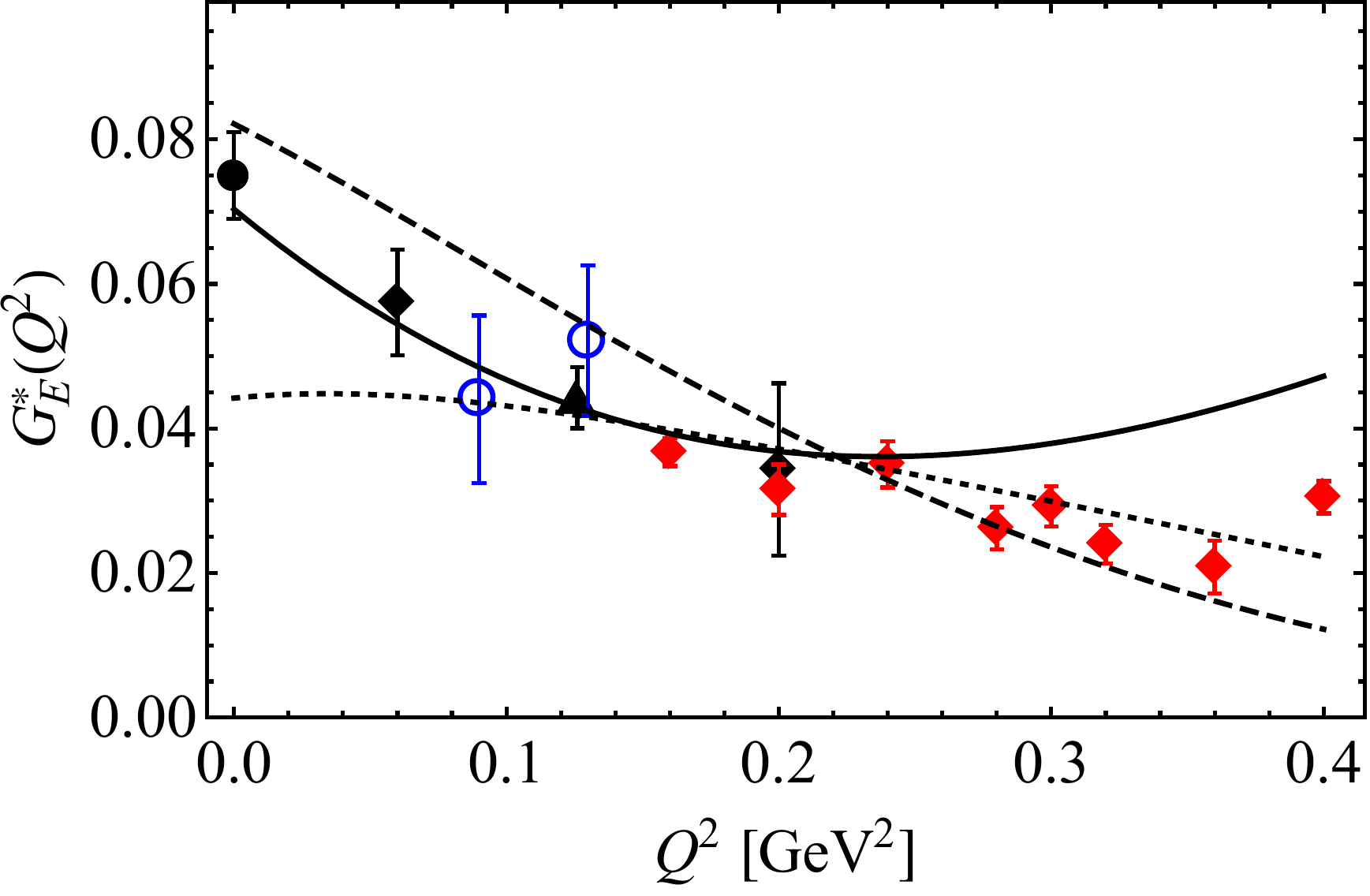}
\hspace*{0.5cm}
\includegraphics[height=4.5cm]{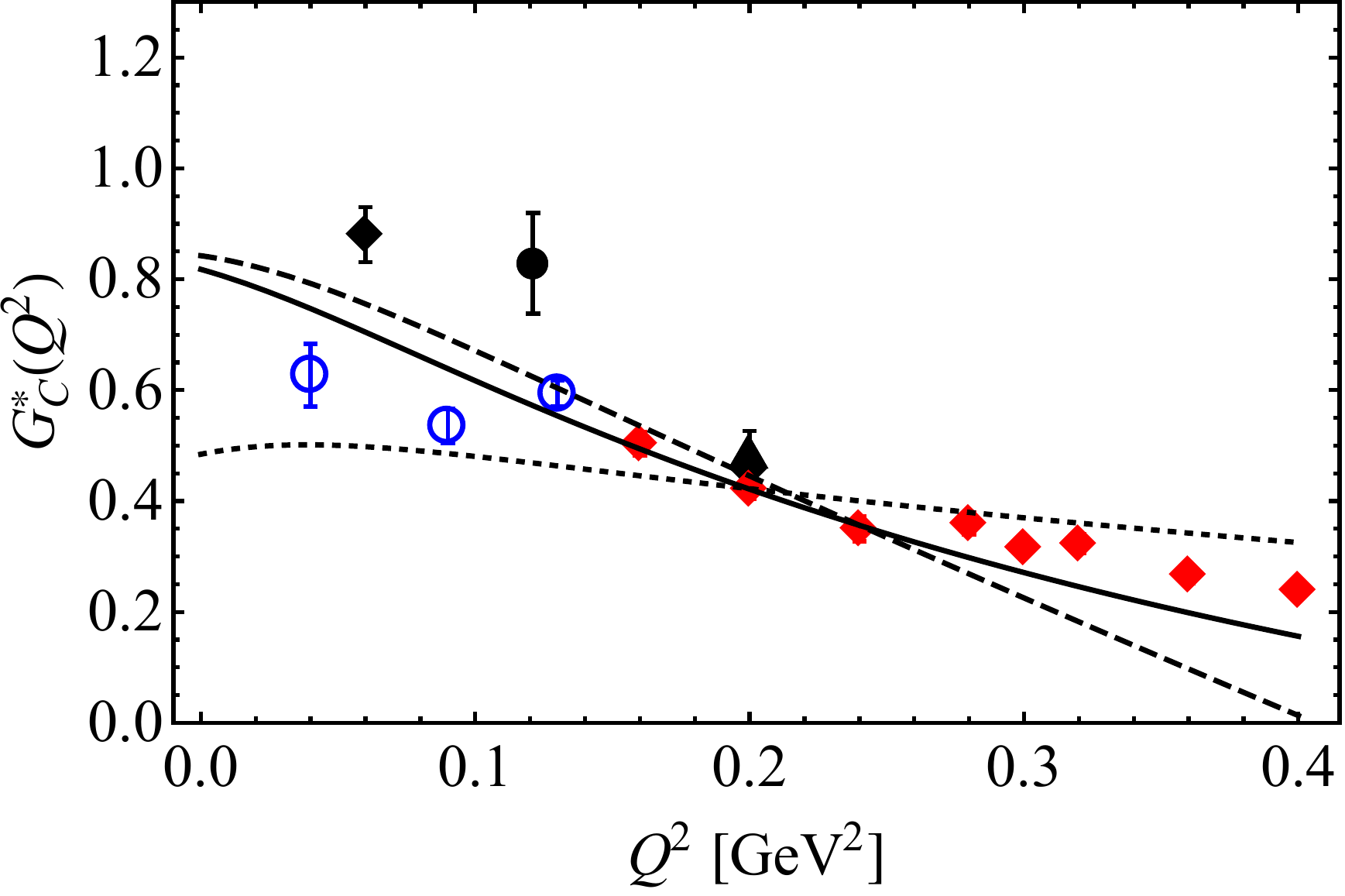}
\caption{Magnetic, electric, and charge transition form factors
$G_M^*$, $G_E^*$, and $G_C^*$ at the Breit-Wigner position
$W=m_\Delta=1232$~MeV.
   The dotted, dashed, and solid lines show Fits I, II, and III, respectively.
   The data is fitted up to $Q^2=0.3$~GeV$^2$.
   For the magnetic form factor, the fits are also compared with experiment for a much
larger range of $Q^2$.
   The data points for $G_M^\ast$ are from Refs.~\cite{Beck:1999ge} (black circle), \cite{Stein:1975yy}
(blue circles), and \cite{Bartel:1968tw} (green triangles);
for $G_E^\ast$ (from $R_{EM}$) from \cite{Beck:1999ge} (black circle),
\cite{Stave:2008aa} (black diamonds), \cite{Blomberg:2015zma} (blue open circles), \cite{Mertz:1999hp} (black triangle)
and \cite{Aznauryan:2009mx} (red diamonds);
and for $G_C^\ast$ (from $R_{SM}$) from \cite{Blomberg:2015zma} (blue open circles), \cite{Stave:2008aa} (black
diamonds), \cite{Pospischil:2000ad} (black circle), \cite{Elsner:2005cz} (black
triangle) and \cite{Aznauryan:2009mx} (red diamonds).
 } \label{fig:gmfit2}
\end{figure}

\begin{figure}[htbp]
    \centering
\includegraphics[height=4.5cm]{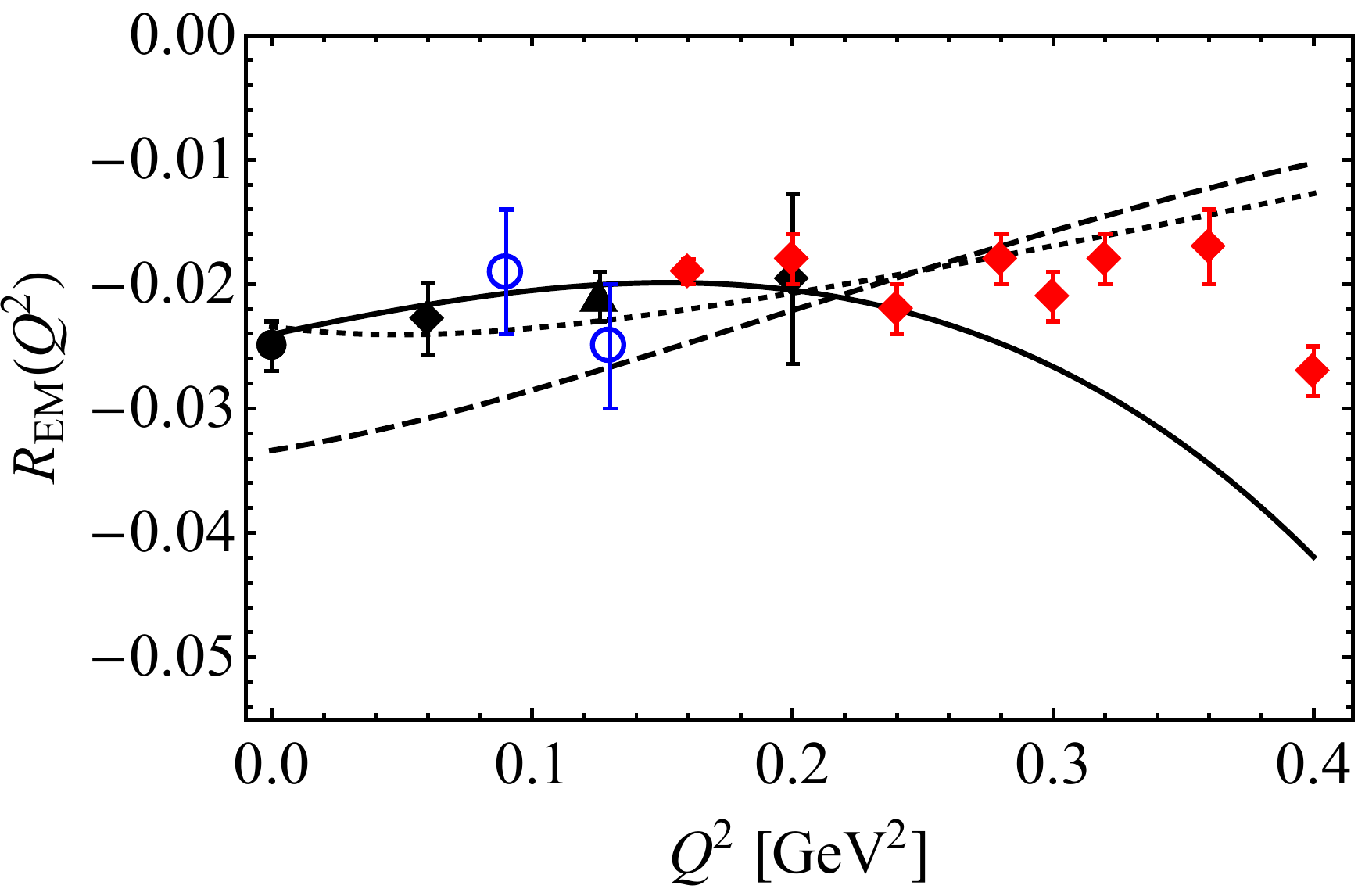}
\hspace*{0.5cm}
\includegraphics[height=4.5cm]{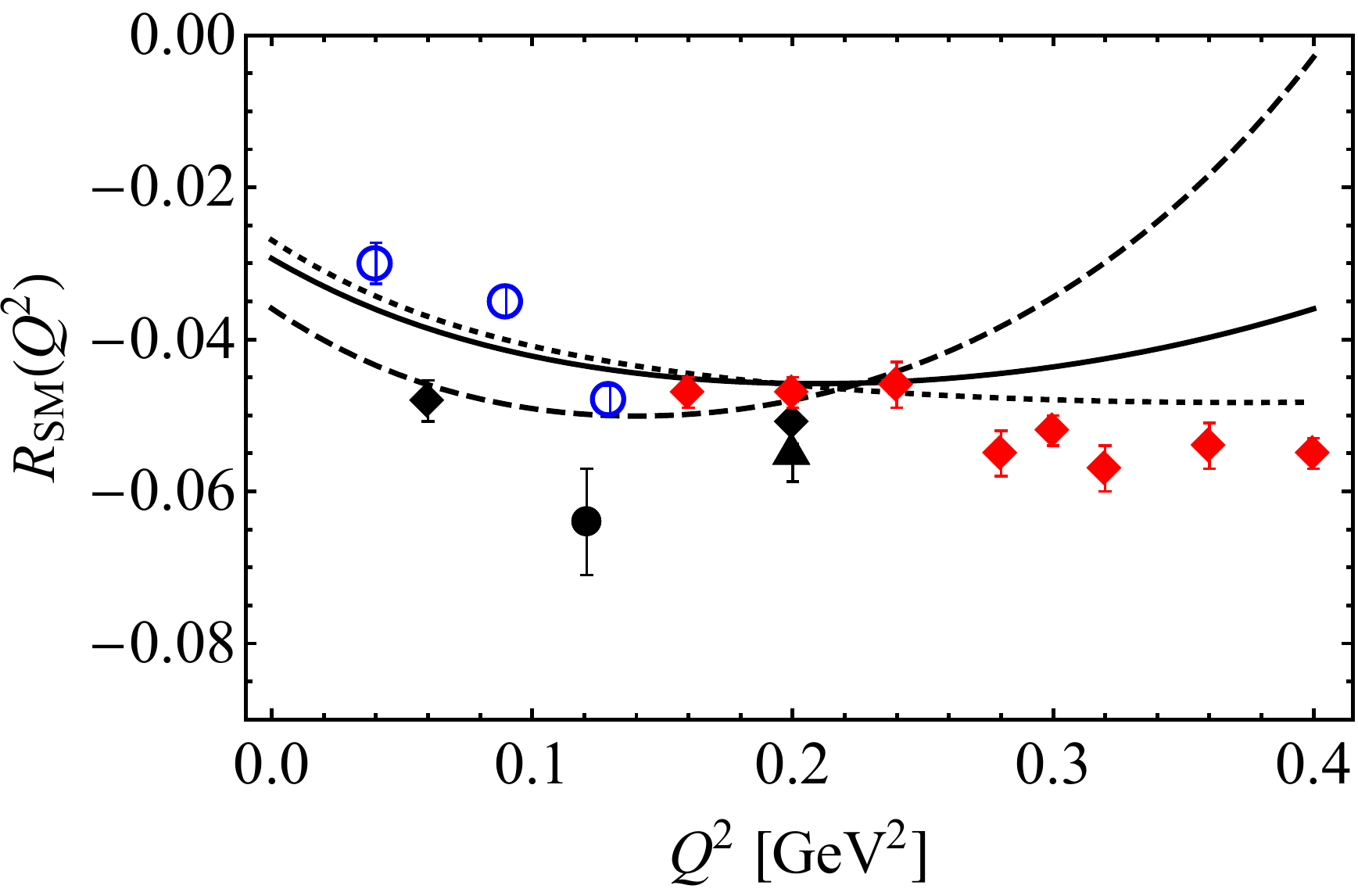}
\caption{Results for the ratios $R_{EM}$ and $R_{SM}$. For further
details, see caption of Fig.~\ref{fig:gmfit2}.}
\label{fig:remfit2}
\end{figure}

   Our results for the magnetic, electric, and charge transition form factors
$G_M^*$, $G_E^*$, and $G_C^*$ at the Breit-Wigner position $W=m_\Delta=1232$~MeV
are shown in Fig.~\ref{fig:gmfit2}.
   The ratios $R_{EM}$ and $R_{SM}$ are displayed in Fig.~\ref{fig:remfit2}.
   Let us first discuss the outcome of the full calculation including the
$\rho$ meson (solid lines).
   For $G_M^*$ we obtain a very good description, once the $\rho$ meson is
included.
   Even though the data were only fitted in the range $[0,0.3]$~GeV$^2$,
our results with the $\rho$ meson are in good agreement with the data up to and including
$Q^2=0.6$~GeV$^2$ (see upper right panel of Fig.~\ref{fig:gmfit2}).
   For $G_E^*$ we obtain a good description up to and including $Q^2=0.25$~GeV$^2$.
   Note, however, that $G_E^*$ is more than an order of magnitude smaller than
$G_M^*$.
   Finally, the description of $G_C^*$ is good over the full range $[0,0.3]$~GeV$^2$.
   The ratios $R_{EM}$  and $R_{SM}$ are rather well described up to and including
$Q^2=0.25$~GeV$^2$ (see Fig.~\ref{fig:remfit2}).
   Without the $\rho$ meson the fit fails dramatically if only
$C_1^\gamma$ and $C_2^\gamma$ are allowed (dotted lines).
   The fit improves with the addition of $C_3^\gamma$ (dashed lines), which is, however, not needed in a fit
including the $\rho$ meson.
   We also checked a description with all six coupling constants, but found very strong
correlations between $C_3^\gamma$ and $C_1^\rho,C_3^\rho$, which can be avoided by fixing
$C_3^\gamma=C_3^\rho=0$.


\begin{figure}[htbp]
    \centering
\includegraphics[height=4.5cm]{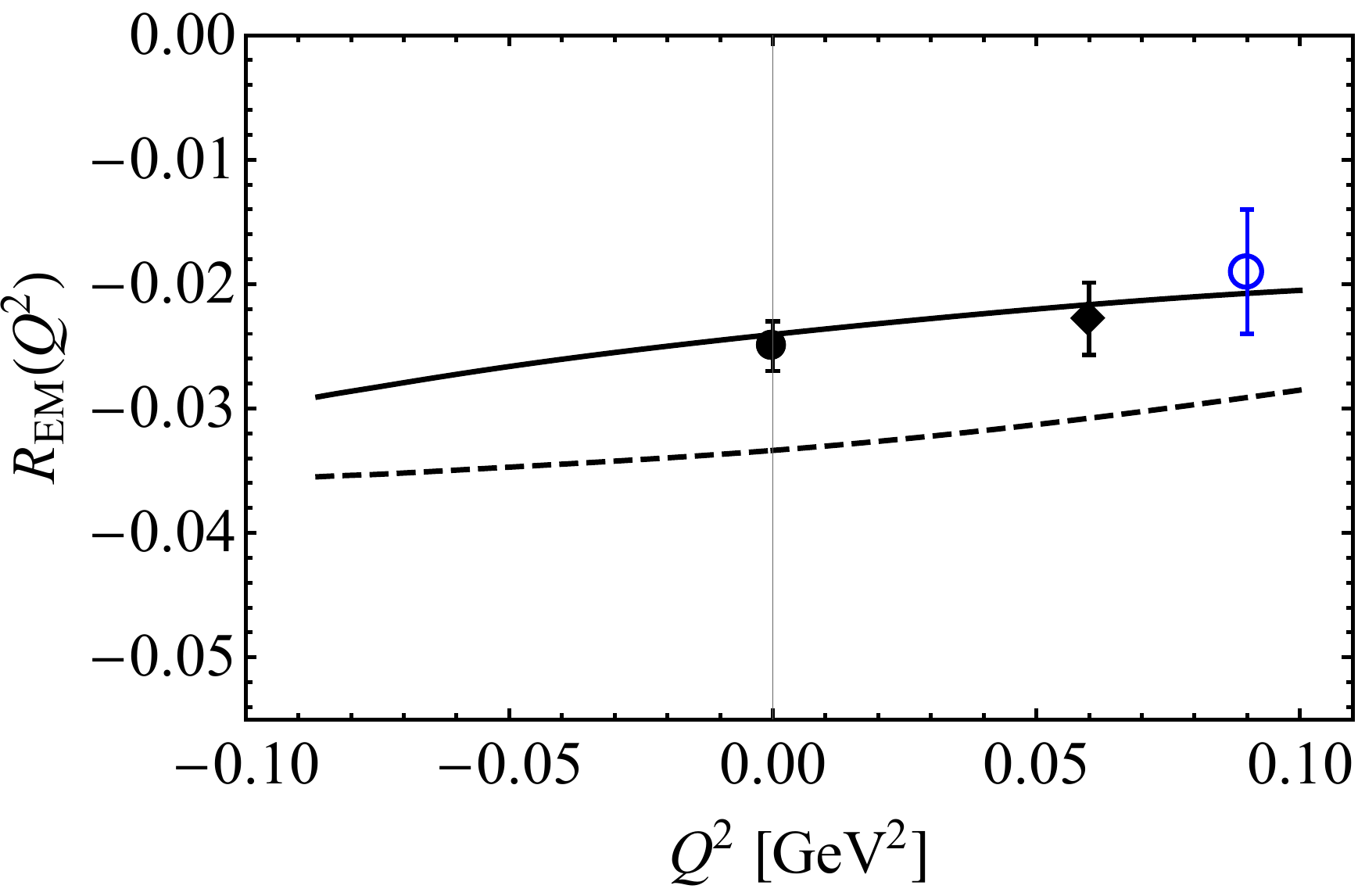}
\hspace*{0.5cm}
\includegraphics[height=4.5cm]{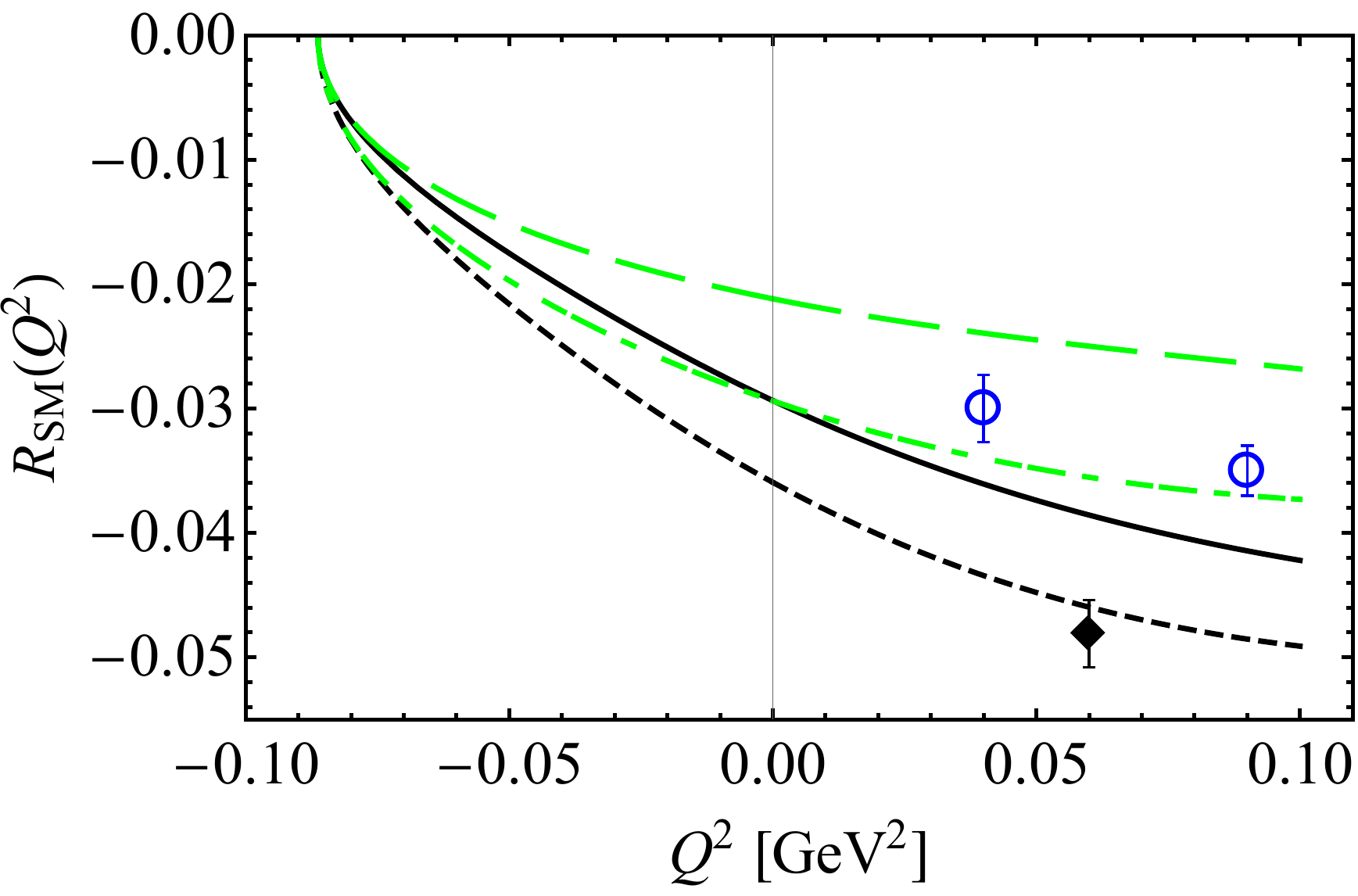}
\caption{Results of the Fits II (dashed black lines) and III (solid black lines) for the ratios $R_{EM}$ and
$R_{SM}$ for low $Q^2$ and extrapolations into the time-like region
down to the Siegert limit.
   The dashed-dotted and long-dashed green lines show the
predictions of the Siegert theorem for the solutions II and III,
respectively (see text). The experimental data are as in
Fig.~\ref{fig:remfit2}. } \label{fig:remfit2Siegert}
\end{figure}

   The Siegert theorem provides a model-independent prediction for relations among different
electromagnetic multipoles and form factors.
   It results from the symmetry that, for very small virtual photon momenta, all
transverse components of the electromagnetic current must be the same (also known as the long-wavelength limit).
   For a detailed introduction, see Refs.~\cite{Drechsel:2007if,Tiator:2016kbr}.
   Recently, the role of the Siegert theorem for low-$Q^2$ transition form factors has been intensively studied
by Ramalho \cite{Ramalho:2016zzo}.
   First of all, in the so-called Siegert limit, $|\vec{q}_{\rm cm}|\rightarrow 0$ with $\vec{q}_{\rm cm}$ being the photon three-momentum
in the center-of-momentum frame, one obtains the following relation:
\begin{equation}
R_{SM}\stackrel{|\vec{q}_{\rm cm}|\rightarrow 0}{\longrightarrow}\frac{|\vec{q}_{\rm cm}|}{m_\Delta-m_N}R_{EM}.
\label{siegertlimit}
\end{equation}
   Using Eqs.~(\ref{Eq:REM}) and (\ref{Eq:RSM}) results in
\begin{equation}
G_C^\ast(Q^2)\stackrel{|\vec{q}_{\rm cm}|\rightarrow 0}{\longrightarrow}\frac{2m_\Delta}{m_\Delta-m_N}G_E^\ast(Q^2).
\label{Eq:GCGESiegert}
\end{equation}
   The corresponding so-called pseudo-threshold, $Q^2_{\rm pt}=-(m_\Delta-m_N)^2=-0.087$~GeV$^2$,
is time-like and thus outside the physical region of electroproduction.
   In the left panel of Fig.~\ref{fig:remfit2Siegert}, we show the results of the Fits II and III for the ratio $R_{EM}$
from the pseudo-threshold $Q^2_{\rm pt}$ to $Q^2=0.1$~GeV$^2$.
   In the right panel of Fig.~\ref{fig:remfit2Siegert}, we then compare the predictions for $R_{SM}$ as obtained from the Siegert
theorem, Eq.~(\ref{siegertlimit}), with the full calculation for the Fits II and III.\footnote{For
that purpose we make use of $|\vec{q}_{\rm cm}|=\sqrt{[Q^2+(m_\Delta+m_N)^2][(Q^2+(m_\Delta-m_N)^2)]}/(2m_\Delta)$.}
   Close to the pseudo-threshold, the ratios $R_{SM}$ (and thus the charge form factors $G_C^*$) follow very well
the predictions of the Siegert theorem, and even for small space-like momentum transfers $Q^2$ it gives, within 30 \%,
a good guideline for the full result.
   Around $Q^2=0.1$~GeV$^2$, the deviations resulting from higher-order terms in the long-wavelength expansion become
more important and the predictions of the Siegert theorem are no longer reliable.

\begin{figure}[htbp]
    \centering
\includegraphics[height=6.5cm]{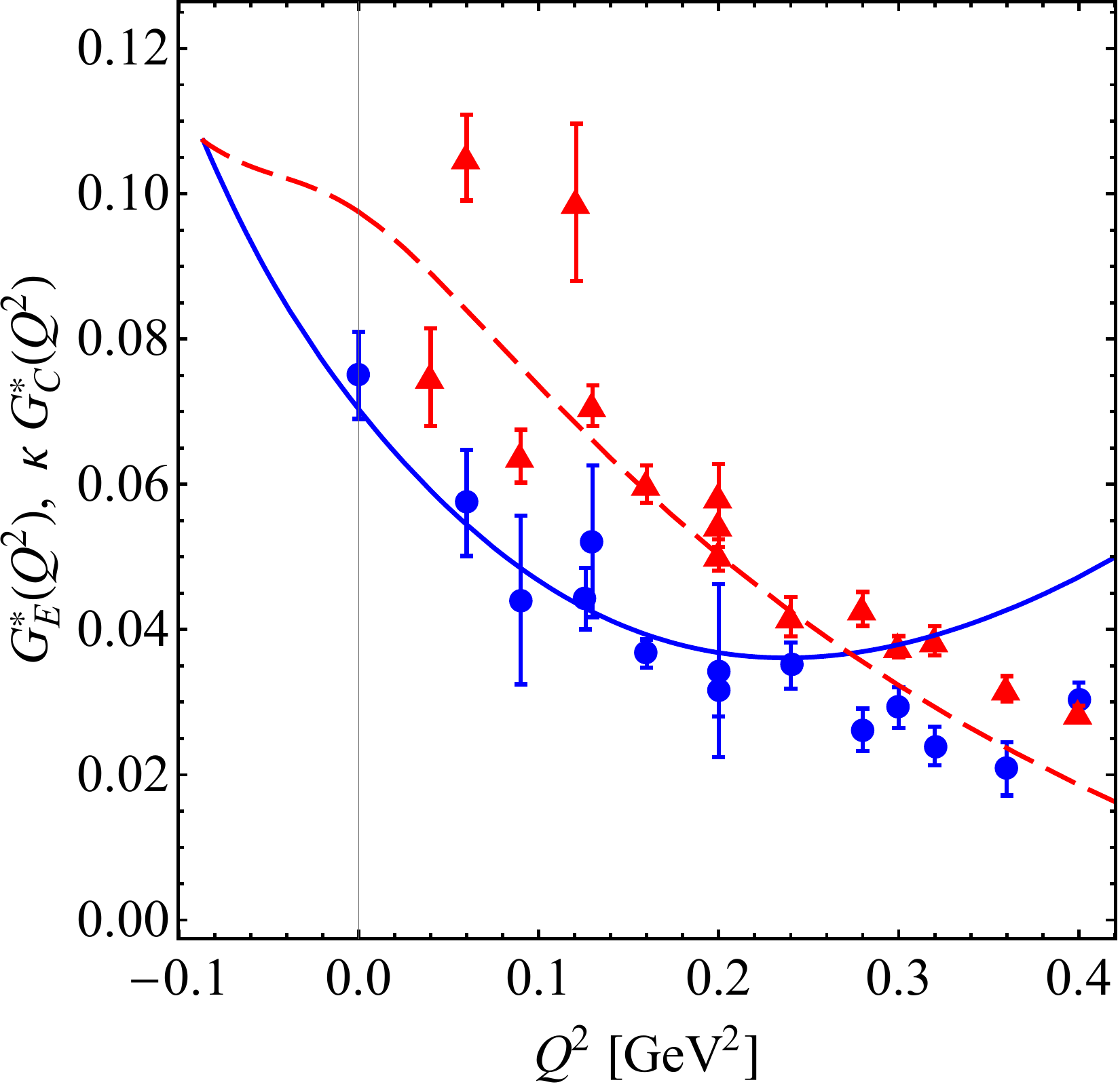}
\caption{Form factors $G_E^*$ (blue solid line) and $\kappa\,G_C^*$ (red dashed line) with
$\kappa=(m_\Delta-m_N)/2m_\Delta$ from time-like to space-like
regions.
   The blue circles and the red triangles show the experimental data for the electric form factor
and the re-scaled charge form factor, respectively.
The data are as in Fig.~\ref{fig:gmfit2}} \label{fig:GCE_Siegert}
\end{figure}

   The consequences of the Siegert theorem for the ratio $R_{SM}$ and
the charge form factor are in fact two-fold, as can be seen from Eq.\ (\ref{siegertlimit}).
   First, the ratio $R_{SM}$ must vanish at pseudo-threshold
and, second, the slope of $R_{SM}$ at pseudo-threshold is related to the slope of $R_{EM}$,
which is not so clearly seen in Fig.~\ref{fig:remfit2Siegert}.
   In Fig.~\ref{fig:GCE_Siegert}, we show the form
factors $G_E^*(Q^2)$ and $\frac{m_\Delta-m_N}{2m_\Delta}G_C^*(Q^2)$
separately, which should be identical in the Siegert limit according
to Eq.~(\ref{Eq:GCGESiegert}).

   At the pole position, the form factors are complex quantities due to
the fact that the $\Delta$(1232) is an unstable particle.
   Similarly as in the previous case, we also performed three fits
(without and with the $\rho$ meson) to the form factor data at the
pole position.
   These data were obtained from the SAID and MAID
partial wave analysis, applying the Laurent-Pietarinen (L+P)
expansion method~\cite{Tiator:2016btt}.
   The results from the SAID and MAID analysis are very similar and we
have taken the average values, showing in the figures the differences of these
analyses as error bars.
   These uncertainties are hardly visible and can not be
used for the statistical weights in the fits of the data.
   Therefore, we have used the same weights for $G_M$ and $G_C$ data, but have
increased the weight for $G_E$ by a factor 100.
   This is comparable to the weights in the fits to the Breit-Wigner data, where the
weight factors are determined from the statistical errors of the
data.
   The results for the fit parameters are given in Table \ref{tab:cms-parameters}.

\begin{table}[htbp]
    \centering
        \begin{tabular}{|l|c|c|c|c|c|}
        \hline
& $C_1^\gamma$ [GeV$^{-1}$] & $C_2^\gamma$ [GeV$^{-2}$]& $C_3^\gamma$ [GeV$^{-3}$] & $C_1^\rho$ [GeV$^{-1}$] & $C_2^\rho$ [GeV$^{-2}$]  \\
\hline
Fit I &  $1.40 -0.21\,i$ & $ 1.60 +0.22\,i$  & 0 & --  & -- \\
\hline
Fit II &  $4.85 +1.24\,i$ & $ 1.62 +0.23\,i$  & $-8.87 -3.72\,i$  & --  & -- \\
\hline
Fit III &  $-2.19 -0.085\,i$ & $ 2.17 -2.60\,i$  & 0 & $-25.7 +1.57\,i$  & $ 4.18 -21.6\,i$ \\
\hline

        \end{tabular}
\caption{The results of the fitting procedure for the parameters of
the tree-level contributions of diagram $(a)$ of Fig.~\ref{fig:diagrams} are labeled $C_i^\gamma$.
The results including the $\rho$ meson also contain the parameters $C_i^\rho$ (see
App.~\ref{tree} for definitions). The data for the fits are taken as
an average of the pole form factors of MAID and SAID~\cite{Tiator:2016btt} with
$Q^2\le 0.3$~GeV$^2$, see text for further details. }
\label{tab:cms-parameters}
\end{table}

\begin{figure}[htbp]
    \centering
\includegraphics[height=4.5cm]{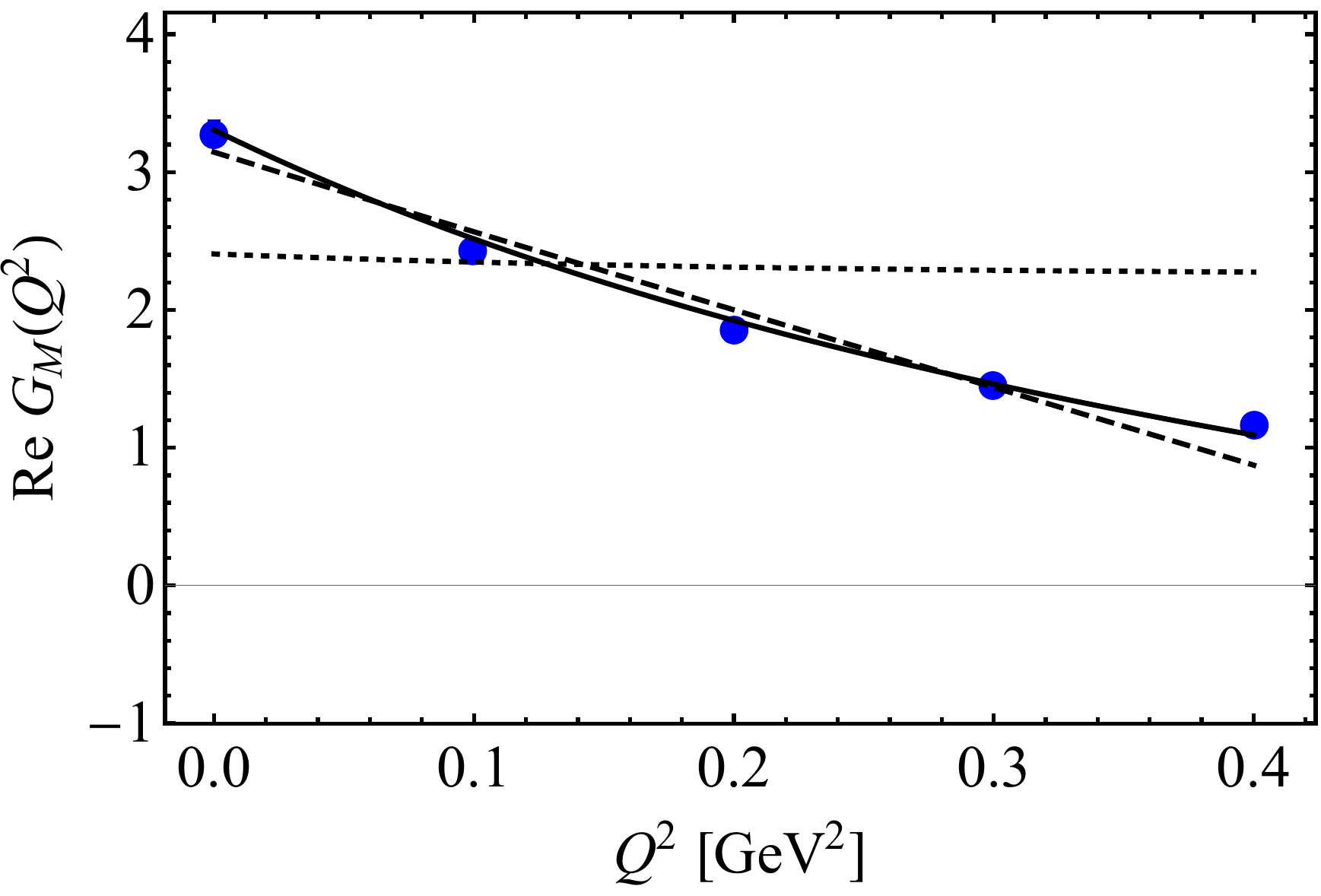}
\hspace*{0.5cm}
\includegraphics[height=4.5cm]{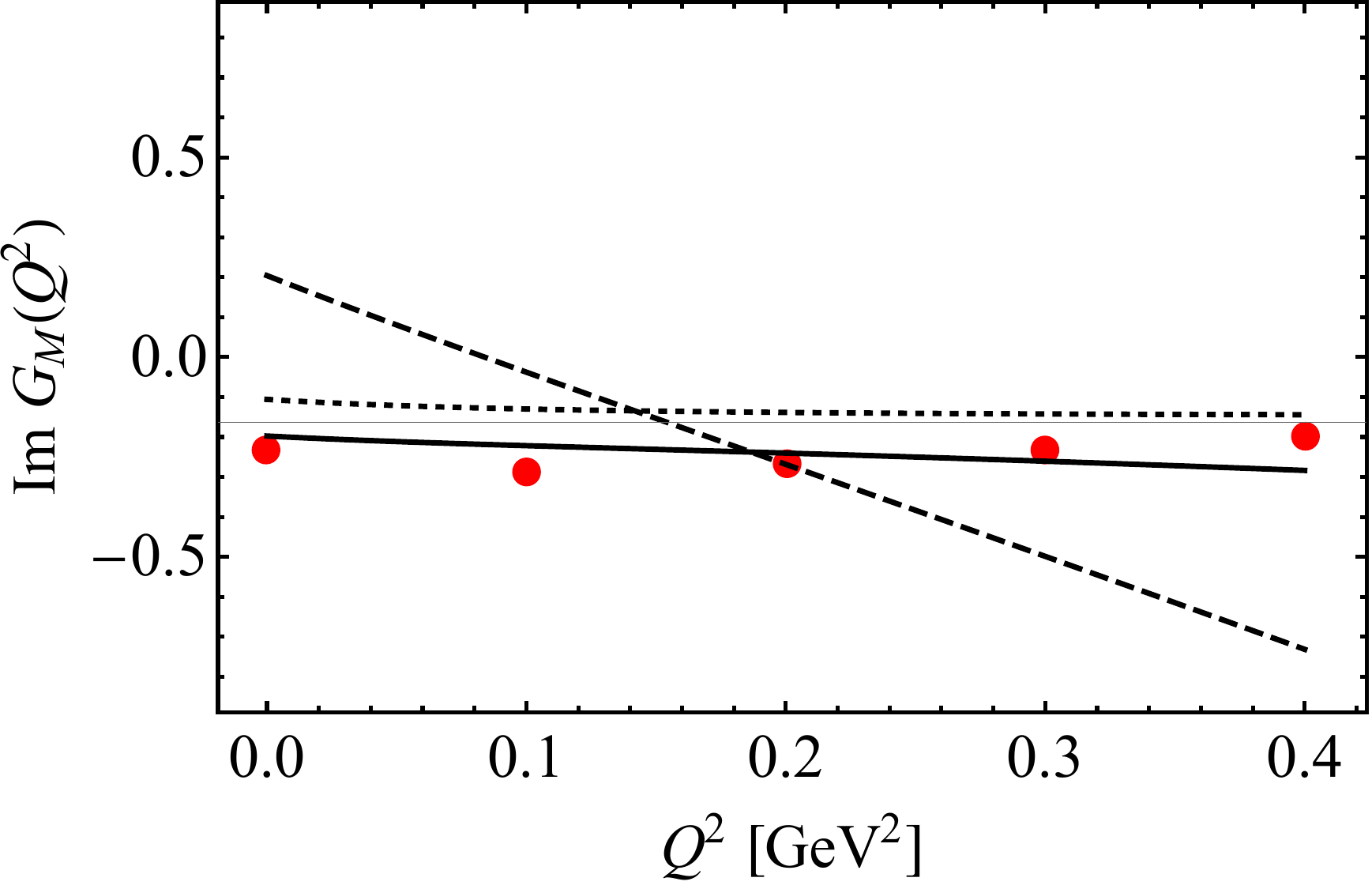}
\vspace*{0.2cm}

\includegraphics[height=4.5cm]{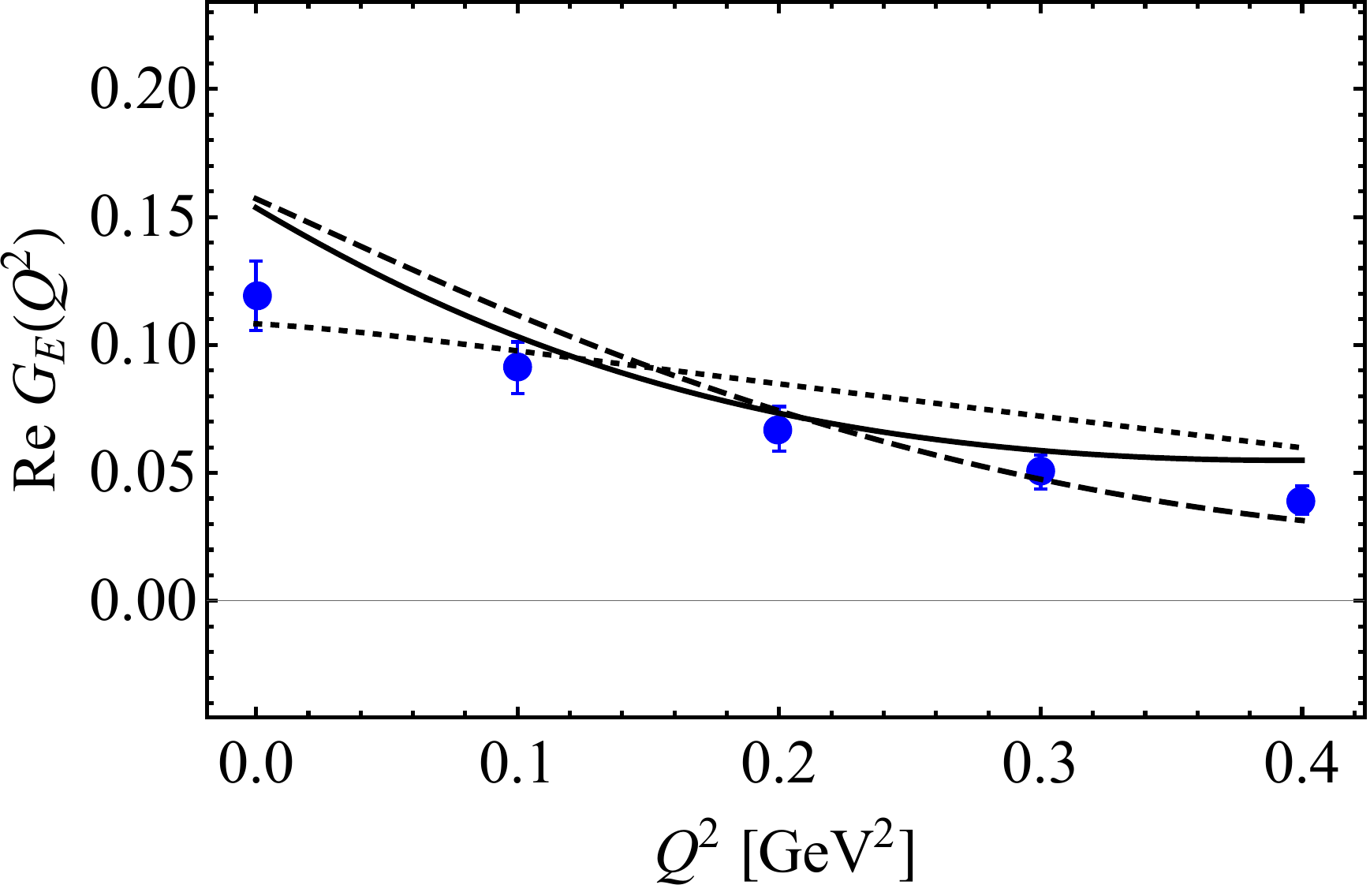}
\hspace*{0.5cm}
\includegraphics[height=4.5cm]{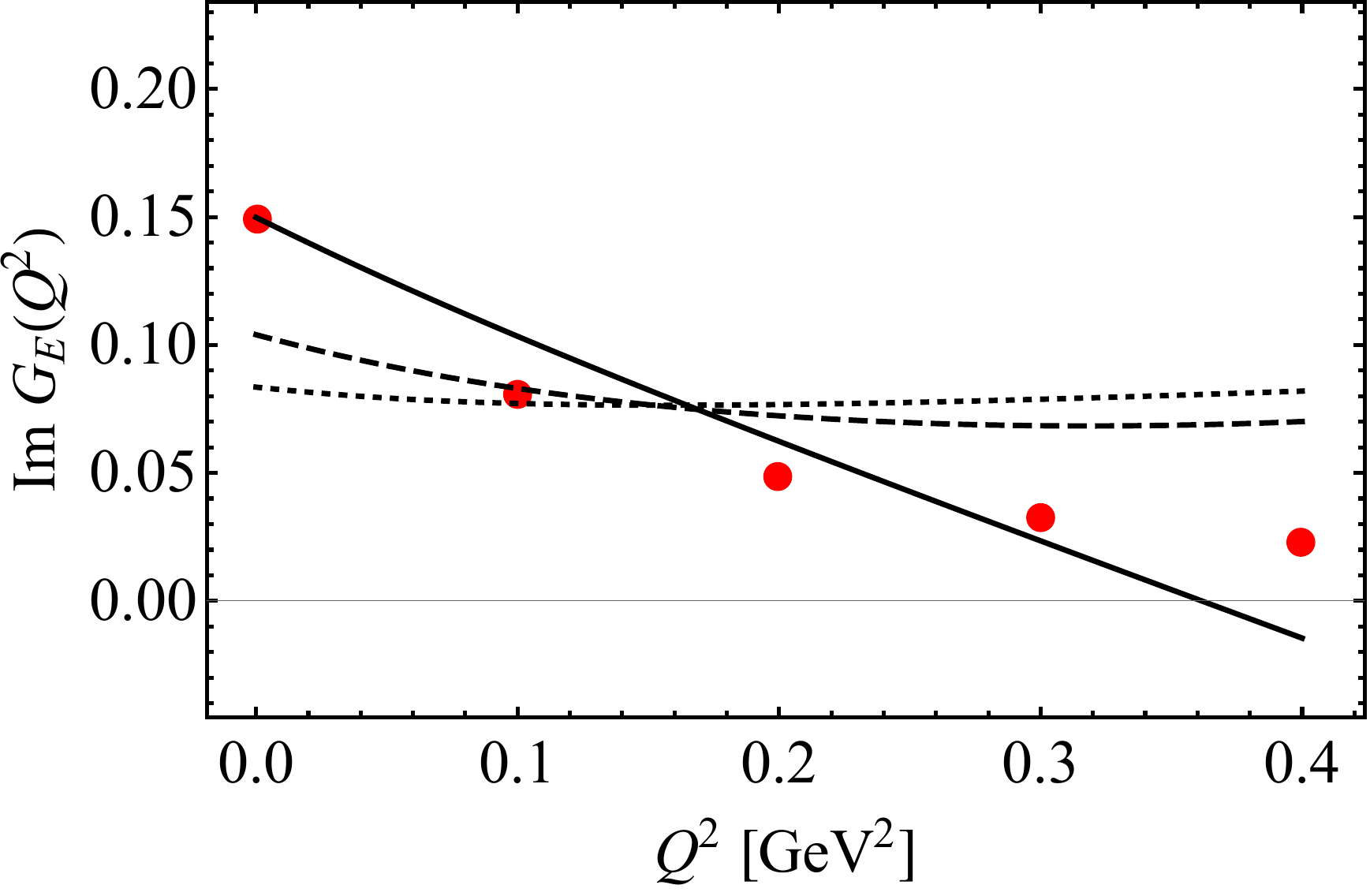}
\vspace*{0.2cm}

\includegraphics[height=4.5cm]{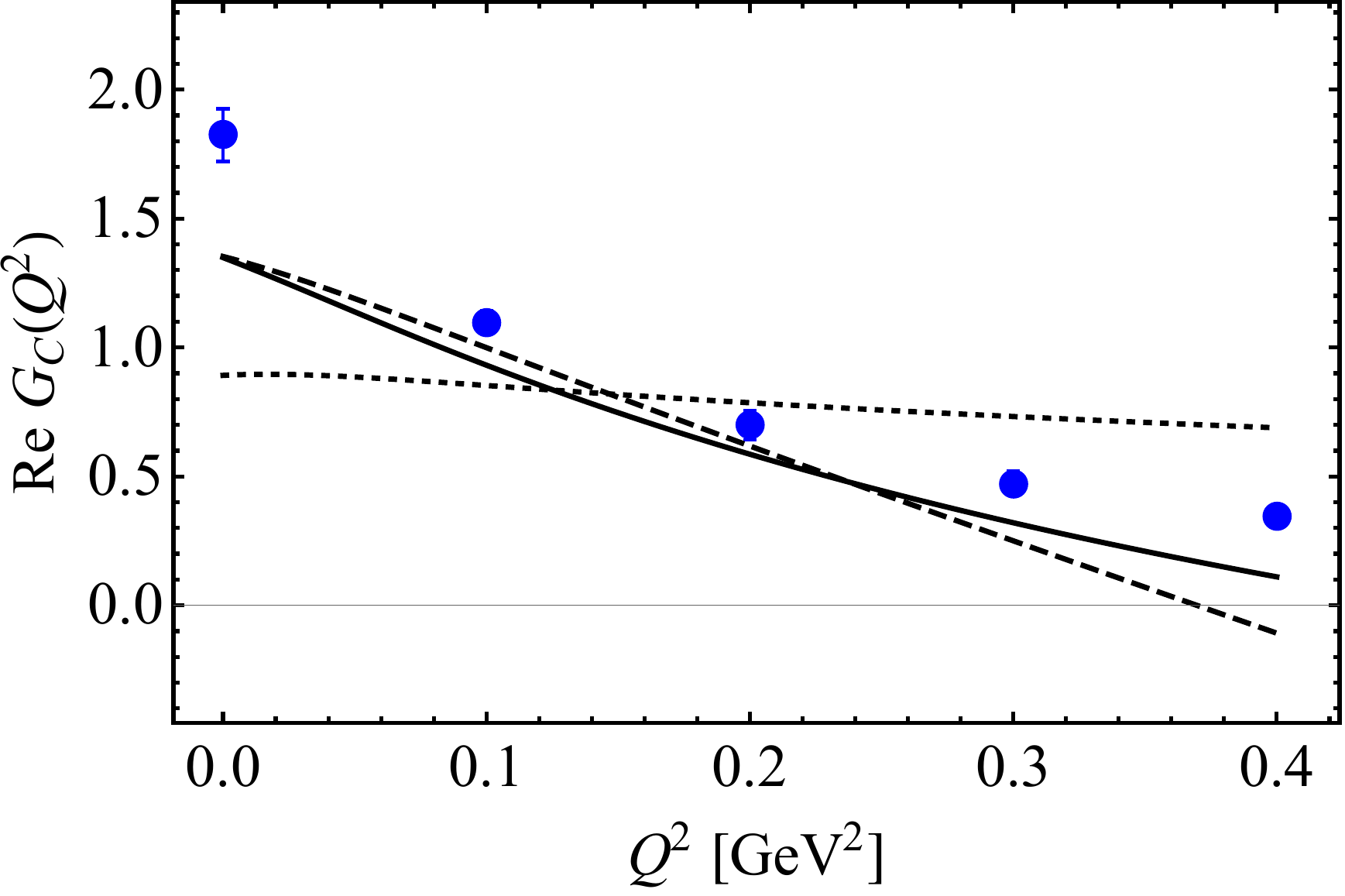}
\hspace*{0.5cm}
\includegraphics[height=4.5cm]{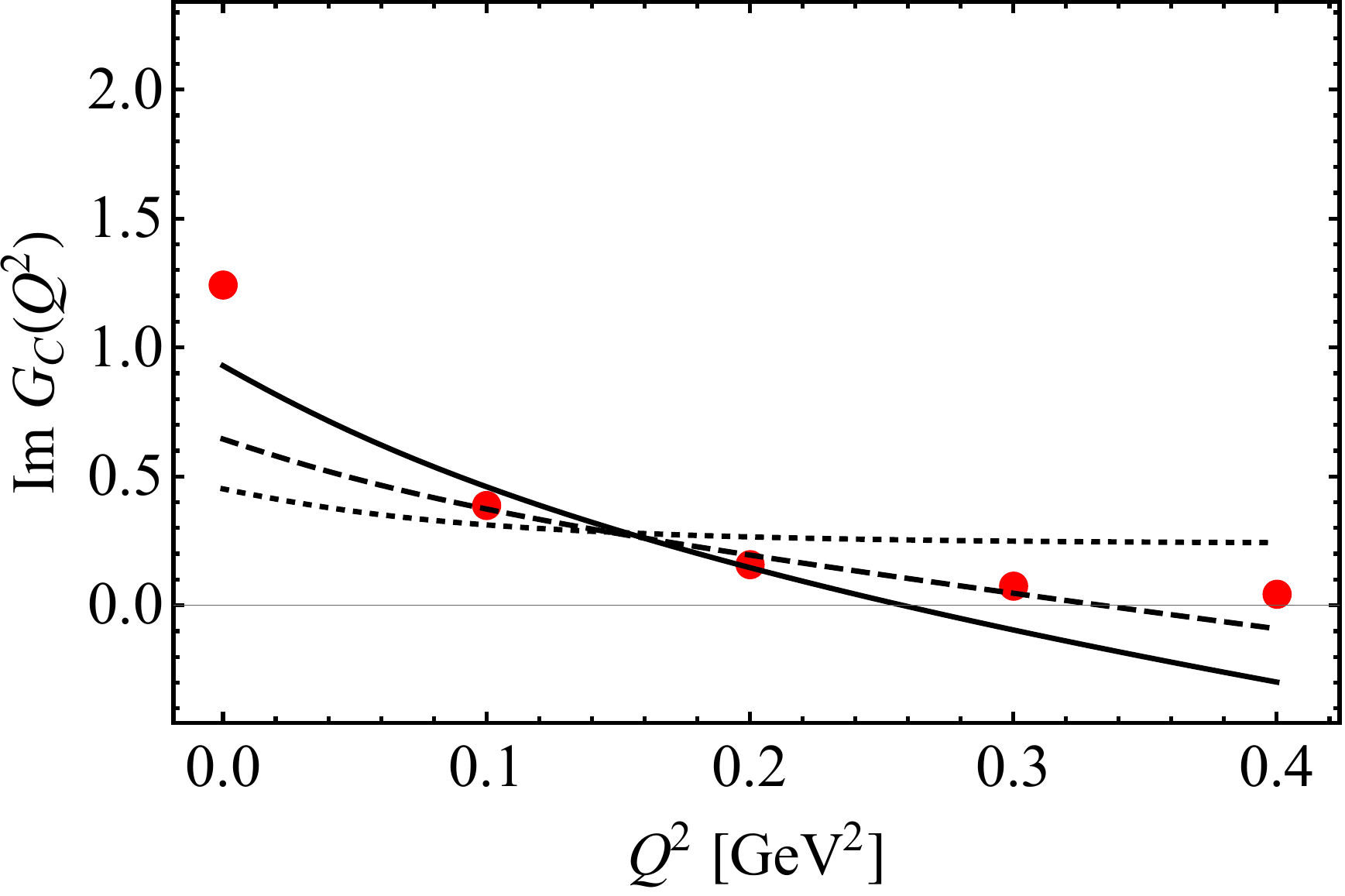}

\caption{Real (left) and imaginary (right) parts of the magnetic,
electric, and charge transition form factors $G_M$, $G_E$, and $G_C$
at the pole position $W_p=(1210-i\,50)$~MeV.
   The solid lines show the results including the $\rho$ meson (Fit III)
and the dotted and dashed lines show Fits I and II without the $\rho$.
   The data points are taken as the averaged MAID and SAID results from
Ref.~\cite{Tiator:2016btt}.}
    \label{fig:gmec-cms}
\end{figure}

   In Fig.~\ref{fig:gmec-cms}, we show the Fits I and II without the $\rho$
meson and Fit III including the $\rho$ meson for the real and
imaginary parts of the form factors $G_M$, $G_E$, and $G_C$  compared to the
data.
   Only in the case of $G_M$, the imaginary part is negligibly small compared to the real part.
   On the other hand, for $G_E^\star$ and $G_C^\star$ the real and imaginary parts are of the same order of magnitude.
   As in the previous case with the Breit-Wigner form factors, a fit without the $\rho$ meson only works
reasonably well with three tree coupling constants.
   However, the fit including the $\rho$ meson describes the
data much better, especially because of the additional curvature in
the $Q^2$ dependence of the $\rho$-meson contribution.

\begin{figure}[htbp]
    \centering
\includegraphics[height=3.5cm]{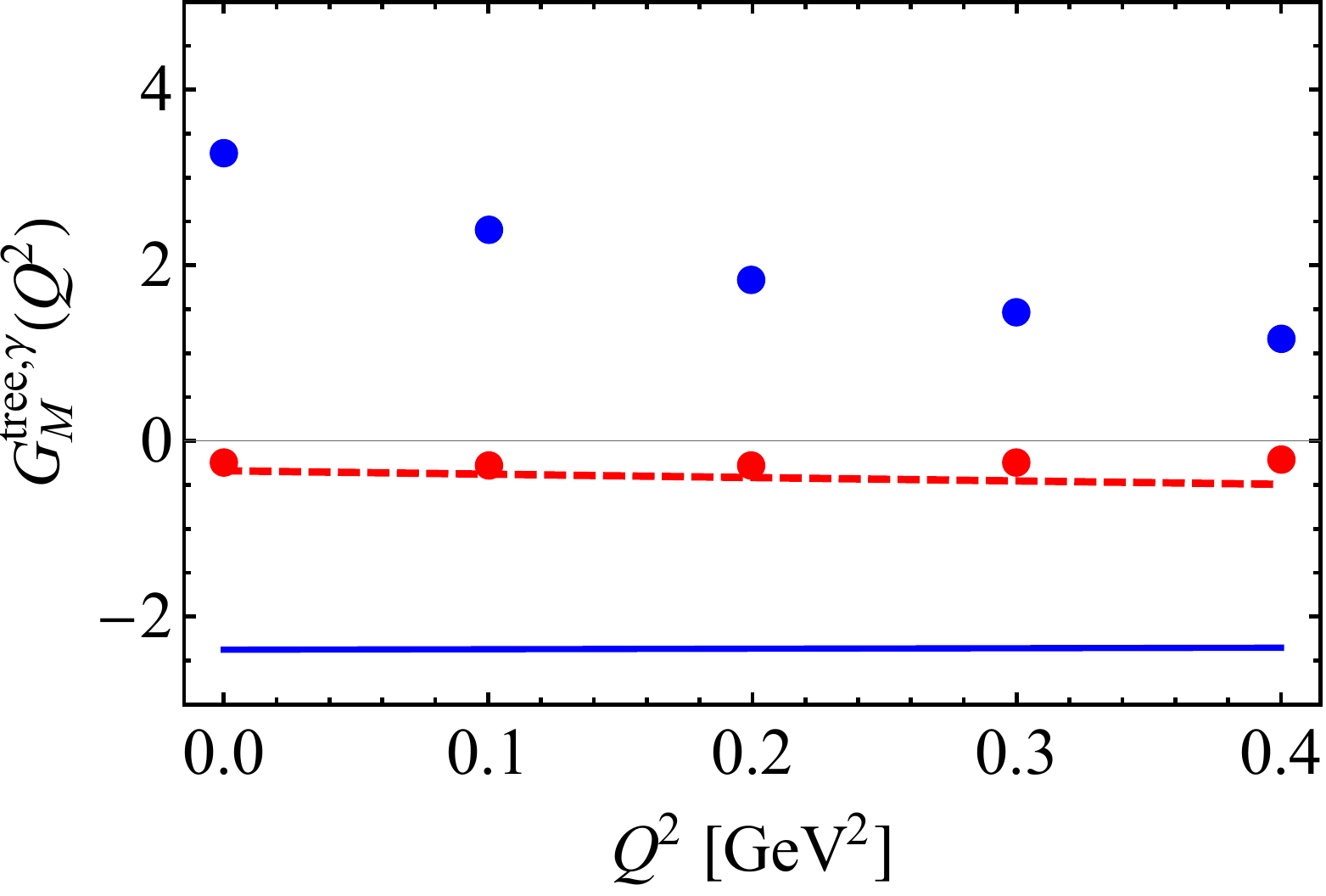}
\hspace*{0.0cm}
\includegraphics[height=3.5cm]{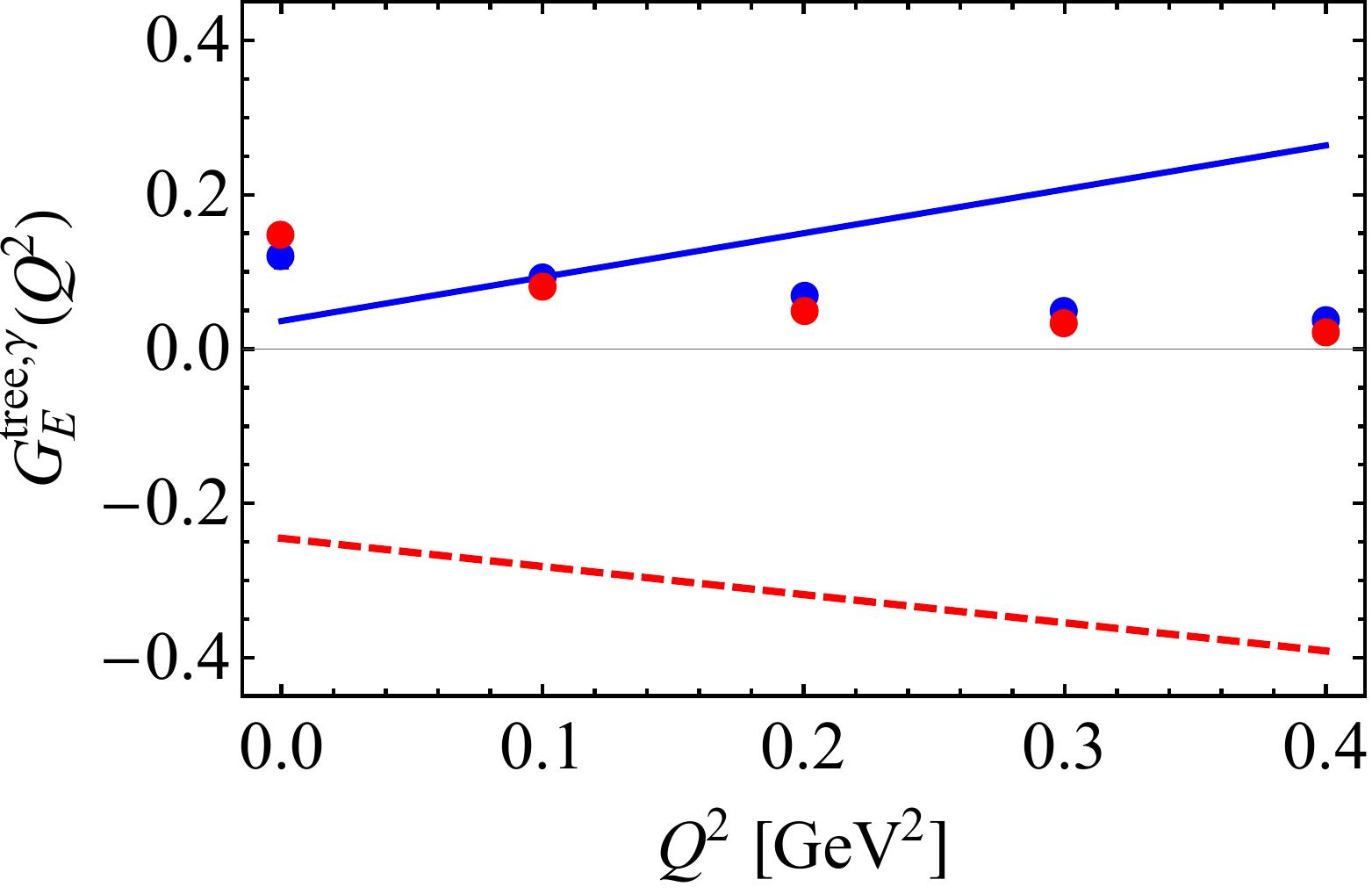}
\hspace*{0.0cm}
\includegraphics[height=3.5cm]{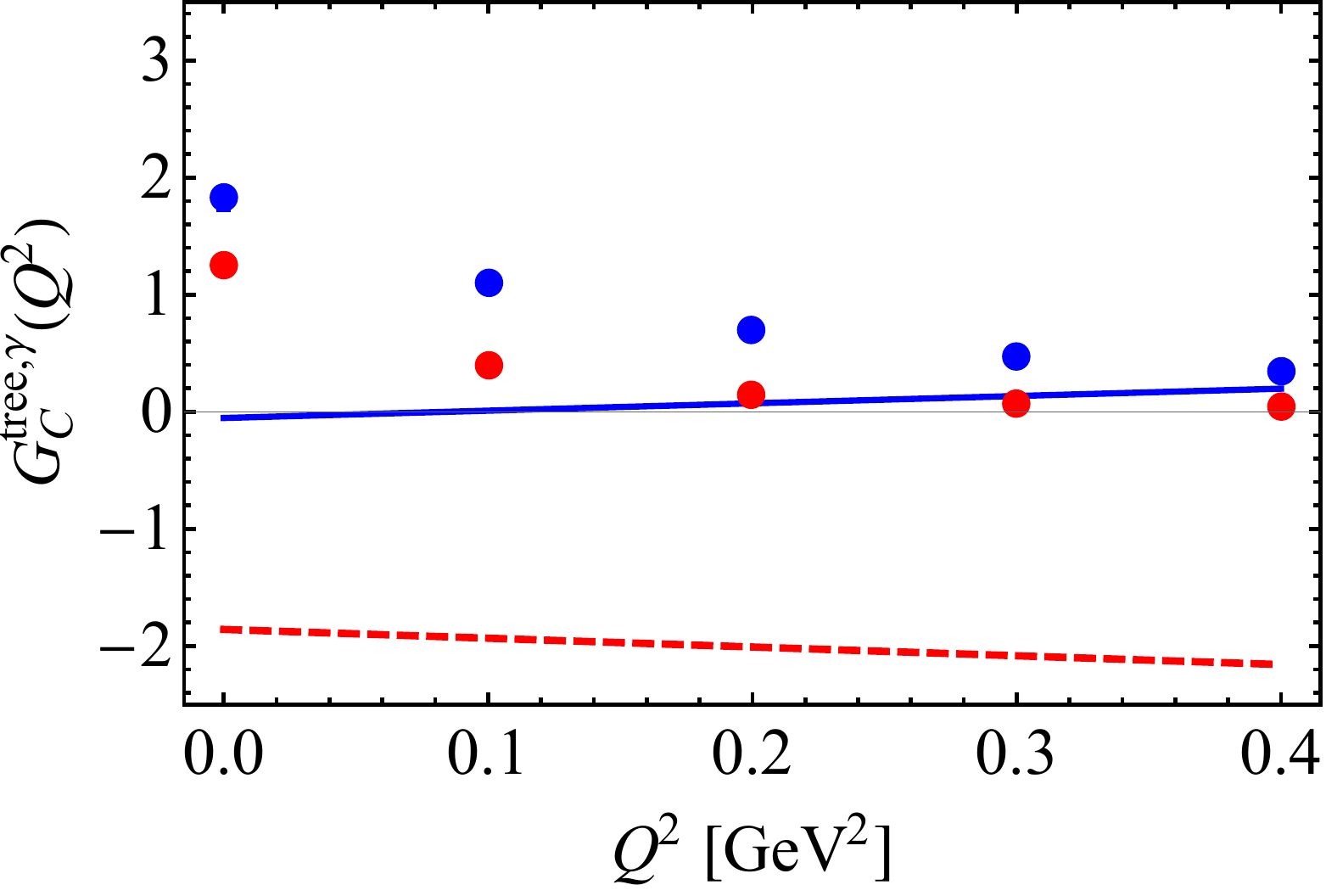}
\vspace*{0.0cm}

\includegraphics[height=3.5cm]{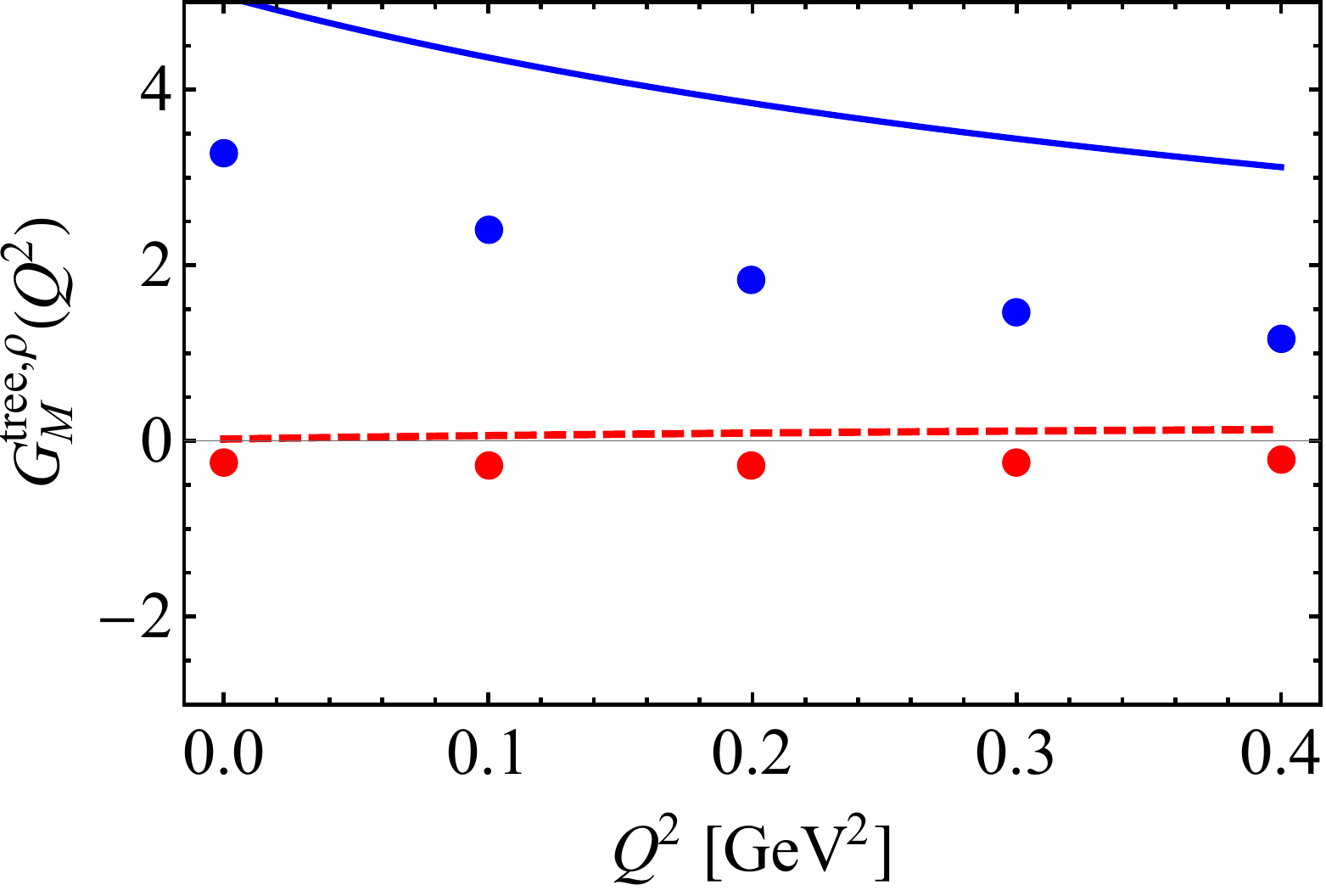}
\hspace*{0.0cm}
\includegraphics[height=3.5cm]{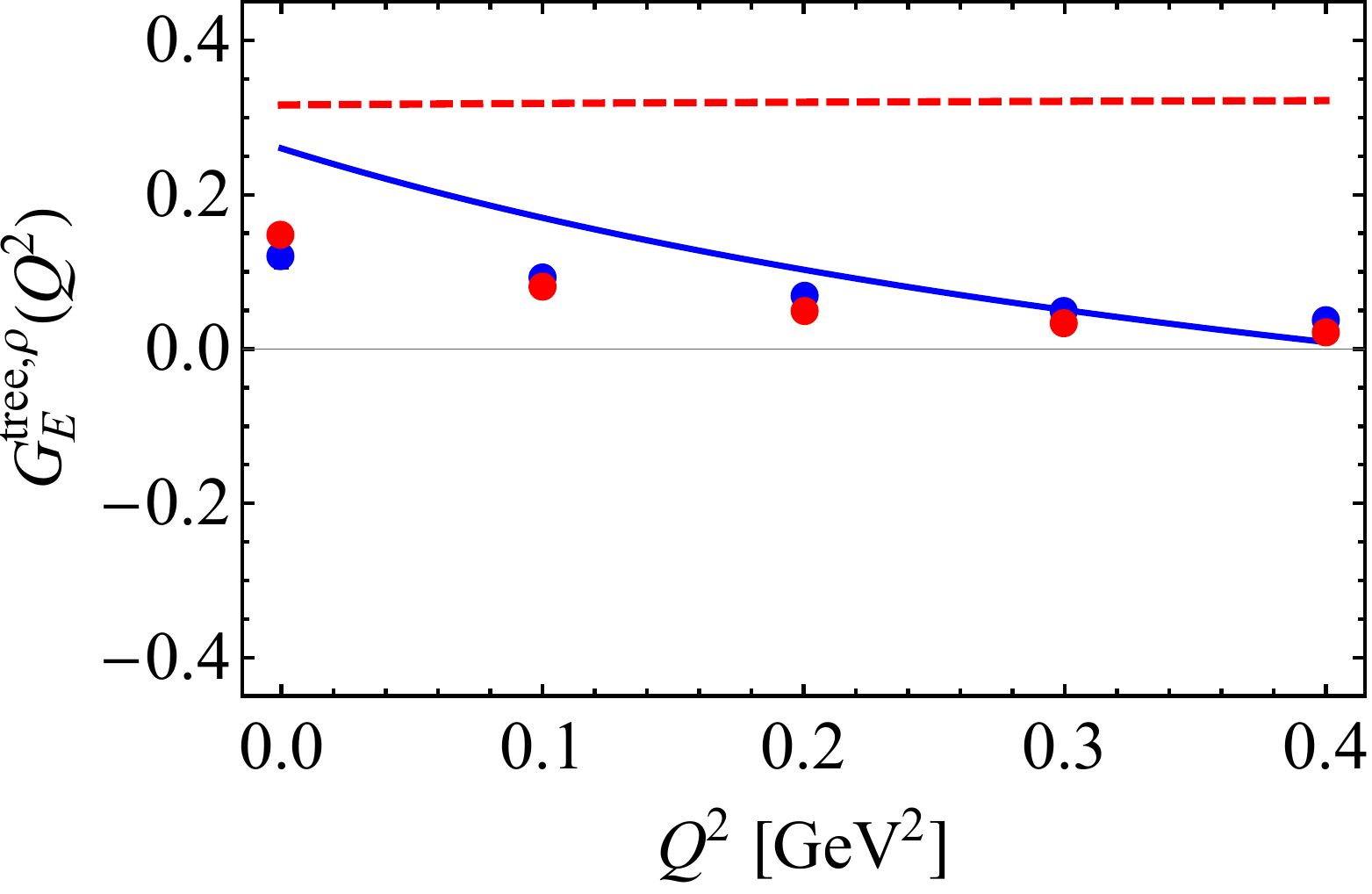}
\hspace*{0.0cm}
\includegraphics[height=3.5cm]{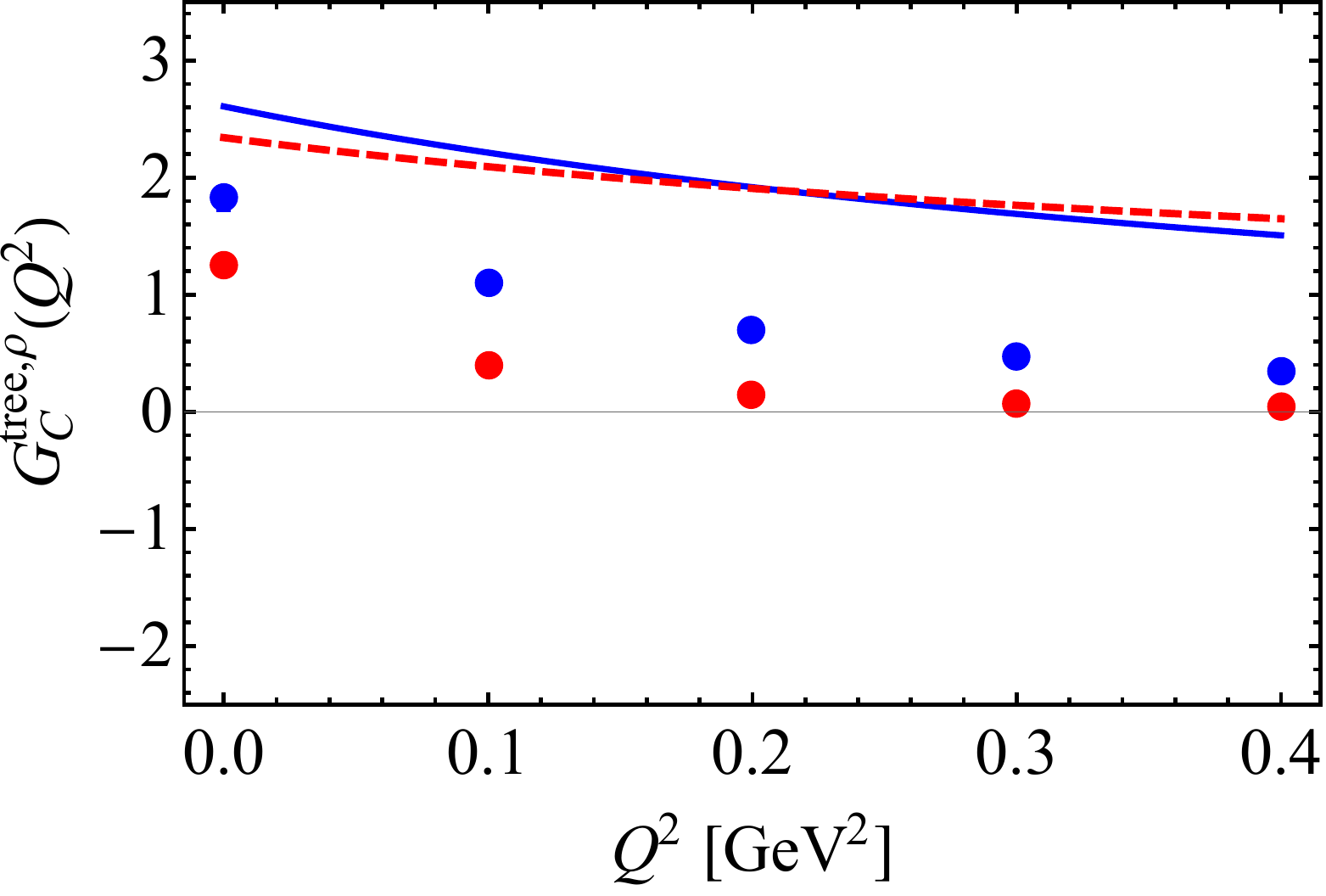}
\vspace*{0.0cm}

\includegraphics[height=3.5cm]{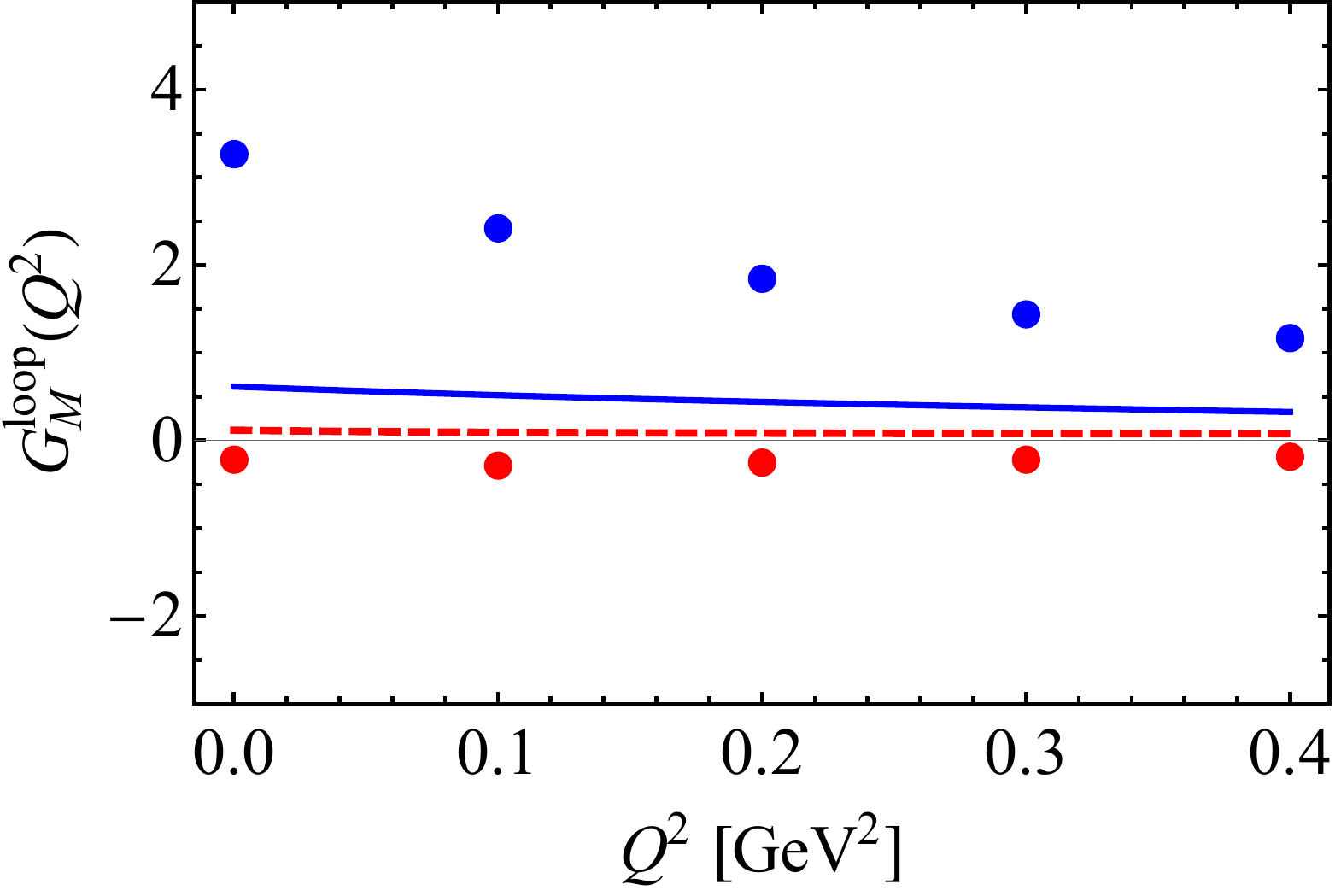}
\hspace*{0.0cm}
\includegraphics[height=3.5cm]{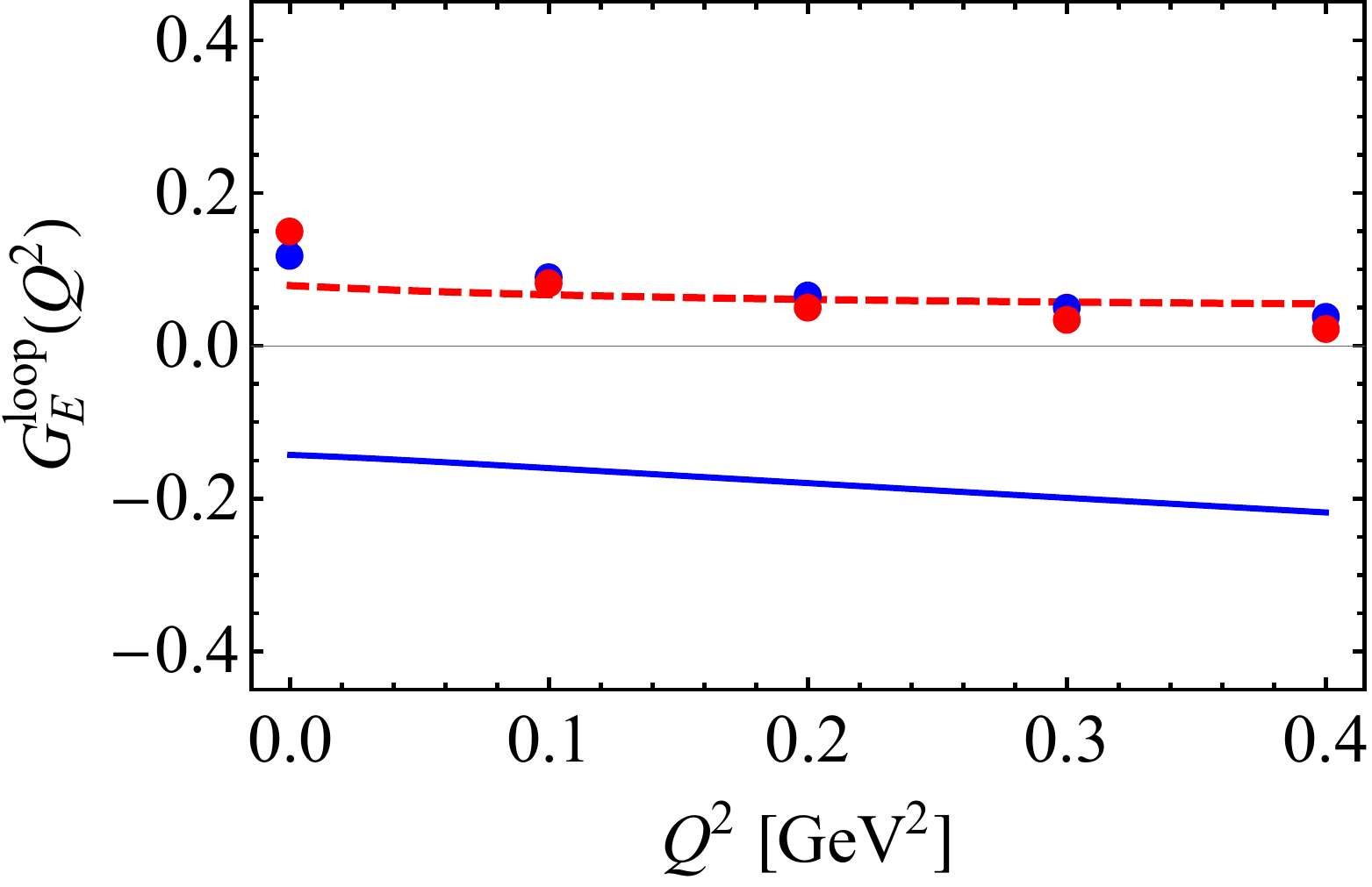}
\hspace*{0.0cm}
\includegraphics[height=3.5cm]{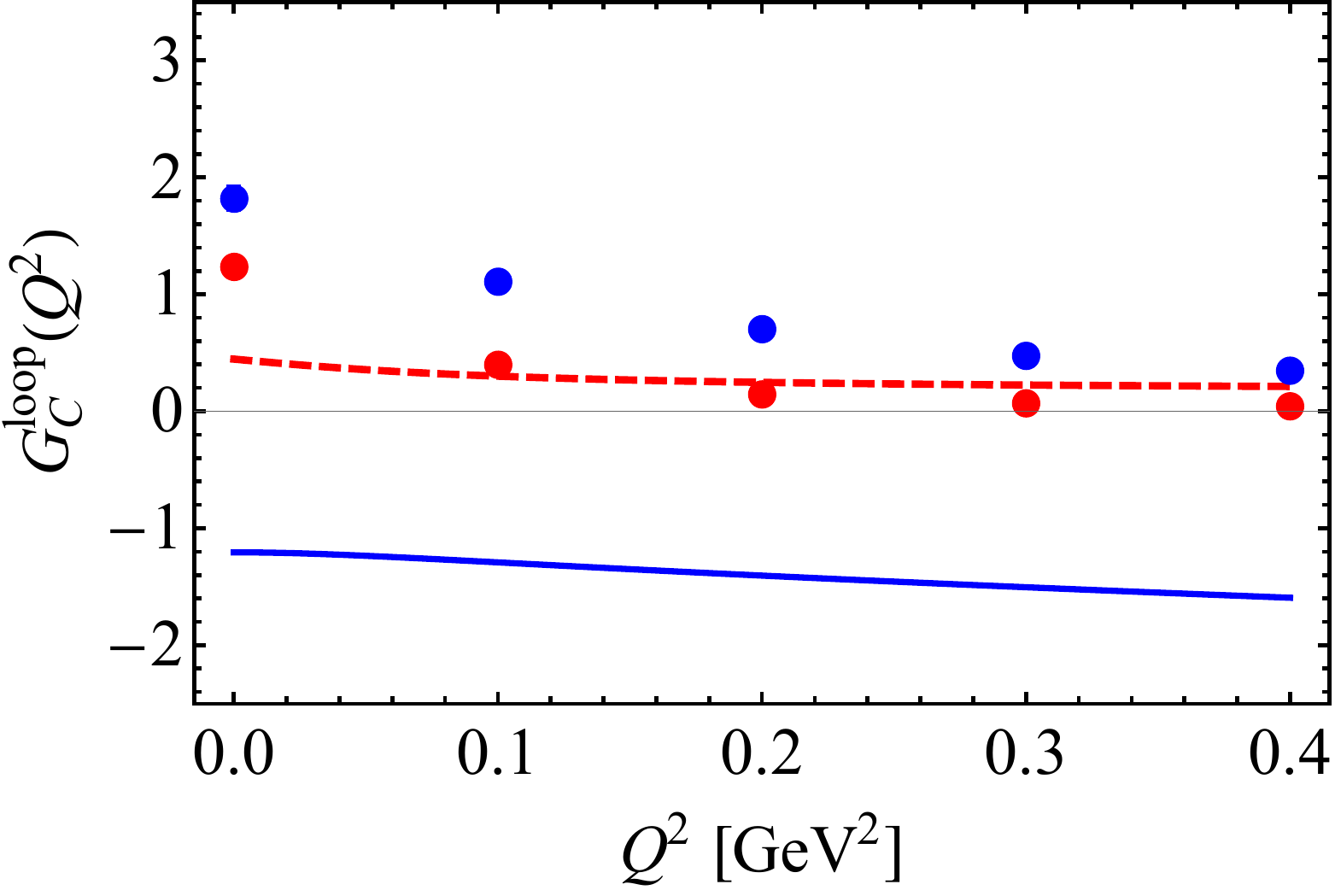}
\vspace*{0.0cm}

\includegraphics[height=3.5cm]{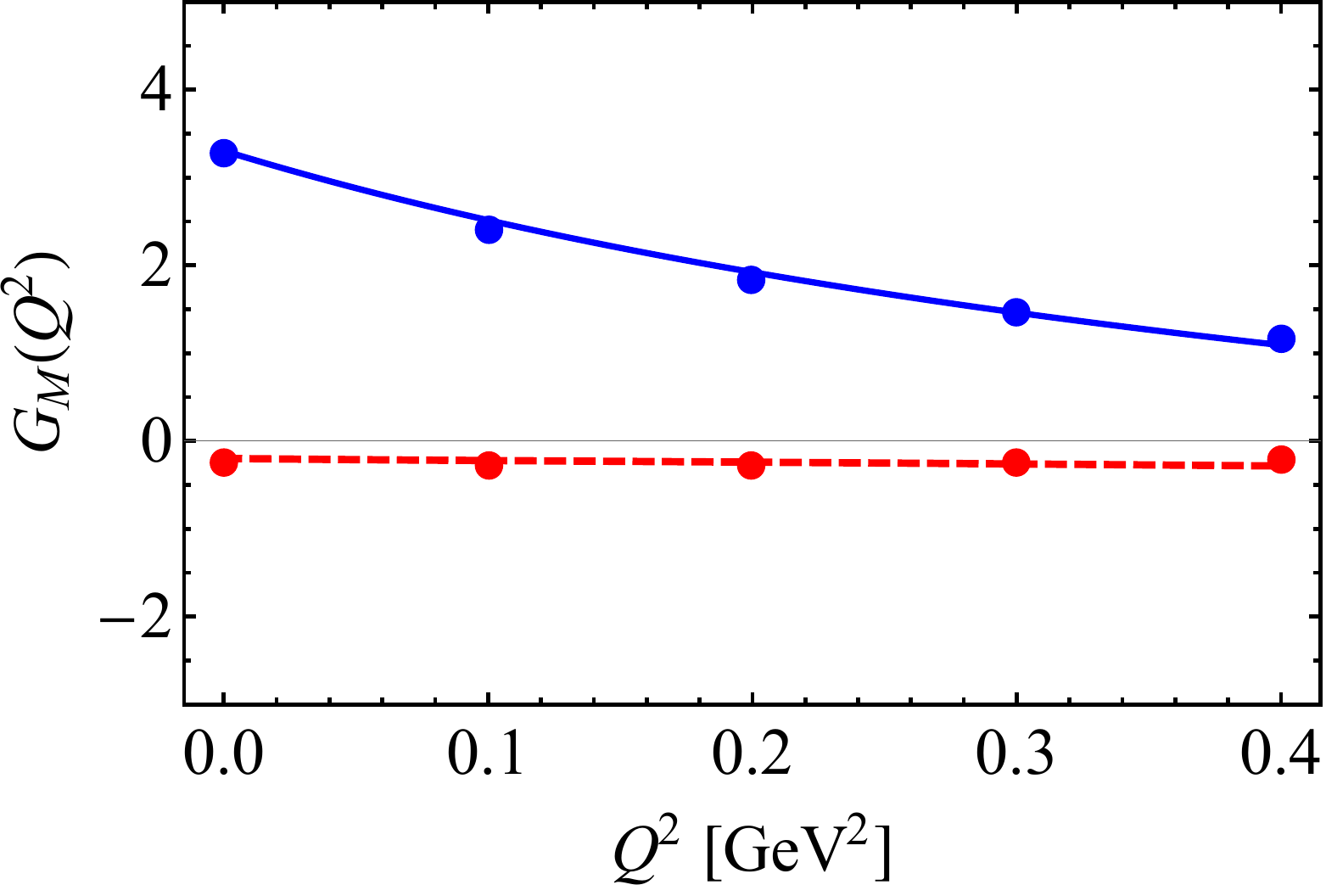}
\hspace*{0.0cm}
\includegraphics[height=3.5cm]{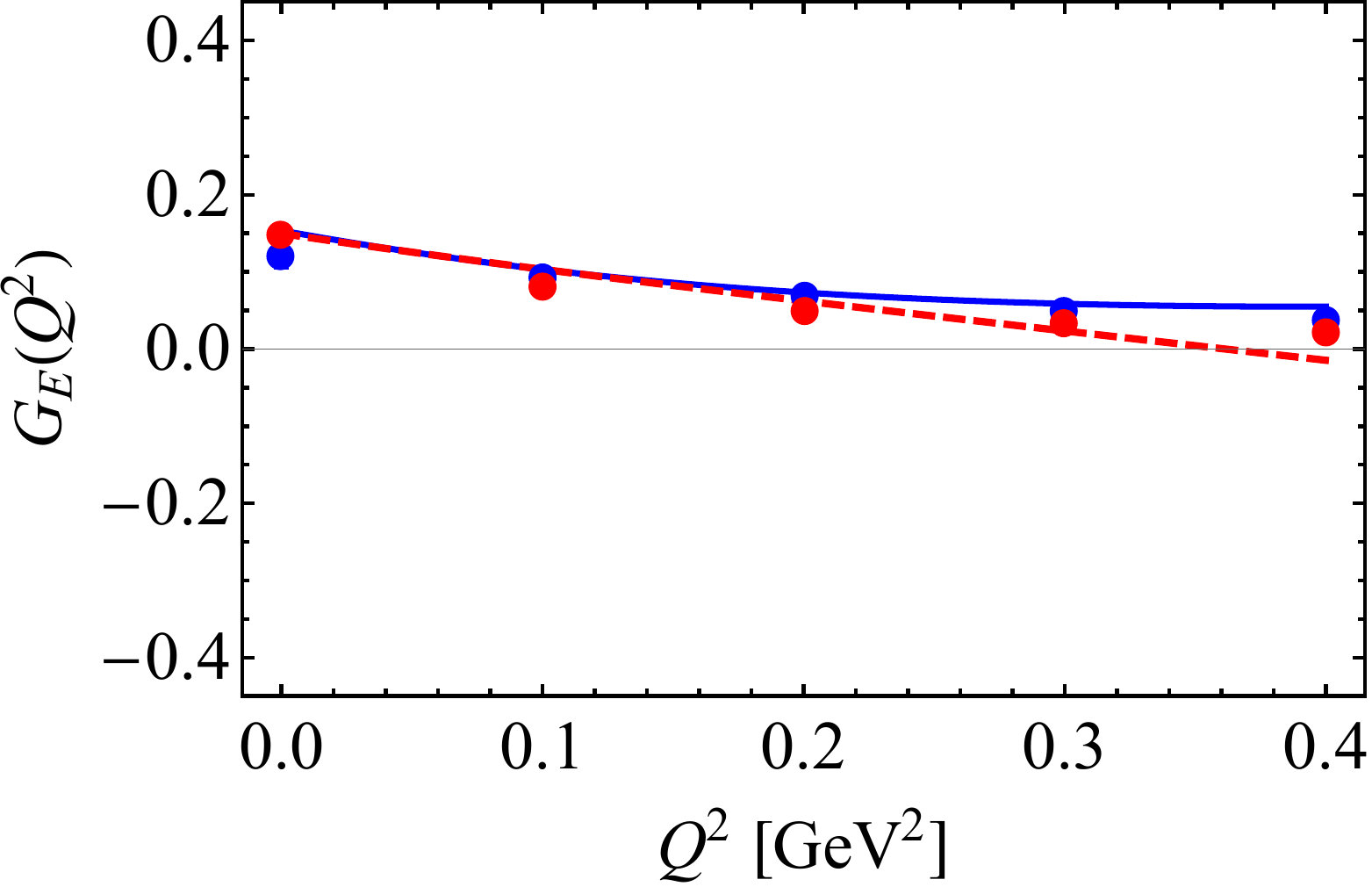}
\hspace*{0.0cm}
\includegraphics[height=3.5cm]{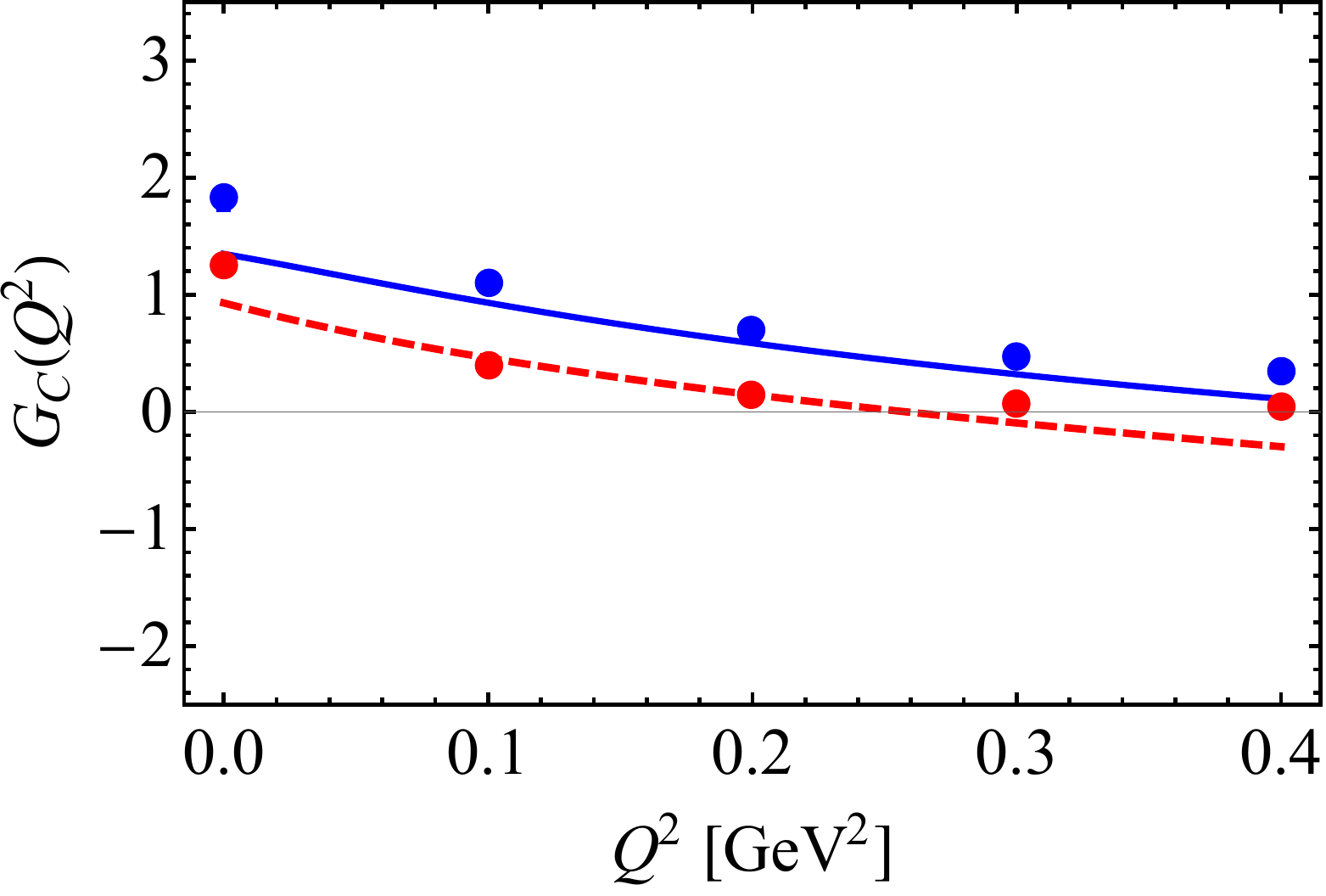}
\vspace*{0.0cm}

\caption{Magnetic, electric, and charge transition form factors
$G_M$, $G_E$, and $G_C$ (Fit III) at the pole position split in tree
diagrams with only photon couplings, tree diagrams with only rho
couplings, loop diagrams and total result.
   The (blue) solid lines and the
blue data points show the real parts and the (red) dashed lines and
the red data points show the imaginary parts of the form factors.
The data points are taken as the averaged MAID and SAID results from
Ref.~\cite{Tiator:2016btt}.}
    \label{fig:tree-loops-all}
\end{figure}

   In Fig.~\ref{fig:tree-loops-all}, we display the individual contributions to the
transition form factors at the pole position for the calculation including the
$\rho$ meson (Fit III).
   The left, middle, and right columns refer to $G_M$, $G_E$, and $G_C$, respectively.
   The first row shows the contribution of the tree-level diagram $(a)$ of Fig.~\ref{fig:diagrams}
(see App.~\ref{tree1} for the detailed expressions).
   The second row displays the $\rho$-meson contribution of Fig.~\ref{fig:rho}, and
the third row refers to the loop contributions of diagrams $(c)$--$(h)$ of
Fig.~\ref{fig:diagrams}.
   The last row contains the total results, i.e., the sum of the individual contributions.
   In each case, the solid lines refer to the real parts and the dashed lines to the imaginary
parts.
   Comparing the first and second rows, we observe the tendency that the tree-level diagram
$(a)$ of Fig.~\ref{fig:diagrams} and the $\rho$-meson contribution of Fig.~\ref{fig:rho}
add destructively.
   Moreover, the loop contribution is relatively small for $G_M$, but sizeable for $G_E$ and
$G_C$, in particular, for their real parts.

\begin{figure}[htbp]
    \centering
\includegraphics[height=4.5cm]{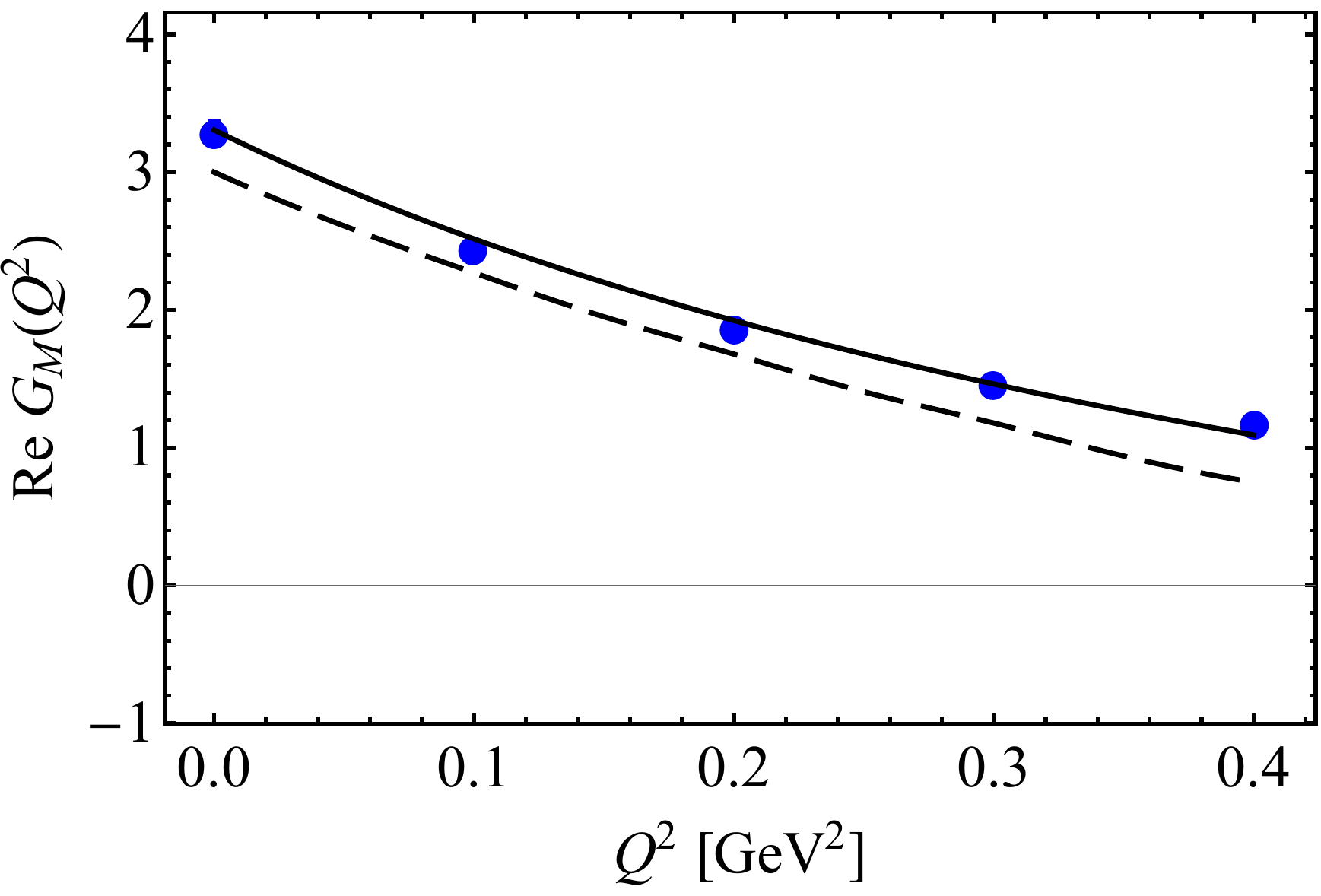}
\hspace*{0.5cm}
\includegraphics[height=4.5cm]{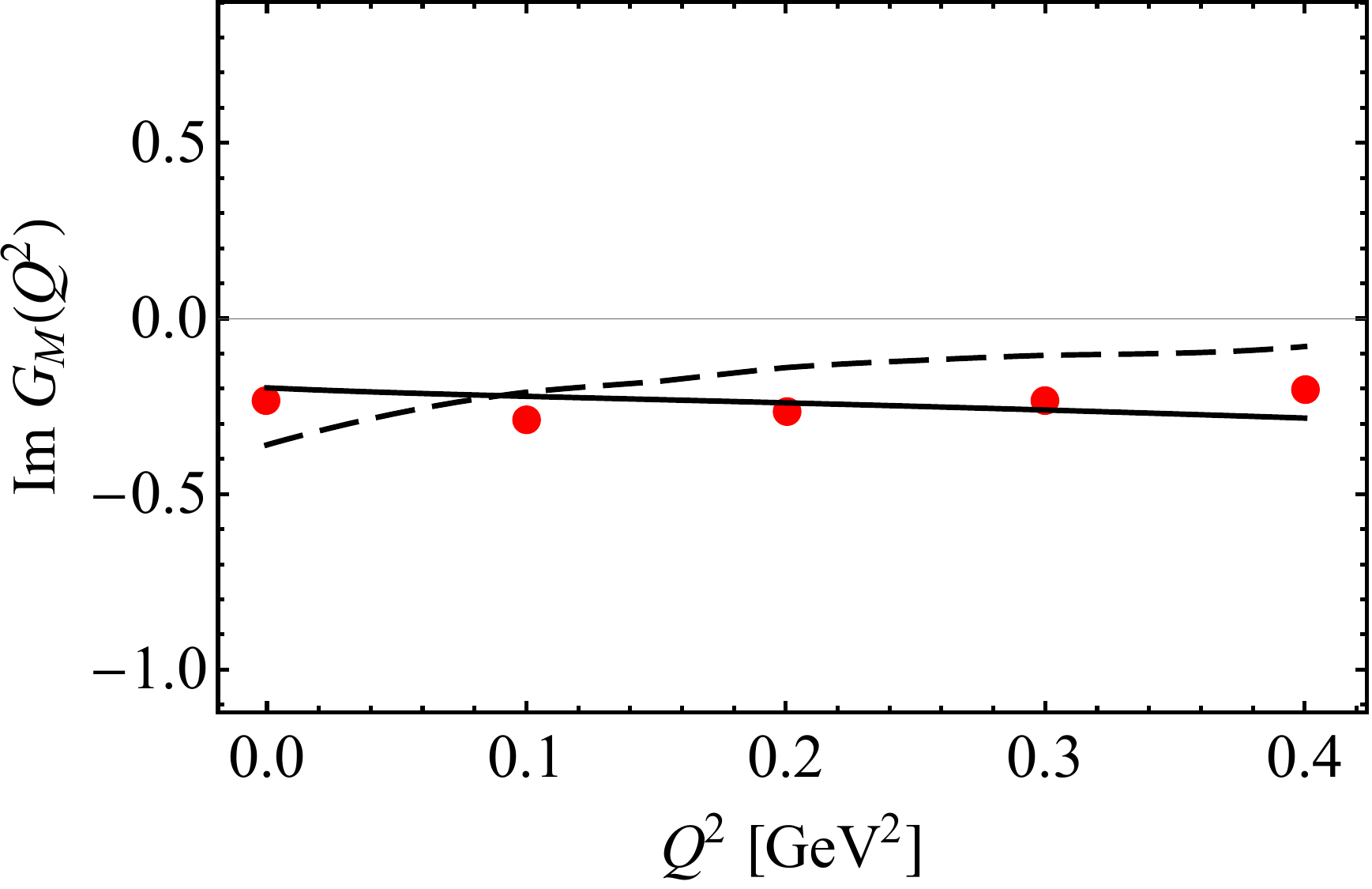}
\vspace*{0.2cm}

\includegraphics[height=4.5cm]{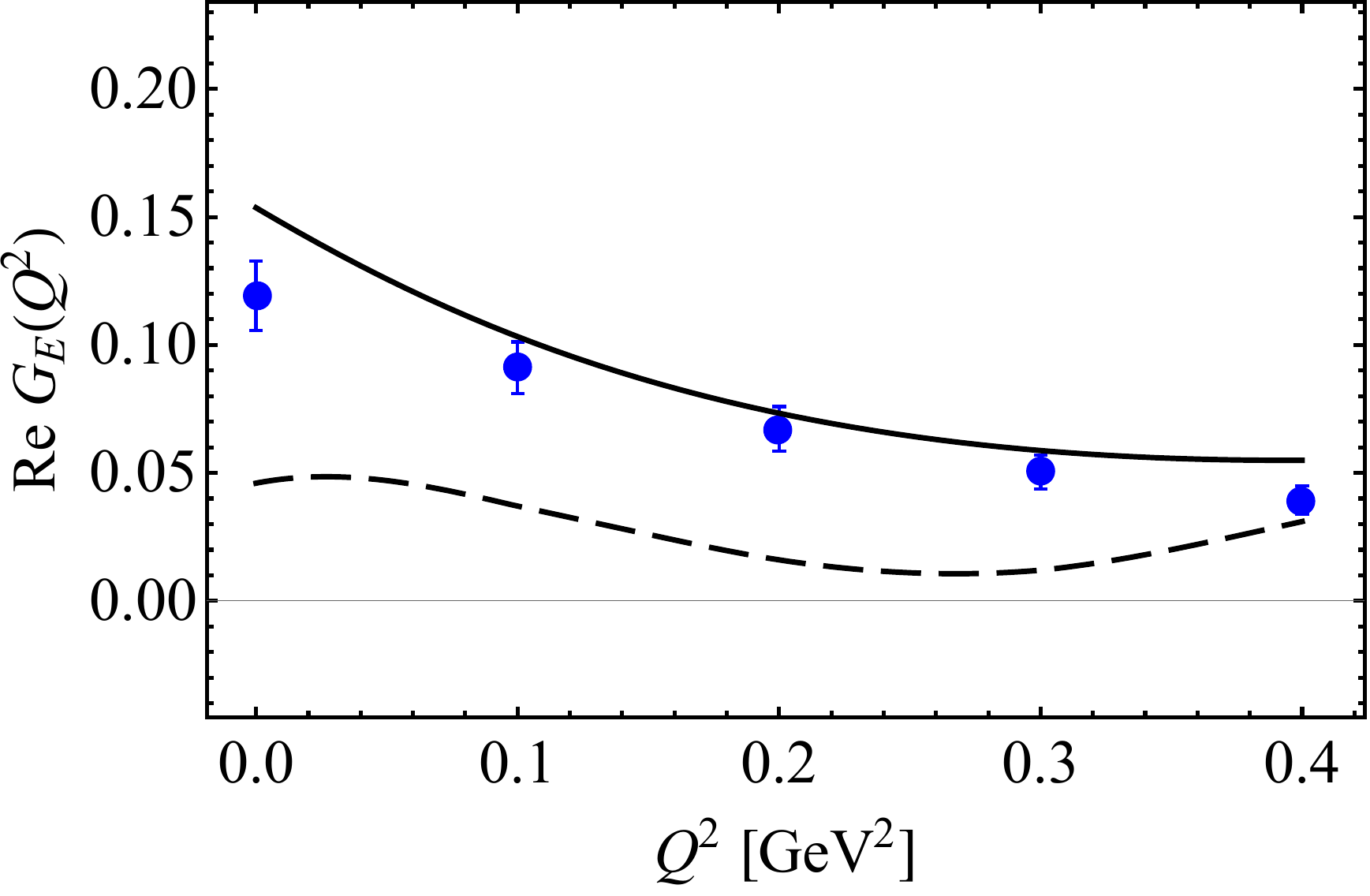}
\hspace*{0.5cm}
\includegraphics[height=4.5cm]{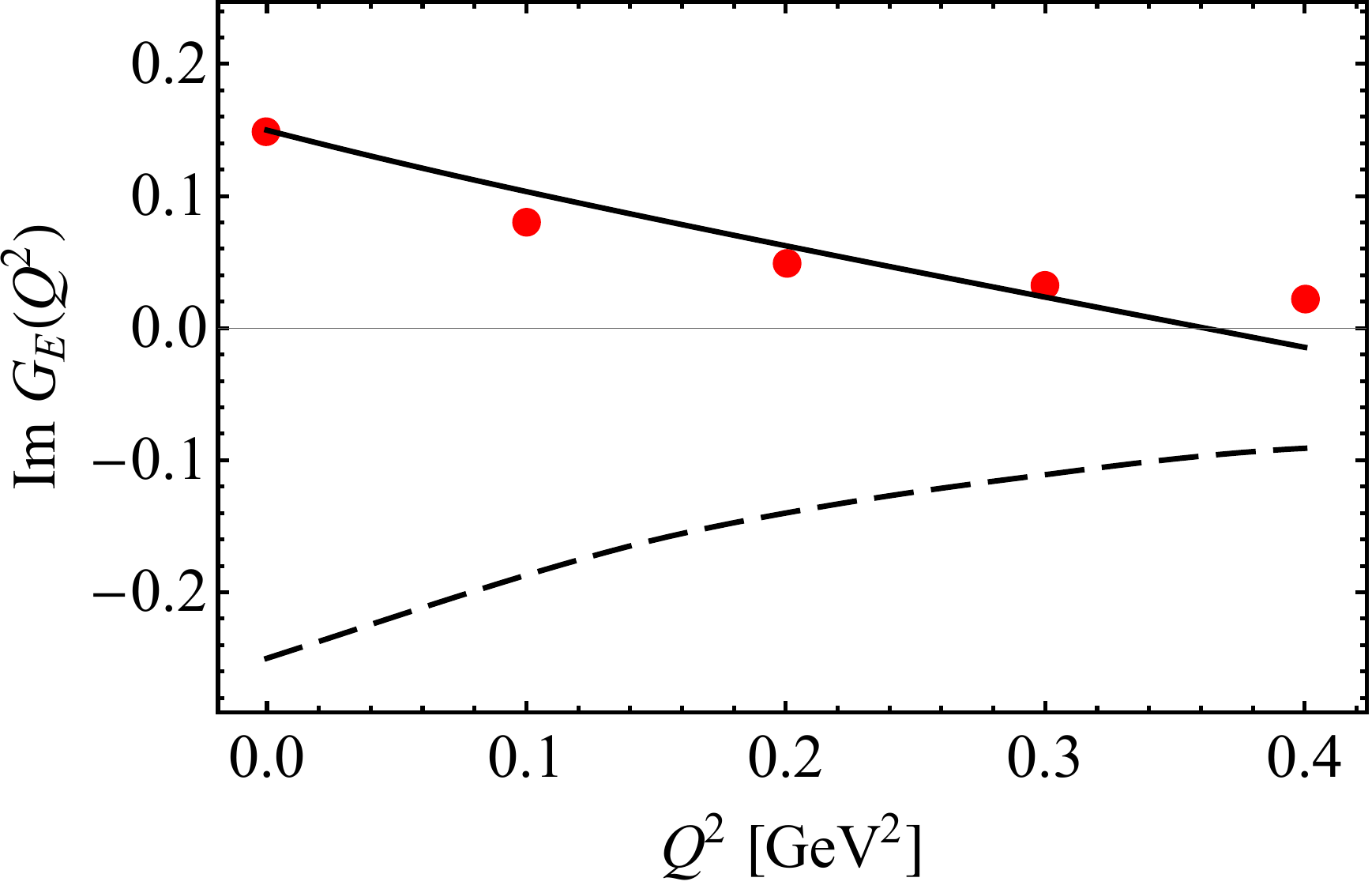}
\vspace*{0.2cm}

\includegraphics[height=4.5cm]{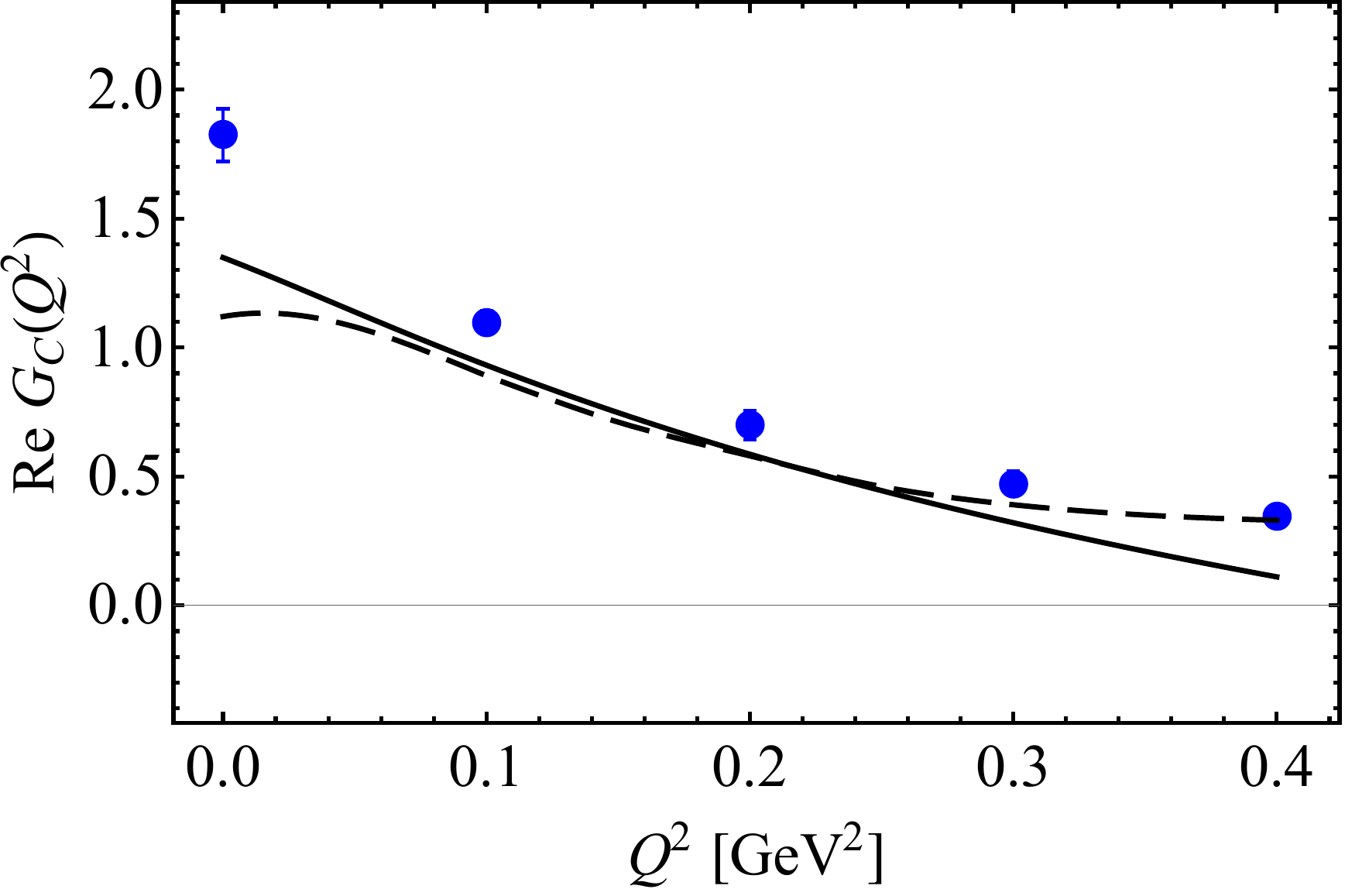}
\hspace*{0.5cm}
\includegraphics[height=4.5cm]{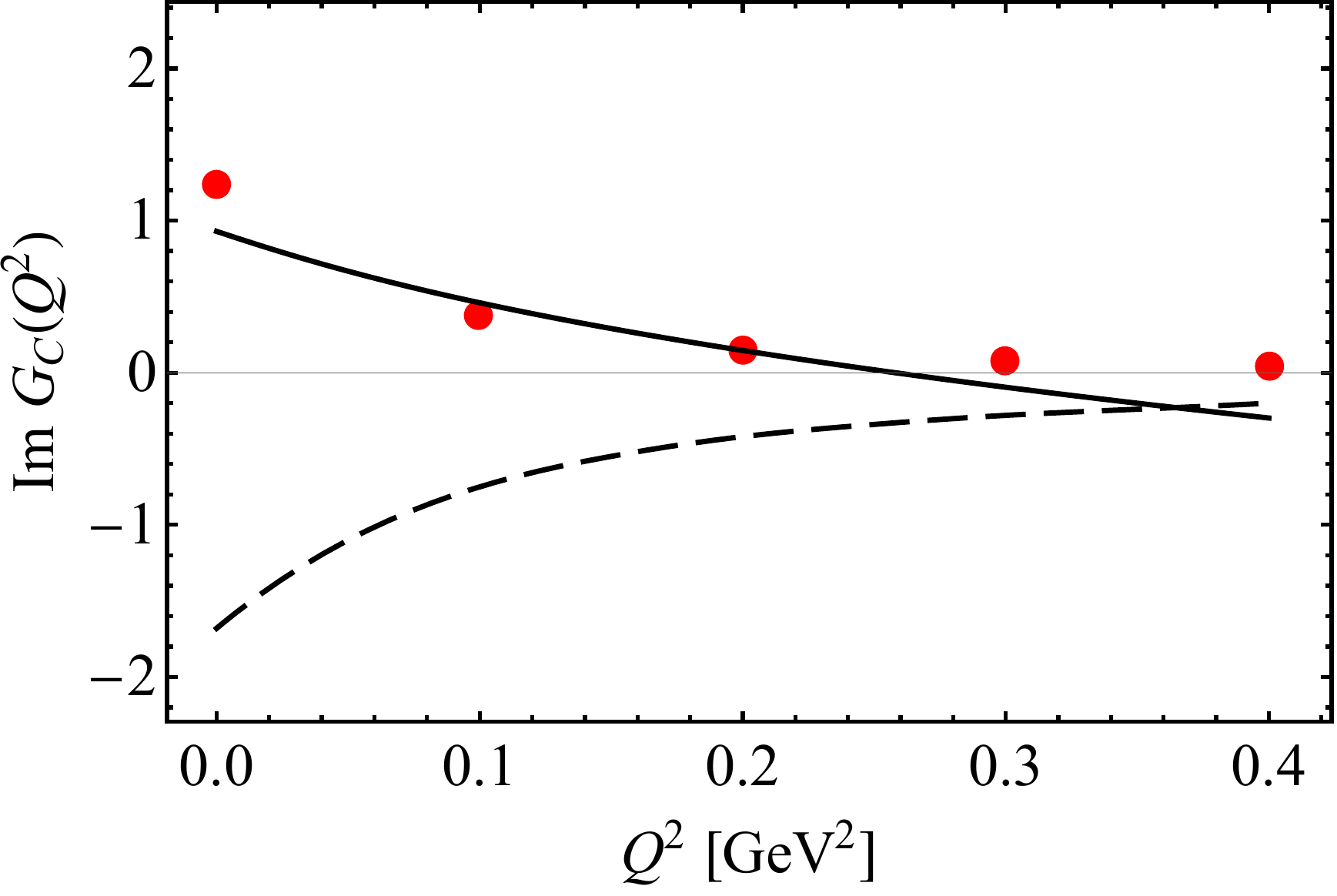}

\caption{Comparison of the transition form factors at the pole
position with heavy-baryon chiral perturbation theory (HBChPT)~\cite{Gail:2005gz}. The real (left)
and imaginary (right) parts of the HBChPT and our calculation are
shown as dashed and solid lines, respectively. The data points are
taken as the averaged MAID and SAID results from
Ref.~\cite{Tiator:2016btt}. }
    \label{fig:comparison-HBChPT}
\end{figure}

   In Fig.~\ref{fig:comparison-HBChPT}, we compare our results
for the transition form factors at the pole position with a calculation
within the framework of heavy-baryon chiral perturbation theory
\cite{Gail:2005gz}.
   The most striking difference consists in the imaginary parts
Im~$G_E$ and Im~$G_C$, because they have opposite signs in the two
calculations.
   In the HBChPT calculation, the imaginary parts originate entirely
from the loop contributions, whereas in our calculation they receive
contributions from all diagrams.
   Nevertheless, also our loop contributions generate in all cases
the opposite sign (see third row of Fig.~\ref{fig:tree-loops-all}).

   Finally, in Table \ref{tab:pole1} we compare our results for the magnetic, electric, and charge
form factors and the ratios $R_{EM}$ and $R_{SM}$ at the real-photon point, $Q^2=0$,
with the MAID and SAID solutions from
Ref.~\cite{Tiator:2016kbr}.
\begin{table}[ht]
\begin{tabular}{|l|rrr|rrr|rrr|}\hline
& \multicolumn{3}{|c|}{MAID} & \multicolumn{3}{|c|}{SAID} & \multicolumn{3}{|c|}{this work}\\
& BW & \multicolumn{2}{c|}{pole} & BW & \multicolumn{2}{c|}{pole} & BW & \multicolumn{2}{c|}{pole}\\
\hline
$G_M$ & $2.97$ & $3.20$ & $-4.7^{\circ} $ & $3.11$ & $3.38$  & $-3.5^{\circ}$ & $2.92$  & $3.31$ & $-3.4^{\circ} $ \\
$G_E$ & $0.064$ & $0.202$ & $49^{\circ} $ & $0.051$& $0.181$ & $54^{\circ}  $ & $0.070$ &$0.215$ & $44^{\circ} $ \\
$G_C$ & $1.18$ & $2.11$ & $35^{\circ} $   & $1.30$ & $2.31$  & $34^{\circ}  $ & $0.82$  & $1.64$ & $35^{\circ} $ \\
\hline
$R_{EM}$ & $-0.022$ & $-0.063$ & $53^{\circ} $ & $-0.016$ & $-0.054$ & $58^{\circ}$ & $-0.024$ & $-0.065$ & $48^{\circ} $ \\
$R_{SM}$ & $-0.042$ & $-0.067$ & $33^{\circ} $ & $-0.044$ & $-0.069$ & $30^{\circ}$ & $-0.029$ & $-0.052$ & $38^{\circ} $ \\
\hline
\end{tabular}
\caption{\label{tab:pole1} Magnetic, electric, and charge transition
form factors and $E/M$, $S/M$ ratios at $Q^2=0$ for the Breit-Wigner
and for the pole position compared with MAID and SAID solutions from
Ref.~\cite{Tiator:2016kbr}. The form factors and ratios are
dimensionless. For the complex values at the pole position, we give
absolute values with the same sign as for the BW values and a phase.
Note that at the leading order, ${\cal O}(q^2)$, we get the
model-independent prediction for the pole ratios
$R_{EM}^{\text{pole}\,(2)}=R_{SM}^{\text{pole}\,(2)}=(-5.98+i\,0.90)\,\%=-0.0605\,e^{-i\,8.6^{\circ}}$.}
\end{table}

\section{Summary and Conclusions}
   We calculated the $\gamma N\rightarrow\Delta$ transition form factors in ChEFT up to and
including chiral order three.
   We made use of a covariant framework and performed the renormalization in terms of the
complex-mass scheme for unstable particles.
   The tree-level contribution at order two was parametrized in terms of the coupling
constant $C_1^\gamma$, whereas at order three two coupling constants $C_2^\gamma$ and
$C_3^\gamma$ enter.
   The coupling constants were fitted to experimental data.
   To improve the description of the form factors, we also investigated the inclusion of
the $\rho$ meson in a semi-phenomenological approach.

   At the leading non-vanishing order, ${\cal O}(q^2)$, the transition form factors
$G_M$, $G_E$, and $G_C$ are proportional to the coupling constant $C_1^\gamma$.
   As a consequence, we obtained a model-independent prediction
for the ratio $R_{EM}(0)$ at the pole, $R_{EM}^{\text{pole}\,(2)}(0)=(-5.98+i\,0.90)\,\%$
[see Eqs.~(\ref{REMO2}) and (\ref{REMO2result})], as well as the relation
$R_{SM}^{\text{pole}\,(2)}(0)=R_{EM}^{\text{pole}\,(2)}(0)$.
   The loop diagrams and the tree-level contributions proportional to
$C_2^{\gamma}$ and $C_3^\gamma$ only enter at ${\cal O}(q^3)$.

   We first discussed the transition form factors $G_M^*$, $G_E^*$, and $G_C^*$
at the Breit-Wigner position $W=m_\Delta=1232$~MeV.
   The unknown parameters of the tree-level diagrams were determined from a simultaneous
fit of all available experimental data for $Q^2\leq 0.3\ \textnormal{GeV}^2$.
   Only after including the $\rho$ meson, we obtained a good description of the data
(see Fig.~\ref{fig:gmfit2}).
   We explicitly verified the Siegert theorem within our calculation and showed that
the prediction of the Siegert limit, Eq.~(\ref{siegertlimit}), provides a good
description of $R_{SM}$ close to the pseudo-threshold (see Fig.~\ref{fig:remfit2Siegert}).
   We then turned to a discussion of the pole transition form factors $G_M$, $G_E$,
and $G_C$ by fitting our results to data of the SAID and MAID partial wave analysis,
which were obtained by applying the Laurent-Pietarinen
expansion method (see Fig.~\ref{fig:gmec-cms}).
   We analyzed the individual contributions originating from the tree-level, $\rho$-meson, and loop diagrams
(see Fig.~\ref{fig:tree-loops-all}).
   Finally, we compared our results with a determination in the framework of
heavy-baryon chiral perturbation theory (see Fig.~\ref{fig:comparison-HBChPT}).
   In conclusion, the CMS is well suited for further examinations of properties of
unstable particles in the framework of chiral effective field theory.

\subsection*{Acknowledgments}
   The authors would like to thank J. Gegelia for useful discussions.
   M.~H.~would like to thank D.~Djukanovic for providing several program routines.
   This work was supported by the Deutsche Forschungsgemeinschaft (SCHE459/4-1).

\begin{appendix}
\section{Relation between the form factors $G_M$, $G_E$, $G_C$ and $G_1$, $G_2$, $G_3$}
\label{relation_form_factors}
   Using Eqs.~(\ref{dirac}) and (\ref{rarita_schwinger}) in combination with
$\gamma^\mu\gamma^\nu+\gamma^\nu\gamma^\mu=2g^{\mu\nu}$ and $\epsilon^{\mu\nu\rho\sigma}\epsilon_{\mu}\,\!^{\nu'\rho'\sigma'}=-\text{det}(g^{\alpha\alpha'})$,
$\alpha=\nu,\rho,\sigma$, $\alpha'=\nu',\rho',\sigma'$, one obtains\footnote{In order to show
Eq.~(\ref{ePqL1L2L3a}), one makes use of
the relation \cite{Scadron:2007qd}
\begin{displaymath}
\gamma^\mu\gamma^\nu\gamma^\rho\gamma^\sigma=g^{\mu\nu}\gamma^\rho\gamma^\sigma-g^{\mu\rho}\gamma^\nu\gamma^\sigma
+g^{\nu\rho}\gamma^\mu\gamma^\sigma+g^{\rho\sigma}\gamma^\mu\gamma^\nu-g^{\nu\sigma}\gamma^\mu\gamma^\rho
+g^{\mu\sigma}\gamma^\nu\gamma^\rho-g^{\mu\nu}g^{\rho\sigma}+g^{\mu\rho}g^{\nu\sigma}-g^{\mu\sigma}g^{\nu\rho}
+i\gamma^5 \epsilon^{\mu\nu\rho\sigma}.
\end{displaymath}
}
\begin{align}
\label{ePqL1L2L3a}
\epsilon^{\lambda\mu}(P,q)&=-z_\Delta L_1^{\lambda\mu}+L_2^{\lambda\mu}+\frac{1}{2}L_3^{\lambda\mu},\\
\label{ePqL1L2L3b}
i\epsilon^\lambda{}_\sigma(P,q)\epsilon^{\mu\sigma}(p_f,q)\gamma_5&=-q\cdot p_f L^{\lambda\mu}_2+P\cdot p_f L^{\lambda\mu}_3,\\
\label{ePqL1L2L3c}
iq^\lambda(q^2P^\mu-q\cdot Pq^\mu)\gamma_5&=q^2L_2^{\lambda\mu}-q\cdot PL_3^{\lambda\mu}.
\end{align}
   Introducing column vectors
\begin{displaymath}
K^{\lambda\mu}=\begin{pmatrix}K^{\lambda\mu}_M\\K^{\lambda\mu}_E\\K^{\lambda\mu}_C\end{pmatrix},\quad
L^{\lambda\mu}=\begin{pmatrix}L^{\lambda\mu}_1\\L^{\lambda\mu}_2\\L^{\lambda\mu}_3\end{pmatrix},\quad
G_{M,E,C}=\begin{pmatrix}G_M\\G_E\\G_C\end{pmatrix},\quad
G=\begin{pmatrix}G_1\\G_2\\G_3\end{pmatrix},
\end{displaymath}
and using Eqs.~(\ref{ePqL1L2L3a})--(\ref{ePqL1L2L3c}), we can write
\begin{displaymath}
K^{\lambda\mu}= M L^{\lambda\mu},
\end{displaymath}
where the $(3\times 3)$ matrix $M$ is given by
\begin{equation}
M=-3[(z_\Delta+m_N)^2-q^2]^{-1}\frac{z_\Delta+m_N}{2m_N}
\begin{pmatrix}
-z_\Delta&1&\frac{1}{2}\\
z_\Delta&-1-4\frac{q\cdot p_f[(z_\Delta+m_N)^2-q^2]}{\Delta(q^2)}&
-\frac{1}{2}+4\frac{P\cdot p_f[(z_\Delta+m_N)^2-q^2]}{\Delta(q^2)}\\
0&2\frac{q^2[(z_\Delta+m_N)^2-q^2]}{\Delta(q^2)}&-2\frac{q\cdot P[(z_\Delta+m_N)^2-q^2]}{\Delta(q^2)}
\end{pmatrix}.
\end{equation}
   The magnetic dipole, electric quadrupole, and
Coulomb quadrupole form factors are then determined from
\begin{displaymath}
G_{M,E,C}={M^{-1}}^T G.
\end{displaymath}
   Using
\begin{displaymath}
q\cdot p_f=\frac{1}{2}(z_\Delta^2-m_N^2+q^2),\quad
q\cdot P=\frac{1}{2}(z_\Delta^2-m_N^2),\quad
P\cdot p_f=\frac{1}{4}(3z_\Delta^2+m_N^2-q^2),
\end{displaymath}
together with Eq.~(\ref{Deltaq2}), results in Eqs.~(\ref{eqn:formfactorsGM})--(\ref{eqn:formfactorsGC}).

\section{Parametrization of the tree-level contribution to the form factors \label{tree}}
\subsection{$\gamma N\Delta$ coupling}
\label{tree1}
   We first want to parametrize the contribution of the $\gamma N\Delta$ interaction Lagrangian of chiral order two and three
to the form factors at tree level.
   As mentioned before, at chiral order one there is no contribution of the Lagrangian.
   As we know the Lorentz structure of the process [see Eqs.~(\ref{eqn:ffmatrix}) and (\ref{eqn:lorentzstructure})],
we assign a chiral order to the tensors and expand the invariant amplitudes.
   Using $\gamma_\mu\gamma_5={\cal O}(q^0)$ and $\gamma_5={\cal O}(q)$ (see, e.g., section 5.2 of Ref.~\cite{Scherer:2002tk})
and the fact that the polarization vector counts as ${\cal O}(q)$, we may parametrize the virtual-photon structure
of the tree-level results as
\begin{eqnarray}
\Gamma^{\lambda\mu(2)}_{\text{tree}}&=&i\left(D_1^{(2)}g^{\lambda\mu}+D_4^{(2)}q^\lambda\gamma^\mu\right)\gamma_5,\nonumber\\
\Gamma^{\lambda\mu(3)}_{\text{tree}}&=&i\left(D_1^{(3)}p_i\cdot q g^{\lambda\mu}+D_2^{(3)}q^\lambda p_i^\mu
+D_4^{(3)}p_i\cdot q q^\lambda\gamma^\mu\right)\gamma_5.
\end{eqnarray}
   The superscripts refer to the chiral order we assign to these expressions.
   Imposing current conservation,  Eq.~(\ref{eqn:cc}), and renaming $C_1^\gamma=D^{(2)}_4$, $C_2^\gamma=D_2^{(3)}$, and $C_3^\gamma=D^{(3)}_4$,
one ends up with the following result:
\begin{align}
\Gamma^{\lambda\mu(2+3)}_{\text{tree}}&=i\left\{\left[-(z_\Delta+m_N)C_1^\gamma-p_i\cdot q C_2^\gamma-p_i\cdot q(z_\Delta+m_N)C_3^\gamma\right]g^{\lambda\mu}\right.\nonumber\\
&\quad\left.
+\left(C_1^\gamma+p_i\cdot q C_3^\gamma\right)q^\lambda\gamma^\mu+C_2^\gamma q^\lambda p_i^\mu\right\}\gamma_5.
\label{Gamma_lambda_mu_2+3}
\end{align}
   Using $p_i\cdot q=(z_\Delta^2-m_N^2-q^2)/2$, the tree-level contributions to the form factors $G_i$
of Eq.~(\ref{eqn:mathews}) read
\begin{equation}
\begin{split}
G_1&=C_1^\gamma+\frac{1}{2}(z_\Delta^2-m_N^2-q^2)C_3^\gamma,\\
G_2&=C_2^\gamma,\\
G_3&=-\frac{1}{2}C^\gamma_2.
\end{split}
\label{Gitree}
\end{equation}
   Introducing dimensionless coupling constants as $\bar{C}_i^\gamma=m_N^i C_i^\gamma$ ($i=1,2,3$)
and using Eqs.~(\ref{eqn:formfactorsGM})--(\ref{eqn:formfactorsGC}),
we obtain the following tree-level contributions to the magnetic dipole, electric quadrupole,
and Coulomb quadrupole form factors:
\begin{align}
\label{eqn:formfactorsGM_tree}
G_M^{\text{tree}}&=\frac{m_N}{3(z_\Delta+m_N)}\left[\frac{(3z_\Delta+m_N)(z_\Delta+m_N)-q^2}{z_\Delta m_N}
\left(\bar{C}_1^\gamma+\frac{1}{2}\frac{z_\Delta^2-m_N^2-q^2}{m_N^2}\bar{C}_3^\gamma\right)\right.\nonumber\\
&\left.\quad+\frac{z_\Delta^2-m_N^2-q^2}{m_N^2}\bar{C}_2^\gamma\right],\\
\label{eqn:formfactorsGE_tree}
G_E^{\text{tree}}&=\frac{m_N}{3(z_\Delta+m_N)}\left[\frac{z_\Delta^2-m_N^2+q^2}{z_\Delta m_N}
\left(\bar{C}_1^\gamma+\frac{1}{2}\frac{z_\Delta^2-m_N^2-q^2}{m_N^2}\bar{C}_3^\gamma\right)
+\frac{z_\Delta^2-m_N^2-q^2}{m_N^2}\bar{C}_2^\gamma\right],\\
\label{eqn:formfactorsGC_tree}
G_C^{\text{tree}}&=\frac{m_N}{3(z_\Delta+m_N)}
\left[4\frac{z_\Delta}{m_N}\left(\bar{C}_1^\gamma+\frac{1}{2}\frac{z_\Delta^2-m_N^2-q^2}{m_N^2}\bar{C}_3^\gamma\right)
+2 \frac{z_\Delta^2+m_N^2-q^2}{m_N^2}\bar{C}_2^\gamma\right].
\end{align}

\subsection{Contribution of the $\rho$ meson at tree level}
\label{tree2}
   To describe the diagram of Fig.\ \ref{fig:rho}, we start with the assumption that the coupling of
the $\rho$ to the $N\Delta$ transition is of the same type as the coupling of the $\gamma$ to the $N\Delta$ transition
[see Eq.~(\ref{Gamma_lambda_mu_2+3})].
   We denote the corresponding coupling constants by $C_i^\rho$.
   The $\gamma\rho$ coupling is obtained from the Lagrangian \cite{Weinberg:1968de,Ecker:1989yg,Bauer:2012pv}
\begin{equation}
M_\rho^2\text{Tr}\left[\left(\rho_\mu-\frac{i}{g}\Gamma_\mu\right)\left(\rho^\mu-\frac{i}{g}\Gamma^\mu\right)\right]
\end{equation}
as
\begin{equation}
{\cal L}_{\gamma\rho}=e\frac{M_\rho^2}{g}{\cal A}^\mu\rho^0_\mu.
\end{equation}
   The coupling constant $g$ is determined from the Kawarabayashi-Suzuki-Riazuddin-Fayyazuddin relation \cite{Kawarabayashi:1966kd,Riazuddin:1966sw},
\begin{equation}
M_\rho^2=2g^2F^2.
\label{ksrf}
\end{equation}
   Combining the $\gamma\rho$ vertex with the $\rho$ propagator yields
\begin{displaymath}
e\epsilon_\mu\frac{M_\rho^2}{g(q^2-M_\rho^2)},
\end{displaymath}
which is of ${\cal O}(q)$.
   Contraction with the $\rho N\Delta$ vertex amounts to the replacement
\begin{equation}
C_i^\gamma\to C_i^\gamma+\frac{M_\rho^2}{g(q^2-M_\rho^2)}C_i^\rho=C_i^\gamma-\frac{C_i^\rho}{g}-\frac{q^2}{g(M_\rho^2-q^2)}C_i^\rho
\label{Cireplacement}
\end{equation}
in Eqs.~(\ref{Gitree}).
   Note that the $d_x$ term of Eq.~(13) of Ref.~\cite{Bauer:2012pv} is of ${\cal O}(q^3)$ and, thus, will start contributing
at ${\cal O}(q^4)$ to the $\gamma N\Delta$ transition.

\subsection{Power-counting-violating contribution}
   The constant $C_1^\gamma$ has to absorb a part from the loop diagrams which violates the power counting.
   Only after renormalization of this constant the counting scheme is consistent.
   For the renormalized constant $C_{1r}^\gamma$ we obtain
\begin{equation}
C_1^\gamma\rightarrow C_{1r}^\gamma=C_1^\gamma+\frac{\texttt{g}\, m_N}{31104 \pi F^2}
\left[\ln\left(\frac{m_N}{\mu}\right)(648\,\texttt{g}_A+1980\,\texttt{g}_1)-324\,\texttt{g}_A-1135\,\texttt{g}_1\right].
\end{equation}

\end{appendix}


\begin{thebibliography}{100}

\bibitem{Anderson:1952nw}
  H.~L.~Anderson, E.~Fermi, E.~A.~Long, and D.~E.~Nagle,
  Phys.\ Rev.\ {\bf 85}, 936 (1952).

\bibitem{Agashe:2014kda}
  K.~A.~Olive {\it et al.} [Particle Data Group Collaboration],
  Chin.\ Phys.\ C {\bf 38}, 090001 (2014).


\bibitem{Bartel:1968tw}
  W.~Bartel, B.~Dudelzak, H.~Krehbiel, J.~McElroy, U.~Meyer-Berkhout, W.~Schmidt, V.~Walther, and G.~Weber,
  Phys.\ Lett.\  {\bf 28B}, 148 (1968).

\bibitem{Baetzner:1972bg}
  K.~Baetzner {\it et al.},
  Phys.\ Lett.\  B {\bf 39}, 575 (1972).

\bibitem{Stein:1975yy}
  S.~Stein {\it et al.},
  Phys.\ Rev.\  D {\bf 12}, 1884 (1975).

\bibitem{Beck:1999ge}
  R.~Beck {\it et al.},
  Phys.\ Rev.\  C {\bf 61}, 035204 (2000).

\bibitem{Pospischil:2000ad}
  T.~Pospischil {\it et al.},
  Phys.\ Rev.\ Lett.\  {\bf 86}, 2959 (2001).

\bibitem{Mertz:1999hp}
  C.~Mertz {\it et al.},
  Phys.\ Rev.\ Lett.\  {\bf 86}, 2963 (2001).

\bibitem{Joo:2001tw}
  K.~Joo {\it et al.}  [CLAS Collaboration],
  Phys.\ Rev.\ Lett.\  {\bf 88}, 122001 (2002).

\bibitem{Sparveris:2004jn}
  N.~F.~Sparveris {\it et al.}  [OOPS Collaboration],
  Phys.\ Rev.\ Lett.\  {\bf 94}, 022003 (2005).

\bibitem{Elsner:2005cz}
  D.~Elsner {\it et al.},
  Eur.\ Phys.\ J.\ A {\bf 27}, 91 (2006).

\bibitem{Kelly:2005jy}
  J.~J.~Kelly {\it et al.},
  Phys.\ Rev.\ C {\bf 75}, 025201 (2007).

\bibitem{Stave:2008aa}
  S.~Stave {\it et al.} [A1 Collaboration],
  Phys.\ Rev.\ C {\bf 78}, 025209 (2008).

\bibitem{Aznauryan:2009mx}
  I.~G.~Aznauryan {\it et al.} [CLAS Collaboration],
  Phys.\ Rev.\ C {\bf 80}, 055203 (2009).

\bibitem{Blomberg:2015zma}
  A.~Blomberg {\it et al.},
  Phys.\ Lett.\ B {\bf 760}, 267 (2016).


\bibitem{Dufner:1967yj}
  A.~J.~Dufner and Y.~S.~Tsai,
  Phys.\ Rev.\  {\bf 168}, 1801 (1968).

\bibitem{Jones:1972ky}
  H.~F.~Jones and M.~D.~Scadron,
  Annals Phys.\  {\bf 81}, 1 (1973).

\bibitem{Davidson:1985wb}
  R.~Davidson, N.~C.~Mukhopadhyay, and R.~Wittman,
  Phys.\ Rev.\ Lett.\  {\bf 56}, 804 (1986).

\bibitem{Wirzba:1986sc}
  A.~Wirzba and W.~Weise,
  Phys.\ Lett.\ B {\bf 188}, 6 (1987).

\bibitem{Bermuth:1988ms}
  K.~Bermuth, D.~Drechsel, L.~Tiator, and J.~B.~Seaborn,
  Phys.\ Rev.\ D {\bf 37}, 89 (1988).

\bibitem{Leinweber:1992pv}
  D.~B.~Leinweber, T.~Draper, and R.~M.~Woloshyn,
  Phys.\ Rev.\ D {\bf 48}, 2230 (1993).

\bibitem{Butler:1993ht}
  M.~N.~Butler, M.~J.~Savage, and R.~P.~Springer,
  Phys.\ Lett.\ B {\bf 304}, 353 (1993).

\bibitem{Cardarelli:1995ug}
  F.~Cardarelli, E.~Pace, G.~Salme, and S.~Simula,
  Phys.\ Lett.\  B {\bf 371}, 7 (1996).

\bibitem{Buchmann:1996bd}
  A.~J.~Buchmann, E.~Hernandez, and A.~Faessler,
  Phys.\ Rev.\ C {\bf 55}, 448 (1997).

\bibitem{Lu:1996rj}
  D.~H.~Lu, A.~W.~Thomas, and A.~G.~Williams,
  Phys.\ Rev.\  C {\bf 55}, 3108 (1997).

\bibitem{Gellas:1998wx}
  G.~C.~Gellas, T.~R.~Hemmert, C.~N.~Ktorides, and G.~I.~Poulis,
  Phys.\ Rev.\ D {\bf 60}, 054022 (1999).

\bibitem{Tiator:2003xr}
  L.~Tiator, D.~Drechsel, S.~S.~Kamalov, and S.~N.~Yang,
  Eur.\ Phys.\ J.\ A {\bf 17}, 357 (2003).

\bibitem{Tiator:2003uu}
  L.~Tiator, D.~Drechsel, S.~Kamalov, M.~M.~Giannini, E.~Santopinto, and A.~Vassallo,
  Eur.\ Phys.\ J.\ A {\bf 19}, 55 (2004).

\bibitem{Alexandrou:2004xn}
  C.~Alexandrou, Ph.~de Forcrand, H.~Neff, J.~W.~Negele, W.~Schroers, and A.~Tsapalis,
  Phys.\ Rev.\ Lett.\  {\bf 94}, 021601 (2005).

\bibitem{Pascalutsa:2005ts}
  V.~Pascalutsa and M.~Vanderhaeghen,
  Phys.\ Rev.\ Lett.\  {\bf 95}, 232001 (2005).

\bibitem{Gail:2005gz}
  T.~A.~Gail and T.~R.~Hemmert,
  Eur.\ Phys.\ J.\  A {\bf 28}, 91 (2006).

\bibitem{Braun:2005be}
  V.~M.~Braun, A.~Lenz, G.~Peters, and A.~V.~Radyushkin,
  Phys.\ Rev.\  D {\bf 73}, 034020 (2006).

\bibitem{Pascalutsa:2006up}
  V.~Pascalutsa, M.~Vanderhaeghen, and S.~N.~Yang,
  Phys.\ Rept.\  {\bf 437}, 125 (2007).

\bibitem{Ramalho:2008dp}
  G.~Ramalho, M.~T.~Pena, and F.~Gross,
  Phys.\ Rev.\ D {\bf 78}, 114017 (2008).

\bibitem{Alexandrou:2010uk}
  C.~Alexandrou, G.~Koutsou, J.~W.~Negele, Y.~Proestos, and A.~Tsapalis,
  Phys.\ Rev.\ D {\bf 83}, 014501 (2011).

\bibitem{Tiator:2016btt}
  L.~Tiator, M.~D\"oring, R.~L.~Workman, M.~Had\v{z}imehmedovic, H.~Osmanovic, R.~Omerovic, J.~Stahov, and A.~\v{S}varc,
  Phys.\ Rev.\ C {\bf 94}, 065204 (2016).



\bibitem{Tiator:2011pw}
  L.~Tiator, D.~Drechsel, S.~S.~Kamalov, and M.~Vanderhaeghen,
  Eur.\ Phys.\ J.\ ST {\bf 198}, 141 (2011).

\bibitem{Aznauryan:2011qj}
  I.~G.~Aznauryan and V.~D.~Burkert,
  Prog.\ Part.\ Nucl.\ Phys.\  {\bf 67}, 1 (2012).


\bibitem{Gegelia:2009py}
  J.~Gegelia and S.~Scherer,
  Eur.\ Phys.\ J.\ A {\bf 44}, 425 (2010).

\bibitem{Hacker:2005fh}
  C.~Hacker, N.~Wies, J.~Gegelia, and S.~Scherer,
  Phys.\ Rev.\  C {\bf 72}, 055203 (2005).

\bibitem{Wies:2006rv}
  N.~Wies, J.~Gegelia, and S.~Scherer,
  Phys.\ Rev.\  D {\bf 73}, 094012 (2006).

\bibitem{Stuart:1990}
R.~G.~Stuart, {\it Pitfalls of radiative corrections near a resonance},
in ${\rm Z}^0$ {\it Physics}, edited by J.~Tran Thanh Van
(Editions Fronti\`eres, Gif-sur-Yvette, 1990) p.\ 41.

\bibitem{Denner:1999gp}
  A.~Denner, S.~Dittmaier, M.~Roth, and D.~Wackeroth,
  Nucl.\ Phys.\ B {\bf 560}, 33 (1999).

\bibitem{Denner:2006ic}
  A.~Denner and S.~Dittmaier,
  Nucl.\ Phys.\ Proc.\ Suppl.\  {\bf 160}, 22 (2006).

\bibitem{Actis:2006rc}
  S.~Actis and G.~Passarino,
  Nucl.\ Phys.\ B {\bf 777}, 100 (2007).

\bibitem{Actis:2008uh}
  S.~Actis, G.~Passarino, C.~Sturm, and S.~Uccirati,
  Phys.\ Lett.\ B {\bf 669}, 62 (2008).

\bibitem{Hemmert:1997ye}
  T.~R.~Hemmert, B.~R.~Holstein, and J.~Kambor,
  J.\ Phys.\ G {\bf 24}, 1831 (1998).

\bibitem{Djukanovic:2009zn}
  D.~Djukanovic, J.~Gegelia, A.~Keller, and S.~Scherer,
  Phys.\ Lett.\  B {\bf 680}, 235 (2009).

\bibitem{Djukanovic:2009gt}
  D.~Djukanovic, J.~Gegelia, and S.~Scherer,
  Phys.\ Lett.\  B {\bf 690}, 123 (2010).

\bibitem{Bauer:2012at}
  T.~Bauer, J.~Gegelia, and S.~Scherer,
  Phys.\ Lett.\ B {\bf 715}, 234 (2014).

\bibitem{Djukanovic:2013mka}
  D.~Djukanovic, E.~Epelbaum, J.~Gegelia, and U.-G.~Mei{\ss}ner,
  Phys.\ Lett.\ B {\bf 730}, 115 (2014).

\bibitem{Bauer:2014cqa}
  T.~Bauer, S.~Scherer, and L.~Tiator,
  Phys.\ Rev.\ C {\bf 90}, 015201 (2014).

\bibitem{Djukanovic:2014rua}
  D.~Djukanovic, J.~Gegelia, A.~Keller, S.~Scherer, and L.~Tiator,
  Phys.\ Lett.\ B {\bf 742}, 55 (2015).

\bibitem{Djukanovic:2015gna}
  D.~Djukanovic, E.~Epelbaum, J.~Gegelia, H.~Krebs, and U.-G.~Mei{\ss}ner,
  Eur.\ Phys.\ J.\ A {\bf 51}, 101 (2015).

\bibitem{Epelbaum:2015vea}
  E.~Epelbaum, J.~Gegelia, U.-G.~Mei{\ss}ner, and D.~L.~Yao,
  Eur.\ Phys.\ J.\ C {\bf 75}, 499 (2015).

\bibitem{Yao:2016vbz}
  D.~L.~Yao, D.~Siemens, V.~Bernard, E.~Epelbaum, A.~M.~Gasparyan, J.~Gegelia, H.~Krebs and U.-G.~Mei{\ss}ner,
  JHEP {\bf 1605}, 038 (2016).

\bibitem{Gegelia:2016pjm}
  J.~Gegelia, U.-G.~Mei{\ss}ner, D.~Siemens, and D.~L.~Yao,
  Phys.\ Lett.\ B {\bf 763}, 1 (2016).

\bibitem{Bauer:2016czc}
  T.~Bauer, Y.~\"Unal, A.~K\"u\c{c}\"ukarslan, and S.~Scherer,
  Phys.\ Rev.\ C {\bf 96}, 025203 (2017).

\bibitem{Bauer:2012gn}
  T.~Bauer, J.~Gegelia, G.~Japaridze, and S.~Scherer,
  Int.\ J.\ Mod.\ Phys.\ A {\bf 27}, 1250178 (2012).

\bibitem{Denner:2014zga}
  A.~Denner and J.~N.~Lang,
  Eur.\ Phys.\ J.\ C {\bf 75}, 377 (2015).

\bibitem{Hemmert:1997ye}
  T.~R.~Hemmert, B.~R.~Holstein, and J.~Kambor,
  J.\ Phys.\ G {\bf 24}, 1831 (1998).

\bibitem{Dirac}
P.~A.~M.~Dirac, {\it Lectures on Quantum Mechanics}
(Dover, Mineola, New York, 2001).

\bibitem{Gitman:1990qh}
D.~M.~Gitman and I.~V.~Tyutin, {\it Quantization of Fields with Constraints}
(Springer, Berlin, 1990).

\bibitem{teitelboim}
M.\ Henneaux and C.\ Teitelboim, {\it Quantization of Gauge
Systems} (Princeton University Press, Princeton, New Jersey, 1992).

\bibitem{Bjorken}
   J.~D.~Bjorken and S.~D.~Drell, {\it Relativistic quantum fields}
   (McGraw-Hill, New York, 1965) Chap.~16.

\bibitem{Nozawa:1989gy}
  S.~Nozawa and T.-S.~H.~Lee,
  Nucl.\ Phys.\ A {\bf 513}, 511 (1990).

\bibitem{Rarita:1941mf}
  W.~Rarita and J.~Schwinger,
  Phys.\ Rev.\  {\bf 60}, 61 (1941).

\bibitem{Kusaka}
   S.~Kusaka,
   Phys.\ Rev.\ {\bf 60}, 61 (1941).

\bibitem{Mathews:1965zz}
  J.~Mathews,
  Phys.\ Rev.\  {\bf 137}, B444 (1965).

\bibitem{Itzykson:1980rh}
  C.~Itzykson and J.~B.~Zuber,
  {\it Quantum Field Theory}
  (McGraw-Hill, New York, 1980).

\bibitem{Scherer:2002tk}
  S.~Scherer,
  Adv.\ Nucl.\ Phys.\  {\bf 27}, 277 (2003).

\bibitem{Scherer:2012zzd}
  S.~Scherer and M.~R.~Schindler,
  Lect.\ Notes Phys.\  {\bf 830}, 1 (2012).

\bibitem{Gasser:1983yg}
  J.~Gasser and H.~Leutwyler,
  Annals Phys.\  {\bf 158}, 142 (1984).

\bibitem{Colangelo:2001sp}
  G.~Colangelo, J.~Gasser, and H.~Leutwyler,
  Phys.\ Rev.\ Lett.\  {\bf 86}, 5008 (2001).

\bibitem{Gasser:1987rb}
  J.~Gasser, M.~E.~Sainio, and A.~\v{S}varc,
  Nucl.\ Phys.\ B {\bf 307}, 779 (1988).

\bibitem{Moldauer:1956zz}
  P.~A.~Moldauer and K.~M.~Case,
  Phys.\ Rev.\  {\bf 102}, 279 (1956).

\bibitem{Gegelia:1999gf}
  J.~Gegelia and G.~Japaridze,
  Phys.\ Rev.\ D {\bf 60}, 114038 (1999).

\bibitem{Gegelia:1999qt}
  J.~Gegelia, G.~Japaridze, and X.~Q.~Wang,
  J.\ Phys.\ G {\bf 29}, 2303 (2003).

\bibitem{Fuchs:2003qc}
  T.~Fuchs, J.~Gegelia, G.~Japaridze, and S.~Scherer,
  Phys.\ Rev.\ D {\bf 68}, 056005 (2003).

\bibitem{Veltman:1963th}
  M.~J.~G.~Veltman,
  Physica {\bf 29}, 186 (1963).

\bibitem{Weinberg:1991um}
  S.~Weinberg,
  Nucl.\ Phys.\ B {\bf 363}, 3 (1991).

\bibitem{Ecker:1995gg}
  G.~Ecker,
  Prog.\ Part.\ Nucl.\ Phys.\  {\bf 35}, 1 (1995).

\bibitem{Scadron:2007qd}
  M.~D.~Scadron, {\it Advanced Quantum Theory (3rd edition)}
  (World Scientific, Singapore, 2007) p.~70.

\bibitem{Kawarabayashi:1966kd}
  K.~Kawarabayashi and M.~Suzuki,
  Phys.\ Rev.\ Lett.\  {\bf 16}, 255 (1966).

\bibitem{Riazuddin:1966sw}
  Riazuddin and Fayyazuddin,
  Phys.\ Rev.\  {\bf 147}, 1071 (1966).

\bibitem{Drechsel:2007if}
  D.~Drechsel, S.~S.~Kamalov, and L.~Tiator,
  Eur.\ Phys.\ J.\  A {\bf 34}, 69 (2007).

\bibitem{Tiator:2016kbr}
  L.~Tiator,
  Few Body Syst.\  {\bf 57}, no. 11, 1087 (2016).

\bibitem{Ramalho:2016zzo}
  G.~Ramalho,
  Phys.\ Rev.\ D {\bf 93}, no. 11, 113012 (2016).


\bibitem{Weinberg:1968de}
  S.~Weinberg,
  Phys.\ Rev.\  {\bf 166}, 1568 (1968).

\bibitem{Ecker:1989yg}
  G.~Ecker, J.~Gasser, H.~Leutwyler, A.~Pich, and E.~de Rafael,
  Phys.\ Lett.\  B {\bf 223}, 425 (1989).


\bibitem{Bauer:2012pv}
  T.~Bauer, J.~C.~Bernauer, and S.~Scherer,
  Phys.\ Rev.\ C {\bf 86}, 065206 (2012).
\end{thebibliography}
\end{document}